\documentclass{lmcs}
\usepackage[multiple]{footmisc}
\usepackage[T1]{fontenc}
\usepackage{lmodern}
\usepackage{graphicx}
\usepackage{DotArrow}
\usepackage{hyperref}
\usepackage{listings}
\usepackage{stmaryrd}
\usepackage{mathtools,mathpartir}
\usepackage{proof}
\usepackage{amsmath,amssymb,amscd,mathrsfs}
\usepackage{epsfig,color,subfigure,enumitem,soul}

\definecolor{darkgreen}{rgb}{0.1, 0.5, 0.1}

\makeatletter
\newcommand{\raisemath}[1]{\mathpalette{\raisem@th{#1}}}
\newcommand{\raisem@th}[3]{\raisebox{#1}{$#2#3$}}
\makeatother

\newcommand{\shortodot}{\!\odot\!}
\usepackage{macrospNets}

\newcommand{\nounderline}[1]{#1}

\newtheorem{property}{Property}
\newtheorem{example}{Example}

\begin{document}

\title{Compositional equivalences based on Open pNets
}
\author{Rab\'ea Ameur-Boulifa}
\address{LTCI, T\'el\'ecom Paris, Institut Polytechnique de Paris, France
}         
\email{rabea.ameur-boulifa@telecom-paris.fr}     

\author{Ludovic Henrio}
\address{Univ Lyon, EnsL, UCBL, CNRS, Inria,  LIP, F-69342, LYON Cedex 07, France.}
\email{ludovic.henrio@cnrs.fr}

\author{Eric Madelaine}
\address{INRIA Sophia Antipolis M\'edit\'erann\'ee, UCA, BP 93, 06902 Sophia Antipolis, France}
\email{eric.madelaine@inria.fr}


\begin{abstract}

Establishing equivalences between programs or systems is crucial both for verifying correctness of programs, by establishing that two implementations are equivalent, and for justifying optimisations and program transformations, by establishing that a modified program is equivalent to the source one. There exist several equivalence relations for programs, and bisimulations are among the most versatile of these equivalences. Among bisimulation relations one distinguishes strong bisimulation, that requires that each action of a program is simulated by a single action of the equivalent program, a weak bisimulation that  is a coarser relation, allowing some of the actions to be
  invisible or internal moves, and thus not simulated by the equivalent program.

pNet is a generalisation of automata that model open systems. They feature variables and hierarchical composition. Open pNets are pNets with holes, i.e. placeholders inside the hierarchical structure  that can be filled later by sub-systems. 

This article defines bisimulation relations  for the comparison of systems specified as pNets.
 We first define a strong bisimulation for open pNets.
  We then define an equivalence relation similar to
  the classical \emph{weak bisimulation}, and study its properties.
Among these properties we are interested in compositionality: if two systems are proven equivalent they will be undistinguishable by their context, and they will also be undistinguishable when their holes are filled with equivalent systems.
We identify sufficient conditions on the automata to ensure compositionality of strong and weak bisimulation.
The article is illustrated with a transport protocol running example; it shows the characteristics of our formalism and our bisimulation relations.
\end{abstract}

\maketitle
%
%
%
%
%
%
%
%
%
%
%
%
%
%
%
%
%
%
%
%
%
%
%
%
%
%
%
%


\section{Introduction}

In the nineties, several 
works extended the basic behavioural models based on labelled
transition systems to address value-passing or parameterised systems, using
various symbolic encodings of the
transitions~\cite{deSimone85,Larsen87}. 
These works use the term \emph{parameter} to designate systems where variables that have a strong influence the system structure and behaviour. In parameterised systems, parameters can typically be the number of processes in the system or the way they interact.
In \cite{IngolfsdottirL:2001,HennessyLin:TCS95}, Lin, Ingolfsdottir and Hennessy developed a full hierarchy of
bisimulation equivalences, together with a proof system, for value passing CCS,
including notions of symbolic behavioural semantics and various symbolic 
bisimulations (early and late, strong and weak, and their congruent versions). They also extended this work to models with explicit assignments \cite{Linconcur96}.
Separately J. Rathke~\cite{HennessyRathke:TCS98} defined another symbolic semantics for 
a parameterised broadcast calculus, together with strong and weak bisimulation
equivalences, and developed a symbolic model-checker based on a tableau
method for these processes. 30 years later, no
practical verification approach and no verification platform are
using this kind of approaches to provide proof methods for
value-passing processes or open process expressions.

This article provides a theoretical background that allows us to implement such a verification 
platform. We build upon the concept of pNets that allowed us to give a behavioural semantics 
of distributed components and verify the correctness of distributed applications in the past 15 years. pNets  is a
low level semantic framework for expressing the behaviour of various
classes of distributed languages, and as a common internal format for
our tools.  pNets allow the
specification of parameterised hierarchical labelled
transition systems:  labelled transition systems with parameters can be
combined hierarchically.

We develop here a semantics for a model of interacting processes with parameters and holes. 
Our approach is originally inspired from Structured Operational Semantics with conditional premisses as in \cite{GROOTE1993263,VANGLABBEEK2004229}.
But we aim at a more constructive and implementable approach to compute the semantics (intuitively transitions including first order predicates) and to 
check equivalences for these open systems.
The main interest of our symbolic approach is to define a method to
prove properties directly on open structures; these properties will then be preserved
by any correct instantiation of the holes.
As a consequence, our model allows us to reason on composition
operators as well as on full-size distributed systems. The parametric
nature of the model and the properties of compositionability of the
equivalence relations are thus  the main
strengths of  our approach. 

\subsubsection*{pNets and open automata}
pNet is a convenient model to model concurrent systems in a hierarchical and parameterised way. The coordination between processes is expressed as synchronisation vectors that allow the definition of complex and expressive synchronisation patterns.
Open pNets are pNets for which some elements in the hierarchy are still undefined, such undefined elements are called \emph{holes}; a hole can be \emph{filled} later by providing another pNet.
 The semantics of pNets can be expressed as a translation to a labelled transition system, but only if the pNet has no parameter and no hole. Adding parameters to a LTS is quite standard but enabling holes inside LTSs is not a well-defined notion. 
We thus define \emph{open automata} that can be seen as LTSs with parameters and holes. The transitions of open automata are much more complex than transitions of an LTS as the firing on a transition depends on parameters and actions that are symbolic. This article defines the notion of \emph{open transition} a transition that is symbolic in terms of parameters and coordinated actions.

Contrarily to pNets, open automata are not hierarchical structures, thus they are more convenient for formal reasoning but not adapted to the definition of a complex and structured system like pNets. Additionally, open transitions are expressed in terms of logics more than in terms of synchronised actions, synchronisation vectors on the contrary make it easy to express synchronisations that exist in process algebra or in specification and high-level languages for distributed systems.

This article defines pNets and illustrates with an example how they can be used to provide the model of a communicating system. Then we introduce open automata to provide a semantics to open pNets and introduce bisimulation relations and their properties.
Open automata can be also seen as an algebra that can be studied independently from its application to pNets but their composition properties make more sense in a hierarchical model like pNets.

\subsubsection*{Previous Works and Contribution}
While most of our previous works relied on closed, fully-instantiated semantics~\cite{BBCHM:article2008,AmeurBoulifa2017,HKM-FASE16}, it is only recently that we could design a first version of a  parameterised semantics for pNets with a strong bisimulation equivalence~\cite{henrio:Forte2016}. This article builds upon this previous parameterised semantics and provides a clean and complete version of the semantics with a slightly simplified formalism that makes proofs easier. It also adds a notion of global state to automata.  Also, in \cite{henrio:Forte2016} the study of compositionability was only partial, and in particular the proof that bisimulation is an equivalence is one new contribution of this article and provides a particularly interesting insight on the semantic model we use.
The new formalism allowed us to extend the work and define weak bisimulation  for open automata, which is entirely new. This allows us to define a weak bisimulation equivalence for open pNets with valuable properties of compositionality. 
To summarise, the contribution of this paper are the following:
\begin{itemize}
\item The definition of open automata: an algebra of parameterised automata with holes, and a strong bisimulation equivalence. This is an adaptation of~\cite{henrio:Forte2016} with an additional property stating that strong bisimulation equivalence is indeed an equivalence relation.
\item A semantics for open pNets expressed as translation to open automata. This is an adaptation of~\cite{henrio:Forte2016} with a complete proof that strong bisimulation is compositional.
\item A theory of weak bisimulation for open automata, and its properties. It relies on the definition of weak open transitions that are derived from transitions of the open automaton by concatenating invisible action transitions with one (visible or not) action transition. The precise and sound definition of the concatenation is also a major contribution of this article.
\item A resulting weak bisimulation equivalence for open pNets and a simple static condition on synchronisation vectors inside pNets that is sufficient to ensure that weak bisimulation is compositional.
\item An illustrative example based on a simple transport protocol, showing the construction of the weak open transitions, and the proof of weak bisimulation.
\end{itemize}

\subsubsection*{What is new about open automata bisimulation?}

Bisimulation over a symbolic and open model like open pNets or open automata is  different from the classical notion of bisimulation because it cannot rely on the equality over a finite set of action labels. Classical bisimulations require to exhibit, for each transition of one system, a transition of the other system that simulates it. Instead, bisimulation for open automata  relies on the simulation of each open transition of one automaton by a set of open transitions of the other one, that should cover all the 
cases where the original transition can be triggered.

Compositionality of bisimulation in our model come from the specification of the interactions, including actions of the holes. In pNets, synchronisation vectors define the possible interactions between the pNet that fills the hole and the surrounding pNets. In open automata, this is reflected by symbolic hypotheses that depend on the actions of the holes. This additional specification is the price to pay to obtain the compositionality of bisimulation that cannot be guaranteed in traditional process algebras. 

This approach also allows us to specify a sufficient condition on allowed transitions to make weak bisimulation compositional; namely it is not possible to synchronise on invisible actions from the holes or prevent them to occur.

\subsubsection*{Structure}
This article is organised as follows. Section~\ref{sec:notations}
provides the definition of pNets and introduces the notations used in
this paper, including the definition of open pNets. 
Section~\ref{sec:OT} defines open automata, i.e. automata
with parameters and transitions conditioned by the behaviour of
``holes''; a strong bisimulation equivalence for open automata is also
presented in this section. 
Section~\ref{section:op-semantics} gives
the semantics of open pNets expressed as open automata, and states
compositional properties on the strong bisimulation for open
pNets. 
Section~\ref{sec:weak} defines a weak bisimulation
equivalence on open automata and derives weak bisimilarity for pNets,
together with properties on compositionality of weak bisimulation for
open pNets. 
Finally, Section~\ref{sec:RW} discusses related works and Section~\ref{section:conclusion} concludes the
paper.

\section{Background and notations}\label{sec:notations}
This section introduces the notations we will use in this article, and  recalls the definition of pNets~\cite{henrio:Forte2016} with an informal semantics  of the pNet constructs. The only significant difference compared to our previous definitions is that we remove here the restriction that was stating that variables should be local to a state of a labelled transition system.

\subsection{Notations}
\subsubsection*{Term algebra.}
Our models rely on a notion of parameterised actions, that are
symbolic expressions using data types and variables. As our model aims
at encoding the low-level behaviour of possibly very different
programming languages, we do not want to impose one specific algebra
for denoting actions, nor any specific communication mechanism. So we
leave unspecified the constructors of the algebra that will allow building
expressions and actions. Moreover, we use a generic {\em action interaction}
mechanism, based on (some sort of) unification between two or more action
expressions, to express various kinds of communication or
synchronisation mechanisms.

Formally, we assume the existence of a term algebra $\AlgT$,
and denote as $\Sigma$ the signature of the data and action constructors. Within $\AlgT$, we distinguish a set of
data expressions $\AlgE$, including a set of boolean
expressions $\AlgE$ ($\AlgB\subseteq\AlgE$), 
and a set of action expressions $\AlgE$ called the action algebra
$\AlgA$, with $\AlgA\subseteq\AlgT,
\AlgE\cap\AlgA=\emptyset$;
naturally action terms will use data expressions as sub-terms.
The function $\vars(t)$ identifies the set of variables in a term
$t\in\AlgT$.

We let $e_i$ range over expressions ($e_i\in\AlgE$), $a$
range over action labels, $\symb{op}$ be operators, and $x_i$ and $y_i$ range over
variable names. 

We define two kinds of parameterised actions. The first kind distinguishes input variables and the second kind does not. 
We first define the set of actions that distinguish input variables, they will be used in the definition of \pLTS\ below:
\[
\begin{array}[l]{rcl@{\quad}p{7.5cm}}
  \alpha\in\AlgA&::=&a(p_1,\ldots,p_n)&\text{action terms}\\
  p_i&::=& ?x~|~e_i&\text{parameters (input variable or expression)}\\
  e_i&::=& \symb{Value}~|~x~|~\symb{op}(\symb{e}_1,..,\symb{e}_n)&\text{Expressions}
\end{array}
\]
The \emph{input variables} in an action term are those marked with a
$\symb{?}$.
We additionally impose that each input variable does not
appear somewhere else in the same action term:
$p_i=?x\Rightarrow\forall j\neq i.\, x\notin \vars(p_j)$.
We define $iv(t)$  as the set of input variables of a term $t$ (without the '?' marker).
Action algebras can encode naturally usual point-to-point message passing calculi (using 
$a(?x_1,...,?x_n)$ for inputs, $a(v_1,..,v_n)$ for outputs), but it also allows
for more general synchronisation mechanisms, like gate negotiation in Lotos, or broadcast
communications. 

The set of actions that do not distinguish input variables is denoted $\AlgAS$, it will be used in synchronisation vectors of pNets:
\[\begin{array}[l]{rcl@{\quad}l}
  \alpha\in \AlgAS &::=&a(e_1,\ldots,e_n)
\end{array}
\]

\subsubsection*{Indexed sets}

In this article, we extensively use indexed structures 
(maps) over some countable indexed sets.   The indices can typically be
integers, bounded or not. We use indexed sets in pNets because we want to consider a set of processes, and specify separately how to synchronise them. Roughly this could also be realised using tuples, however indexed sets are more general, can be infinite, and give a compact representation than using the position in a possibly long tuple.

An indexed family is denoted as
follows: $t_i^{i\in I}$ is a family of elements $t_i$ indexed over the
set $I$. Such a family
is equivalent to the mapping $(i\mapsto t_i)^{i\in I}$, and we will also use mapping 
notations to manipulate indexed sets.
To specify the set over which the structure is indexed, 
indexed structures are always denoted with an exponent of the form $i\in I$.

Consequently, $t_i^{i\in I}$ defines first $I$ the set over which the
family is indexed, and then $t_i$ the elements of the family.
For example $t_i^{i\in\{3\}}$ is
the mapping with a single entry $t_3$ at index $3$; exceptionally, for mappings with
only a few entries we use the notation $(3\mapsto t_3)$ instead.
In this article, sentences of the form ``there exists $t_i^{i\in I}$'' means there exists $I$ and a function that maps each element of $I$ to a term $t_i$.

When this is not ambiguous, we shall use abusive notations 
for sets, and typically write ``indexed set over I'' when  
formally we should speak of multisets, and ``$x\in
A_i^{i\in I}$'' to mean $\exists i\in I.\, x=A_i$.
To simplify equations, an indexed set can be denoted $\set{t}$
instead of $t_i^{i\in I}$ when $I$ is irrelevant.

The disjoint union on sets is $\uplus$. We extend it to  disjoint union  of indexed 
sets defined by the merge of the 
two sets provided they are indexed on disjoint families.
The elements
of the union of two indexed sets are then accessed by using an index of one of the two
joined families.
The standard subtraction operation on indexed sets is $\setminus$, with $\dom(A\setminus B)=\dom(A)\setminus B$.


\subsubsection*{Substitutions}
\label{def:substitutions}

This article also uses substitutions. Applying a substitution inside a term $t$ is denoted $t\subst{y_i\gets e_i}^{i\in I}$ and consists in replacing in parallel all the occurrences of variables $y_i$ in the term $t$ by the terms $e_i$. Note that a substitution is defined by a partial function that is applied on the variables inside a term. We let $\Post$ range over partial functions that are used as substitution and use the notation $\{y_i\gets e_i\}^{i\in I}$ to define such a partial function\footnote{When using this notation, we suppose, without loss of generality that each $y_i$ is different.}. These partial functions are sometimes called \emph{substitution functions} in the following. Thus,
 $\subst{\Post}$ is  the operation that applies,  in a parallel manner,  the substitution defined by the partial function $\Post$. $\odot$ is a composition operator on these partial functions, such that for any term $t$ we have: $t\subst{\Post\shortodot\Post'} = (t\subst{\Post'})\subst{\Post}$.
This property must also be valid when the substitution does not operate on all variables.
We thus define a composition operation as follows: 
%

\[(x_{k}\gets e_{k})^{k\in K}\shortodot (x'_{k'}\gets e'_{k'})^{k'\in K'} =  
(x_{k}\gets e_{k}\subst{(x'_{k'}\gets e'_{k'})^{k'\in K'}})^{k\in K} \cup (x'_{k'}\gets e'_{k'})^{k'\in K''}\]
where $K''=\{k'\in K'|x'_{k'}\not\in\{x_k\}^{k\in K}\}$
%
%
%

\subsection{Parameterised Networks (pNets)}
\label{section:pnets}

pNets are tree-like structures, where the leaves are either \emph{parameterised labelled transition systems (pLTSs)}, expressing the
behaviour of basic processes, or \emph{holes}, used as placeholders
for unknown processes. 
Nodes of the tree (pNet nodes) are synchronising artefacts, using a
set of \emph{synchronisation vectors} that express the possible
synchronisation between the parameterised actions of a subset of the sub-trees.


A pLTS is a labelled transition system with variables; variables can be
used inside states, actions, guards, and
assignments. 
Note that we make no assumption on finiteness of the set of states nor
on finite branching of the transition relation. Compared to our previous works~\cite{henrio:Forte2016,AmeurBoulifa2017} we extend the expressiveness of the model by making variables global.

\begin{defi}[pLTS]
\label{pLTS}
A pLTS is a tuple
$\pLTS\triangleq\mylangle S,s_0, V, \to\myrangle$ where:
\begin{itemize}
\item[$\bullet$]
$S$ is a set of states.
\item[$\bullet$]
$s_0 \in S$ is the initial state.
\item[$\bullet$] $V$ is a set of global variables for the pLTS.
\item[$\bullet$] $\to \subseteq S \times L \times S$ is the transition relation and 
$L$ is the set of labels of the form:\\
$\langle \alpha,~e_b,~(x_j\!:= {e}_j)^{j\in J}\rangle$,
where $\alpha \in\AlgA$ is a parameterised action, $e_b \in
\AlgB$ is a guard, and the variables $x_j$ 
are assigned the expressions $e_j\in \AlgE$.
If 
$s \xrightarrow{\langle \alpha,~e_b,~(x_j\!:= {e}_j)^{j\in
		J}\rangle} s'\in \to $ then 
		$\vars(\alpha)\backslash \iv(\alpha)\!\subseteq\! V$, 
		$\vars(e_b)\!\subseteq\! \vars(s)\cup\vars(\alpha)$, and
		$\forall j\!\in\! J .\,\left(\vars(e_j)\!\subseteq\! V\cup\iv(\alpha)\land 
		x_j\!\in V \right)$. 

\end{itemize}
\end{defi}

The semantics of the assignments is that a set of assignments between two states is performed in parallel so that their order do not matter and they all use the values of variables before the transition (or the values received as action parameters).

Now we define
pNet nodes as constructors for hierarchical behavioural structures.
A pNet has a set of sub-pNets that can be either pNets or pLTSs, and a
set of holes, playing the role of process parameters. A pNet is thus a composition operator that can receive processes as parameters; it expresses how the actions of the sub-processes synchronise.

Each sub-pNet exposes
a set of actions, called \emph{internal actions}. The synchronisation between global actions exposed by the pNet and
internal actions of its sub-pNets is given by  \emph{synchronisation vectors}: a
synchronisation vector synchronises one or several internal actions, and
exposes a single resulting global action.

We now define the structure of pNets, the following definition relies on the definition 
of holes, leaves and sorts formalised below in Definition~\ref{def-sortholeleave}. Informally, holes are process parameters, leaves provide the set of pLTSs at the leaves of the hierarchical structure of a pNet, and sorts give the signature of a pNet, i.e. the actions it exposes.

\begin{defi}[pNets]\label{def-pnets}
A pNet $\pNet$ is a hierarchical structure where leaves are pLTSs and holes\\
$\pNet\triangleq \pLTS~|~\mylangle \pNet_i^{i\in I}, \Sort_j^{j\in J}, \symb{SV}_k^{k\in 
K}\myrangle$
where:
\begin{itemize}
\item[$\bullet$] $I$ is a set of indices and $\pNet_i^{i\in I}$ is the family of sub-pNets indexed over $I$. $vars(P_i)$ and $vars(P_j)$ must be disjoint for $i\neq j$.

\item[$\bullet$] $J$ is a set of indices, called \emph{holes}.
$I$ and $J$ are \emph{disjoint}: $I\!\cap\! J=\emptyset$,  $I\!\cup\! J\neq\emptyset$.
\item[$\bullet$] $\Sort_j \subseteq \AlgAS$  is a set of action terms, denoting the 
\emph{sort} of
hole $j$.

\item[$\bullet$] $\symb{SV}_k^{k\in K}$ is a set of
  synchronisation vectors. $\forall k\!\in\! K.\,
  \symb{SV}_k\!=\!\SV{\alpha_{l}^{l\in I_k \uplus J_k}}{\alpha'_k}{e_k}$ where
  $\alpha'_k\in \AlgAS$, $I_k\subseteq I$, $J_k\subseteq J$,
  $\forall i\!\in\!
  I_k.\,\alpha_{i}\!\in\!\Sort(\pNet_i)$,  $\forall j\!\in\!
  J_k.\,\alpha_{j}\!\in\!\Sort_j$, and $\vars(\alpha'_k)\subseteq \bigcup_{l\in I_k\uplus 
  J_k}{\vars({\alpha_l})}$. The global action of a vector $\symb{SV}_k$ is
$\alpha'_k$. $e_k \in \AlgB$ is a guard associated to the vector such that
$\vars(e_k)\subseteq \bigcup_{l\in I_k\uplus J_k}{\vars({\alpha_l})}$.
\end{itemize}
Synchronisation vectors are identified modulo renaming of variables that appear in their 
action terms.
\end{defi}

The preceding definition relies on the auxiliary functions defined below:

\begin{defi}[Sorts, Holes, Leaves, Variables of pNets]\label{def-sortholeleave}~~

  \begin{itemize}
  \item The sort of a pNet is its signature, i.e. the set of actions in $\AlgAS$ it can
perform, where each action signature is an action 
label plus the arity of the action.
\[
\begin{array}{l}
\Sortop(\mylangle S,s_0,V, \to\myrangle) = \{\Sortop(\alpha)|s \xrightarrow{\langle \alpha,~e_b,~(x_j\!:= {e}_j)^{j\in
    J}\rangle} s'\in \to \} \\
\Sortop(\mylangle \set{\pNet}\!, 
\set{\Sort},
\set{\symb{SV}\,}\myrangle)
=\{\Sortop(\alpha') |\, \SV{\set{\alpha}}{\alpha'}{e_b}\in\set{\symb{SV}}\,\}\\
\Sortop(\alpha(p_1,..,p_n))=(\alpha,n)
\end{array}
\]

\item The set of variables of a pNet $P$, denoted $vars(P)$ is disjoint union the set of variables of  all pLTSs that compose $P$.

\item
The set of holes $\Holes(\pNet)$ of a pNet is the indices of the holes of the pNet 
itself plus the indices of all the holes of its sub-pNets.
It is defined inductively (we suppose those indices 
disjoints):
  \[\begin{array}{l}
\Holes(\mylangle S,s_0,V, \to\myrangle) \!=\! \emptyset\\
\Holes(\mylangle \pNet_i^{i\in I}\!,\set{\Sort}, \set{\symb{SV}}\myrangle) 
=J\uplus{\displaystyle \bigcup_{i\in 
I}\Holes(\pNet_i)}\\
\forall i\in I.\, \Holes(\pNet_i)\cap J=\emptyset\\
\forall i_1,i_2\in I.\,i_1\neq i_2\Rightarrow  \Holes(\pNet_{i_1})\cap\Holes(\pNet_{i_2})=\emptyset
\end{array}\]

\item
The set of leaves of a pNet is the set of all pLTSs occurring in the structure, as an 
indexed family of the form $\Leaves(\pNet)= \mylangle \pNet_i \myrangle^{i \in L}$.
\[\begin{array}{l}
\Leaves(\mylangle S,s_0,V, \to\myrangle) \!=\!\emptyset\\
\Leaves(\mylangle \pNet_i^{i\in I}\!,
\set{\Sort}\!, \set{\symb{SV}\,}\myrangle) = {\displaystyle \biguplus_{i\in 
I}\Leaves(\pNet_i)\uplus\{i\mapsto \pNet_i|\pNet_i \text{ is a }\pLTS\}}
\end{array}\]
\end{itemize}

A pNet $Q$ is \emph{closed} if it has no hole: $\Holes(Q)=\emptyset$; else it
is said to be \emph{open}.
Sort comes naturally with a compatibility relation that is similar to a type-compatibility check. We simply say that two sorts are compatible if they consist of the same actions with the same arity. In practice, it is sufficient to check the equality of the two sets of action signatures of the two sorts\footnote{A more complex compatibility relation could be defined, but this is out of the scope of this article.}.
\end{defi}
  
The informal semantics of pNets is as follows. pLTSs behave more or less like  classical automata with conditional branching and variables. The actions on the \pLTS s can send or receive values, potentially modifying the value of variables. 
pNets are synchronisation entities: a pNet node composes several sub-pNets and  define how the sub-pNets interact, where a sub-pNet is either a pNet or a pLTS. The synchronisation between sub-pNets is defined by synchronisation vectors (originally introduced by \cite{Arnold1982}) that express how an action of a sub-pNet can be synchronised with actions of other sub-pNet, and how the resulting synchronised action is visible from outside of the pNet. The synchronisation mechanism is very expressive, including pattern-matching/unification between the parameterized actions within the vector, and an additional predicate over their variables.
Consider a pNet node that assembles several pLTSs, the synchronisation vectors specify the way that transitions of the composed pNet are built from the transitions of the sub-nets. This can be seen as "conditional transitions" of the pNet, or alternatively, as a syntax to encode structural operational semantics (SOS rules) of the system: each vector expresses not only the actions emitted by the pNet but also what transitions of the composed pLTSs must occur to trigger this global transition.
Synchronisation vectors can also express the exportation of an action of a sub-pNet to the next level, or to hide an interaction and make it non-observable. Finally, a pNet can leave sub-pNets undefined and instead declare holes with a well-defined signature. Holes can then be filled with a sub-pNet. This is defined as follows.

\begin{defi}[pNet composition]
	An open pNet: $\pNet = \mylangle \pNet_i^{i\in I}, \Sort_j^{j\in J}, 
	\set{\symb{SV}}\,\myrangle$
 can be (partially) filled by providing  a pNet $\pNetQ$ to fill one of  its holes.	
	Suppose $j_0\in J$ and $\Sortop(\pNetQ) \subseteq \Sort_{j_0}$, then: 
	\[\pNet\left[\pNetQ\right]_{j_0}= \mylangle 
	\pNet_i^{i\in I}\uplus\{j_0\mapsto \pNetQ\},\Sort_j^{j\in J\setminus \{j_0\}},
	\set{\symb{SV}}\,\myrangle
	\]
\end{defi}

pNets are composition entities equipped with a rich synchronisation mechanism: synchronisation vectors allow the expression of synchronisation between any number of entities and at the same time the passing of data between processes. Their strongest feature is that the data emitted by processes can be used  inside the synchronisation vector to do addressing: it is easy to synchronise a process indexed by $n$ with the action $a(v,n)$ of another process. This is very convenient to model systems and encode futures or message routing. 

pNets have been used to model GCM distributed component systems, illustrating the expressiveness of the model~\cite{AmeurBoulifa2017}. 
These works show that pNets are convenient to express the behaviour of the system in a compositional way, which is crucial for the definition of the semantics, especially when dealing with a hierarchical component system. 
Unfortunately, the semantics of pNets and the existing tools at this point were only able to deal with a closed system completely instantiated: pNets could be used as composition operator in the definition of the semantics, which was sufficient to perform finite-state model checking on a closed system, but there was no theory for the use of pNets as operators and no tool for proving properties on open system. 
Consequently, much of the formalisation efforts did not use holes and the interplay between holes, sorts, and synchronisation vector was not formalised.
In previous works~\cite{AmeurBoulifa2017}, only closed pNets were equipped with a semantics, it was defined as labelled transition systems which are instantiations of pNets. 
The theory of pNets as operators able to fully take into account open systems is given in the following sections. Comparing formally the existing direct operational semantics and the semantics derived from open automata in the current article would be an interesting partial proof of soundness for our semantics. The proof could only be partial as the formal semantics that exists only consider closed and fully instantiated pNets. Proving 
an equivalence between the semantics presented in this article and the operational one shown in~\cite{AmeurBoulifa2017} is outside the scope of this article.

\subsection{Running Example}
To illustrate this work, we use a simple communication protocol, that provides safe transport of data between two processes, over unsafe media.

\begin{figure}[t]
   \includegraphics[width=.37\textwidth]{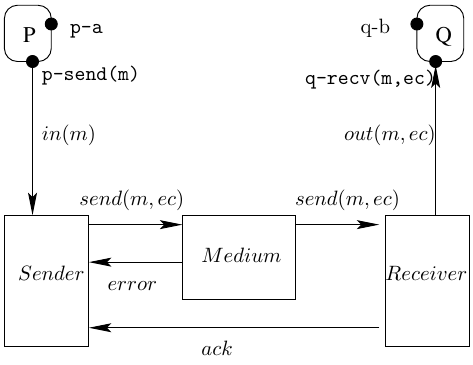}
   \includegraphics[width=.62\textwidth]{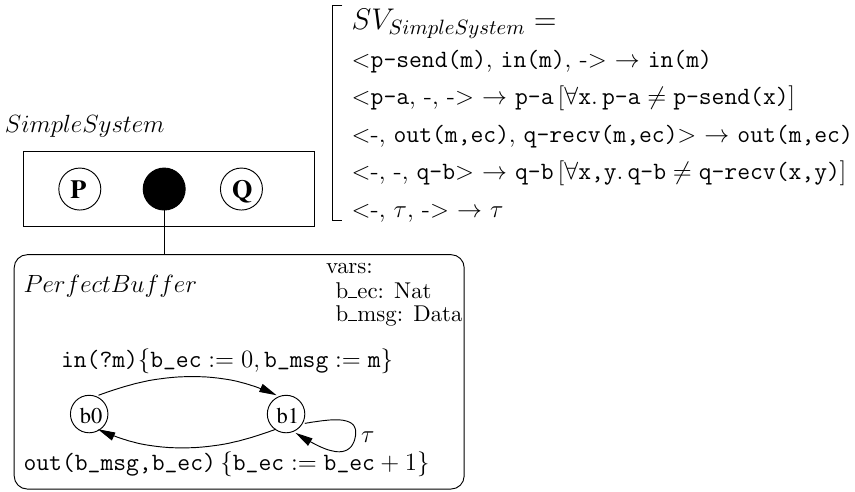}
   \caption{pNet structure of the example and its specification expressed as a pNet called $\symb{SimpleProtocolSpec}$ }
   \label{SimpleProt:Spec}
\end{figure}

Figure \ref{SimpleProt:Spec} (left) shows the example principle, which corresponds to the hierarchical structure of a pNet: two unspecified
processes $P$ and $Q$ (holes) communicate messages, with a data value
argument, through the two protocol entities. Process $P$ sends an $\texttt{p-send(m)}$ message to the $\symb{Sender}$; this communication is denoted as $\texttt{\nounderline{in(m)}}$
.
At the other end, process $Q$ receives the message from the $\symb{Receiver}$. The holes $P$ and $Q$ can also have other interactions with their environment, represented here by actions $\texttt{p-a}$ and $\texttt{q-b}$. The underlying network is modelled by a medium entity transporting messages from the sender to the receiver, and that is able to detect transport errors and signal them to the sender. The return $\symb{ack}$ message from $\symb{Receiver}$ to $\symb{Sender}$ is supposed to be safe. The final transmission of the message to the recipient (the hole $Q$) includes the value of the ``error counter'' $ec$.

Figure \ref{SimpleProt:Spec} (right) shows a graphical view of the pNet $\symb{SimpleProtocolSpec}$ that specifies the system. The pNet is made of the composition of two pNets: a $\symb{SimpleSystem}$ node, and a $\symb{PerfectBuffer}$ sub-pNet. 
The full system implementation should be equivalent (e.g.  weakly bisimilar)
to this $\symb{SimpleProtocolSpec}$.  
The pNet has a tree-like structure.  The root node of the tree {\it SimpleSystem} is the top level of the pNet structure. It acts as the parallel operator. It consists of three nodes: two holes $P$ and $Q$ and one sub-pNet, denoted {\it PerfectBuffer}. Nodes of the tree are synchronised using four synchronisation vectors, that  express the possible synchronisations between the parameterised actions of a subset of the nodes. For instance, in the vector $< \texttt{p-send(m)},\texttt{in(m)},\_> \rightarrow \texttt{\nounderline{in(m)}}$ only $P$ and {\it PerfectBuffer} nodes are involved in the synchronisation. The synchronisation between these processes occurs when  process $P$  performs $\texttt{p-send(m)}$  action sending a message, and  the {\it PerfectBuffer} accepts the message through an $\texttt{in(m)}$ action  at the same time; the result that will be returned at upper level is the action  $\texttt{\nounderline{in(m)}}$.

\begin{figure}[t]
  \centerline{\includegraphics[width=.95\textwidth]{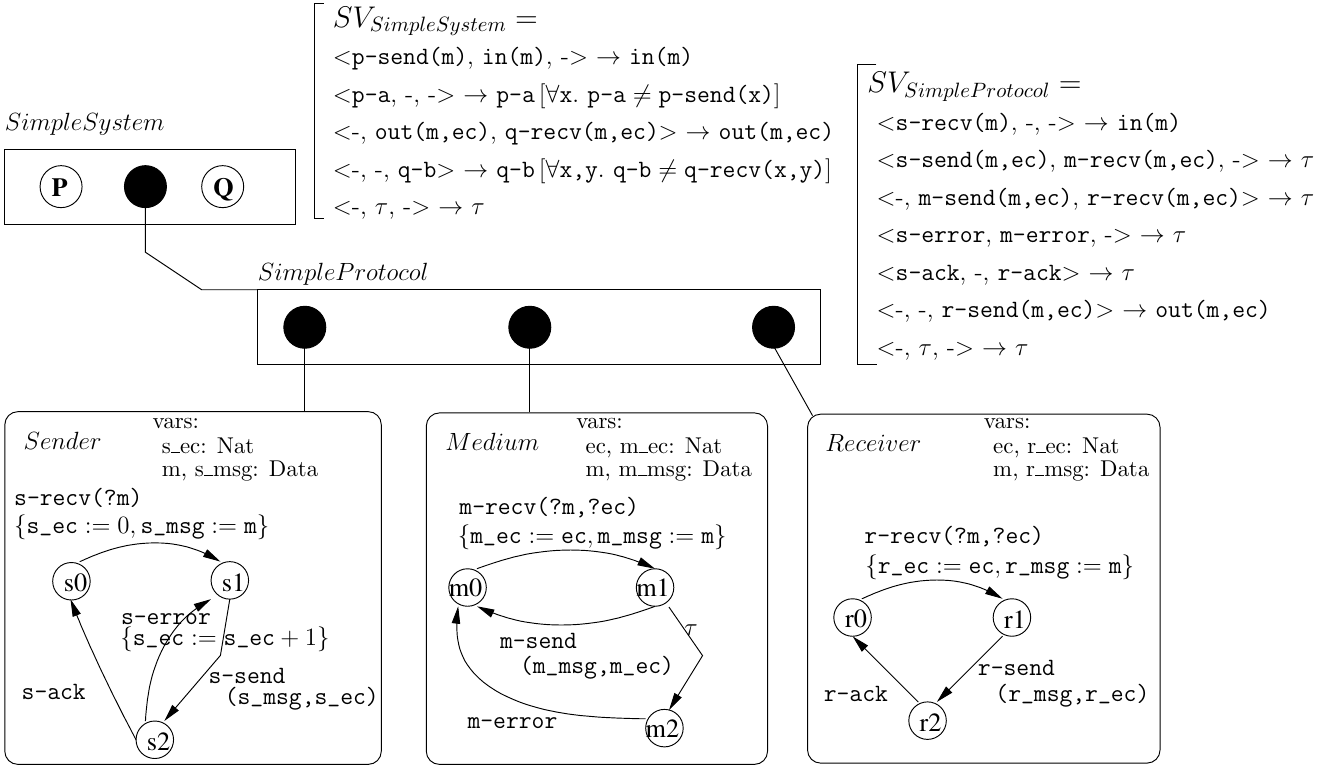}}
  \caption{The $\symb{SimpleProtocolImpl}$ pNet resulting from the composition of  the $\symb{SimpleSystem}$  and the $\symb{SimpleProtocol}$ pNets.}  \label{SimpleProt:Impl}
\end{figure}

Figure \ref{SimpleProt:Impl} shows the pNet model of the protocol implementation, called $\symb{SimpleProtocolImpl}$. When the $\symb{Medium}$ detects an error (modelled by a local $\tau$ action), it sends back a $\texttt{m-error}$ message to the $\symb{Sender}$. The $\symb{Sender}$ increments its local error counter $ec$, and resends the message (including $ec$) to the $\symb{Medium}$, that will, eventually, transmit $m,ec$ to the $\symb{Receiver}$. 

\section{A model of process composition}\label{sec:OT}

The semantics of open pNets will be defined  as an open automaton. An open
automaton is an automaton where each transition composes transitions of several LTSs with
action of some holes, the transition occurs if some predicates hold, and can involve a 
set of state modifications. This section defines open automata and a bisimulation theory for them. This section is an improved version of the formalism described in \cite{henrio:Forte2016}, extending the automata with a notion of global variable, which makes the state of the automaton more explicit. We also adopt a semantics and logical interpretation of the automata that intuitively can be stated as follows: ``if a transition belongs to an open automaton, any refinement of this transition also belongs to the automaton''.

\subsection{Open Automata}
 Open automata (OA) are not composition structures but they are made of transitions that are dependent of the actions of the holes, and they can reason on a set of variables (potentially with only symbolic values). 
\begin{defi}[Open transitions]\label{def:OT}
	\label{def:OpenTransitions}
	An \emph{open transition} (OT) over a
	set $J$ of holes with sorts $\Sort_j^{j\in J}$, a set $V$ of variables, and a set of states $\mathcal{S}$ is 
	a structure of the form:	
	\begin{mathpar}
	\openrule
	{	\beta_j^{j\in J'}, \Pred, \Post}
	{s \OTarrow {\alpha}s'}
	\end{mathpar}
	Where $J'\subseteq J$ is the set of holes involved in the transition; $s, s'\in\mathcal{S}$ are states of the automaton; and $\beta_j$
        is a transition of the hole $j$, with $\Sortop(\beta_j)\in\Sort_j$. $\alpha$  is the resulting action of this open transition.
        \Pred\ is a predicate, \Post\ is a set of 
	assignments that are effective after the open transition, they are
        represented as a substitution function: $({x_k\gets e_k})^{k\in K}$.
Predicates and expressions of an open transition can refer to the variables in $V$, and
        in the different terms
        $\beta_j$ and $\alpha$. More precisely:

\begin{multline*}
 \vars(\Pred) \subseteq V\cup\vars(\alpha)\cup
			{\displaystyle \bigcup_{j\in J'}\!\!\vars(\beta_j)}\quad\land\\
\forall k.\, x_k\in V \quad\land\quad \forall k.\,\vars(e_k)\subseteq V\cup\vars(\alpha)\cup
			{\displaystyle \bigcup_{j\in J'}\!\!\vars(\beta_j)}
\end{multline*}
The assignments are applied simultaneously because the variables in $V$ can be in both sides ($x_k$s are distinct).

 Open transitions are identified
        modulo logical equivalence on their predicate. 
\end{defi}

It is important to understand the difference between the red dotted rule and a classical 
inference rule. They correspond to two different logical levels.
On one side, classical (black) inference rules  use  an expressive logic (like any other computer science article).
 On the other side, open transition rules (with dotted lines) are logical implications, but using a  logic with a specific syntax and that can be mechanized (this logic includes the boolean expressions $\AlgB$, boolean operators, and term equality).

An open automaton is  an automaton where each transition is an open transition.
\begin{defi}[Open automaton]
	\label{def:open-automaton}
	An \emph{open automaton} is a structure\\ $A =
	\mylangle J,\mathcal{S},s_0,V,\mathcal{T}\myrangle$ where:
	\begin{itemize}
		\item[$\bullet$]   $J$ is a  set of indices.
		\item[$\bullet$]   $\mathcal{S}$ is a set of states and $s_0$ an initial state
		  among $\mathcal{S}$.
 \item[$\bullet$] $V$ is a set of variables of the automaton
		and each $v\in V$ may have an initial value $init(v)$.
		\item[$\bullet$] $\mathcal{T}$ is a set of open transitions and for each
		$t\in \mathcal{T}$ there exists  $J'$ with  $J'
		\subseteq J$, such that $t$ is an open transition over  $J'$
		and  $\mathcal{S}$.
		
	\end{itemize}

While the definition and usage of the open transition can be formalised and taken in a pure syntactic acceptance, we take in this article a semantics and logical understanding of open automata. 
Formally, the open transition sets in open automata are closed by a simple form of refinement that
allows us to refine the predicate, or substitute any free variable by
an expression as expressed below.

For all predicate $\Pred$ for all partial function $\Post$,if $V\!\cap \!\dom(\Post)=\emptyset$, we have: 
		 \begin{mathpar}
    \openrule
         {
           \set{\beta}, \Pred\,',\Post\,'}
          {t \OTarrow {\alpha} t'}\in\mathcal{T}
\quad\implies\quad
    \openrule
         {
           \set{\beta}\subst{\Post}, \Pred\,'\subst{\Post}\land\Pred,\Post\shortodot\Post\,'}
         {\raisemath{-2pt}{t \OTarrow {\alpha\subst{\Post}} {t'}}{}}
 \in\mathcal{T}
\end{mathpar}
\end{defi}

Because of the semantic interpretation of open automata, the set of open transition of an open automaton is infinite (for example because every free variable can be renamed). However an open automaton is characterized by a  subset of these open transitions which is sufficient to generate, by substitution the other ones. In the following, we will abusively write that we define an ``open automaton'' when we provide only the set of open transitions that is sufficient to generate a proper open automaton by saturating each open transition by all possible substitutions and refinements.

Another aspect of the logical interpretation of the
formulas is that we make no distinction between the equality and the
equivalence on boolean formulas, i.e. equivalence of two predicates
$\Pred$ and $\Pred\,'$ can be denoted $\Pred=\Pred\,'$, where the $=$ symbol is not interpreted in a syntactical way.

Though the definition is simple, the fact that transitions are complex structures relating events must not be underestimated. The first element of theory for open automata, i.e. the definition of a strong bisimulation, is given below.

\subsection{Bisimulation for open Automata}
\label{section:bisimulation}

The equivalence we need is a strong bisimulation between
open automata having exactly the same holes (same indices and same sorts), but using a
flexible matching 
between open transitions, this will allow us to compare pNets
with different architectures.

We define now a bisimulation relation tailored to open automata and their parametric nature. This relation relates states of the open automata and guarantees that the related states are observationally equivalent, i.e. equivalent states can trigger transitions with identical action labels. Its key characteristics are 1) the introduction of predicates in the bisimulation relation: the relation between states may depend on the value of the variables; 2) the bisimulation property relates elements of the open transitions and takes into account predicates over variables, actions of the holes, and state modifications.
 We name it FH-bisimulation,
 as a short cut for the ``Formal Hypotheses'' over the holes behaviour manipulated in the
 transitions, but also as a reference to the work of De Simone~\cite{deSimone85},
 that pioneered this idea.

One of the original aspects of FH-bisimulation is due to the symbolic nature of open automata. Indeed, a single state of the automaton represents a potentially infinite number of concrete states, depending on the value of the automaton variables, and a single open transition of the automaton may also be instantiated with an unbounded number of values for the transition parameters. Consequently it would be too restrictive to impose that each transition of one automaton is matched by exactly one transition of the bisimilar automaton. Thus the definition of bisimulation requires that, for each open transition of one automaton, there exists a matching  set of open transitions covering the original one, indeed depending on the value of action parameters or automaton variables, different open transitions might simulate the same one.

The parametric nature of the automata entails a second original aspect of FH-bisimulation: the nature of the bisimulation relation itself.
 A classical relation between states can be seen as a function mapping pairs of state to a boolean value (true if the states are related, false if they are not). An FH-bisimulation relation maps pairs of states to boolean expressions that use variables of the two systems.  
Formally, a relation over the states of two open automata  $\mylangle J,\mathcal{S}_1, s_0,V_1,
   \mathcal{T}_1 \myrangle$ and $\mylangle J,\mathcal{S}_2,t_0,V_2, \mathcal{T}_2 \myrangle$ has the signature $\mathcal{S}_1\times\mathcal{S}_2\to\AlgB$. 
We suppose without loss of generality that the variables of the two open automata are disjoint.
We adopt a notation similar to standard relations and denote it
 $\mathcal{R}=\{(s,t|\Pred_{s,t})\}$, where: 1) For any pair $(s,t)\in \mathcal{S}_1\times \mathcal{S}_2$, there is a 
   single
      $(s,t|\Pred_{s,t})\in\mathcal{R}$  stating that $s$ and $t$ are related 
      if $\Pred_{s,t}$       is 
      True, i.e. the states are related when the value of the automata variables
  verify the predicate $\Pred_{s,t}$. 2) The free variables of  $\Pred_{s,t}$ belong to $V_1$ and $V_2$, i.e. $\vars(\Pred_{s,t})\subseteq V_1\cup V_2$.
 FH-bisimulation is defined formally\footnote{In this article, we denote $\beta_{j x}$  a double indexed set, instead of the classical $\beta_{j,\, x}$. Indeed the standard notation would be too heavy in our case.}: 
 \begin{defi}[Strong FH-bisimulation]\label{def-FH-bisim} ~\\
\noindent
	Suppose 
   $A_1 = \mylangle J,\mathcal{S}_1, s_0,V_1,
   \mathcal{T}_1 \myrangle$ and $A_2 = \mylangle J,\mathcal{S}_2,t_0,V_2, \mathcal{T}_2 \myrangle$
   are open automata with identical holes of the same sort, with disjoint sets of variables ($V_1\cap V_2=\emptyset$).  

 Then 
$\mathcal{R}$ is an FH-bisimulation if and only if for any  states
$s\in\mathcal{S}_1$ and $t\in\mathcal{S}_2$, $(s,t|\Pred_{s,t})\in\mathcal{R}$, we 
have
the following:

 \begin{itemize}
 \item   
For any open transition $OT$ in $\mathcal{T}_1$:\\
\begin{minipage}{0.67\linewidth}  \begin{mathpar}
     \openrule
         {
           \beta_j^{j\in J'},\Pred_{OT},\Post_{OT}}
         {s \OTarrow {\alpha} s'}

\end{mathpar}
 there exists an indexed set of  open transitions $OT_x^{x\in X} \subseteq \mathcal{T}_2$:
 \begin{mathpar}
    \openrule
         {
           \beta_{j x}^{j\in J_{x}}, \Pred_{OT_x},\Post_{OT_x}}
         {t \OTarrow {\alpha_x} t_x}
\end{mathpar}

\end{minipage}
\hspace{2mm}
\begin{minipage}{0.30\linewidth}
\vspace{-2em}
{	\includegraphics[width=\linewidth]{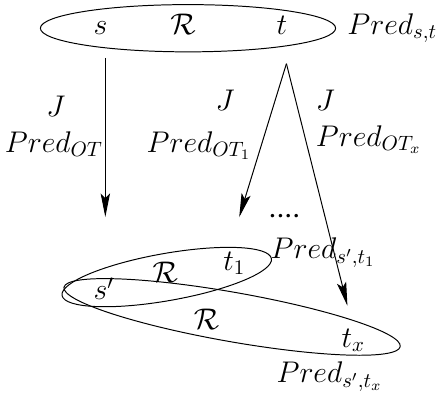}}
\end{minipage}
 such that  $\forall x.\, J'=J_{x}$ and there exists $\Pred_{s',t_x}$ such that $(s',t_x|\Pred_{s',t_x})\in 
 \mathcal{R}$
 and  
\begin{multline*}
 \Pred_{s,t} \land \Pred_{OT}\implies\\
 \displaystyle{\bigvee_{x\in X}
   \left( \forall j. \beta_j=\beta_{jx}  \land \Pred_{OT_x}
     \land \alpha\!=\!\alpha_x \land  
     \Pred_{s',t_x}\subst{\Post_{OT}\!\uplus\!\Post_{OT_x}}\right)}
\end{multline*}
%

     
 \item  and symmetrically any open transition from $t$ in $\mathcal{T}_2$ can be 
      covered by a set of transitions from $s$ in $\mathcal{T}_1$.
 \end{itemize}

 

 \end{defi}
Classically, $\Pred_{s',t_x}\subst{\Post_{OT}\uplus\Post_{OT_x}}$
applies in parallel the  
substitution defined by the partial functions $\Post_{OT}$ and $\Post_{OT_x}$ (parallelism is crucial
inside each $\Post$ set but not between  $\Post_{OT}$ and
$\Post_{OT_x}$ that are independent), applying the assignments of the involved rules.
We can prove that such a bisimulation is an equivalence relation.

\begin{thm}[FH-Bisimulation is an equivalence]\label{thm-equiv} Suppose $\mathcal{R}$ 
is an FH-bisimulation. Then $\mathcal{R}$ is an equivalence, that is, $\mathcal{R}$ is 
reflexive, symmetric and transitive.
\end{thm}

The proof of this theorem can be found in Annex~\ref{thm-equiv-proof}. The
only non-trivial part of the proof is the proof of transitivity. It
relies on the following elements. First,  the transitive composition
of two relations with predicate is defined; this is not exactly
standard as it requires to define the right predicate for the
transitive composition and producing a single predicate to relate any
two states. Then the fact that one open transition is simulated by a
family of open transitions leads to a doubly indexed family of
simulating open transition; this needs particular care, also because
of the use of renaming (\Post) when proving that the predicates
satisfy the definition (property on $\Pred_{s,t} \land \Pred_{OT}$ in
the definition).  


\medskip

\subsubsection*{Finite versus infinite open automata, and decidability:} 
As mentioned in Definition \pageref{def:open-automaton}, we adopt here a semantic view on open automata. More precisely, in \cite{hou:hal-02406098}, we  define
\emph{semantic open automata} (infinite as in Definition \ref{def:open-automaton}),
and \emph{structural open automata} (finite) that can be generated as
the semantics of pNets (see Definition \ref{def:operationalSemantics}), and used in the implementation. Then we define
an alternative version of our bisimulation, called
structural-FH-Bisimulation, based on structural open automata, and
prove that the \emph{semantic} and \emph{structural} FH-Bisimulations coincide.
In the sequel, all mentions of finite automata, and algorithms for
bisimulations, implicitly refer to their \emph{structural} versions.

If we assume that everything is finite (states and transitions in the
open automata), then it is easy to
prove that it is decidable whether a relation is a 
FH-bisimulation, provided the logic of the predicates is
decidable (proof can be found in \cite{henrio:Forte2016}). Formally:

\begin{thm}[Decidability of FH-bisimulation]
Let $A_1$ and $A_2$ be finite open automata
and $\mathcal{R}$ a relation over their states $\mathcal{S}_1$ and
$\mathcal{S}_2$ constrained by a set of predicates. Assume that
the predicates inclusion is decidable over  
the action algebra $\AlgA$. Then it is decidable whether the relation 
$\mathcal{R}$ is an FH-bisimulation.
  
\end{thm}

\section{Semantics of Open pNets}
\label{section:op-semantics}

This section defines the semantics of an open pNet as a translation into an open automaton. 
In this translation, the states of the open automata are obtained as products of 
the states of the pLTSs at the leaves of the composition. The
predicates on the transitions are obtained both from the predicates on
the transitions of the pLTSs, and from the synchronisation vectors involved in
the transition. 

The definition of bisimulation for open automata allows us to derive the characterization and properties of a
bisimulation relation for open pNets. As pNets are composition
structures, it then makes sense to prove composition lemmas: we prove
that the composition of strongly bisimilar pNets are themselves
bisimilar. 

\subsection{Deriving an open automaton from an open pNet}
To derive an open automaton from a pNet, we first describe the set of states of the automaton. Then we show the construction rule for transitions of the automaton, this  relies on the derivation of predicates unifying synchronisation vectors and the actions of the pNets involved in a given synchronisation.


States of open pNets are tuples of states. We denote them
 as $\triangleleft\ldots\triangleright$ for distinguishing tuple 
states from other tuples.
\begin{defi}[States of open pNets]\label{def-states}
  A state of an open pNet is a tuple (not necessarily finite) of the
  states of its leaves.

  For any pNet \pNet, let $\Leaves(\pNet) = \mylangle S_i,{s_i}_0,V, \to_i\myrangle^{i \in L}$ be 
  the set of pLTS at its leaves,
  then $States(\pNet) = \{\triangleleft s_i^{i\in L}
  \triangleright| \forall i\in L. s_i \in S_i\}$.
A pLTS being its own single leave:\\
  $States(\mylangle S,s_0,V, \to\myrangle) = \{\triangleleft s \triangleright| s \in S \}$.  

The initial state is defined as:
$InitState(\pNet) = \triangleleft {{s_i}_0}^{i\in L}  \triangleright$.
\end{defi}
To be precise, the state of each pLTS is entirely characterized by both the state of the automaton, and the value of its variables $V$. Consequently, the state of a pNet is not only characterized the tuple of pLTS states but also contains the value of its variables $vars(P)$.


\subsubsection*{Predicates} 
 We 
define a
predicate $\Predsv$ relating a synchronisation vector (of the form $\SV{{(\alpha'_i)}^{i\in I}, {(\beta'_j)}^{j\in J}}{\alpha'} {e_b}$),
the actions of the involved sub-pNets and the resulting actions.

 This predicate verifies:
\begin{multline*}
\Predsv \Big(
\big({\SV{{(\alpha'_i)}^{i\in I}, {(\beta'_j)}^{j\in J}}{\alpha'}{e_b}}\big)
, \alpha_i^{i\in I}, \beta_j^{j\in J}, \alpha\Big)\Leftrightarrow \\
\forall i\in I.\, \alpha_i=\alpha'_i\land \forall j \in J.\, \beta_j=\beta'_j \land 
\alpha=\alpha' 
\land e_b
\end{multline*}

Somehow, this predicate entails a verification of satisfiability in the sense that if the 
predicate $\Predsv$ is not satisfiable, then the transition associated with the 
synchronisation will not occur in the considered state, or will occur with a \False\ precondition which is equivalent.
If the action families do not match or if there is no valuation of
variables such that the above formula can be ensured then the predicate is undefined.

The definition of this predicate is not constructive but it is easy to build the predicate constructively by brute-force unification of the sub-pNets actions with the corresponding vector actions, possibly followed by a simplification step.

\begin{example}[An open-transition]
  \label{OT:SimpleProt}
At the upper level, the $\symb{SimpleSystem}$ pNet of Figure \ref{SimpleProt:Impl} has 2 holes and $\symb{SimpleProtocol}$ as
a sub-pNet, itself containing 3 pLTSs. One of its possible open transitions
(synchronizing the hole $P$
with the $\symb{Sender}$ within the \emph{SimpleProtocol}) is:

 \smallskip\noindent
 \[  OT_1  = \openrule{
      \{\texttt{P}\mapsto \texttt{p-send(m)}\},  [\texttt{m=m'}],
        (\texttt{s\_msg}\gets \texttt{m})
                      }
    {\ostate{s_0,m_0,r_0} \OTarrow{\nounderline{\texttt{in(m')}}} \ostate{s_1,m_0,r_0}}
    \]

    \smallskip
    The global states here are triples, the product of states of the 3 pLTSs (holes have no state). The assignment
    performed by the open transition uses the variable $\texttt{m}$ from the action of hole $\texttt{P}$ to set the
    value of the sender variable named $\texttt{s\_msg}$.







    \smallskip

\end{example}

We build the semantics of open pNets as an open automaton over the states  given by 
Definition~\ref{def-states}. The open transitions first
 project the global state into states of the leaves, then apply
pLTS transitions on these states, and compose them with the sort of the holes. 
The semantics    instantiates fresh variables using the predicate $\fresh(x)$, additionally, for an action 
$\alpha$, $\fresh(\alpha)$ means all variables in $\alpha$ are fresh.

\begin{defi}[Semantics of open pNets]
	\label{def:operationalSemantics} The semantics of a pNet $\pNet$ is an open automaton $A\!= 
	\mylangle \symb{Holes}(\pNet),\symb{States}(\pNet),\symb{InitState}(\pNet),\symb{vars}(P),\mathcal{T} \myrangle$ where $\mathcal{T}$   is the smallest set of open transitions such that $\mathcal{T}=\{OT\,|\,\pNet \models OT \}$ and	$\pNet \models OT$	is defined by the following  rules:
	

\begin{itemize}
\item The rule for a pLTS  checks that the guard is verified and transforms assignments into post-conditions:		
\begin{mathpar}\inferrule
		{ s \xrightarrow{\langle \alpha,~e_b,~(x_j\!:= {e}_j)^{j\in
					J}\rangle} s'\in \to  }
		{ \mylangle  S,s_0, \to \myrangle
			\models
			\openrule
			{\emptyset ,
			e_b,\left\{x_j\gets e_j\right\}^{j\in J}}
			{\ostate{s} \OTarrow{\alpha} \ostate{s'}}
		}\quad {\TrUn}
\end{mathpar}
	
\item	The second rule deals with pNet nodes: for each possible
	synchronisation vector (of index $k$) applicable to the rule subject, the premisses
	include one {\em open transition} for each sub-pNet involved, one possible
	{\em action} for each hole involved, and the predicate relating these
	with the resulting action of the vector. The sub-pNets involved are split between two 
	sets, $I_2$ for sub-pNets that are pLTSs (with open transitions obtained by rule \textbf{Tr1}), and $I_1$ for the sub-pNets that are not pLTSs (with open transitions obtained by rule \textbf{Tr2}), $J$ is the set of 
	holes involved in the transition\footnote{Formally, if $SV_k \!=\! \SV{({\alpha'})_m^{m 
	\in M}}{\alpha'}{e_b}$ is a synchronisation vector  of \pNet\  then $J=M\cap 
	\Holes(\pNet)$, $I_2=M\cap \Leaves(\pNet)$,  $I_1=M\setminus J \setminus 
	I_2$}\footnote{We could replace $I_1$ and $I_2$ by their formal definition in \textbf{Tr2} but the rule would be more difficult to read.}.                                                                    
\begin{mathpar}
    \mprset {vskip=.45ex}
\inferrule
    {
\Leaves(\mylangle {\pNet}_m^{m\in I}, \set{\Sort}, \symb{SV}_k^{\,k\in 
    	K}\myrangle) \!=\! \pLTS_l^{\,l\in L} \quad  	
k\!\in\! K \quad SV_k \!=\! \SV{(\alpha'_m)^{m \in I_1\uplus I_2\uplus J}}{\alpha'}{e_b} 
\\
\\     	
	\forall m\!\!\in\!\! I_1. {\pNet_m 
	\models\openrule
    	{\beta_{j}^{j\in J_m}, \Pred_m, \Post_m}
    	{\ostate{s_{i}^{i \in L_m}} \OTarrow {\alpha_m}
    		\ostate{(s_i^\prime)^{i\in L_m}}} }	
  \qquad
\forall m\!\!\in\!\! I_2.		{ \pNet_m 
    	 \models
    	\openrule
    	{\emptyset, \Pred_m, \Post_m}
    	{\ostate{s_m} \OTarrow {\alpha_m}
    		\ostate{s_m'}} }\\\\
    J' = \biguplus_{m\in I_1}\!\! J_m \uplus J	\\
    	\Pred = \bigwedge_{m\in I_1\uplus I_2}\!\! \Pred_m \land
    	\Predsv(SV_k,\alpha_m^{m\in I_1\uplus I_2},\beta_j^{j\in J},\alpha)\\ 
    	\forall i\in	L\backslash \left(\biguplus_{m\in I_1}\!\! L_m \uplus I_2\right).\,s'_i=s_i \\
    \fresh(\alpha'_m,\alpha',\beta_j^{j\in J},\alpha) 
    }
    {\mylangle {\pNet}_m^{m\in I}, \set{\Sort}, \symb{SV}_k^{\,k\in K}\myrangle
    	\models
    	{\openrule
    		{
    		\beta_j^{j\in J^\prime}, \Pred,  \biguplus_{m\in I_1\uplus I_2} 
    		\Post_m}
    		{\ostate{s_i^{i\in L}} \OTarrow {\alpha}
    			\ostate{(s_i^\prime)^{i\in L}}}
    	}
    }\quad {\TrDeux}
\end{mathpar} 
\end{itemize} 
	\medskip
\end{defi}
        	A key to understand this rule is that the open transitions are
	expressed in terms of the leaves and holes of the whole pNet structure,
	i.e. a flatten view of the pNet. For example, $L$ is the index set of the
	Leaves, $L_m$ the index set of the leaves of one sub-pNet indexed $m$, so all $L_m$
	are disjoint subsets of $L$. Thus the states in the open transitions,
	at each level, are tuples including states of all the
	leaves of the pNet, not only those involved in the chosen
	synchronisation vector.

Note that  the construction is symbolic, and each open transition deduced expresses a whole family of
behaviours, for any possible value of the variables.

In \cite{henrio:Forte2016}, we have  shown a detailed example of the construction of a complex open transition, building a deduction tree using rules \TrUn ~and \TrDeux.
We have also shown in \cite{henrio:Forte2016} that an open pNet
with finite synchronisation sets, finitely many leaves and
holes, and each pLTS at leaves having a finite number of states and
(symbolic) transitions, has a finite automaton. The algorithm for building such an automaton can be found in~\cite{QBMZ-AVOCS18}.







\begin{figure}[ht]
   \centerline{\includegraphics[width=13cm]{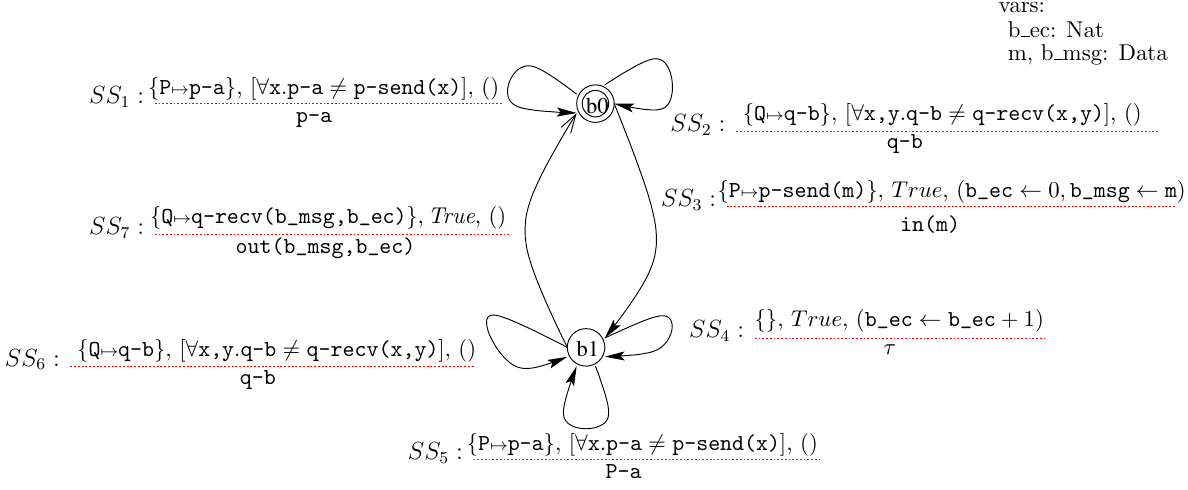}}
   \caption{Open automaton for  $\symb{SimpleProtocolSpec}$}
   \label{SimpleProtCounter:SpecOA}
\end{figure}

\subsection*{Example}  Figure~\ref{SimpleProtCounter:SpecOA} shows the open automaton computed from the \symb{SimpleProtocolSpec} pNet  given in Figure \ref{SimpleProt:Spec}. 
For later references, we name $SS_i$ the transitions of this (strong)
specification automaton while transitions of the
 \symb{SimpleProtocolImpl} pNet are labelled $SI_i$.
 In the figures we
annotate each open automaton with the set of its variables.

 \begin{figure}[ht]
  \centerline{\includegraphics[width=15cm]{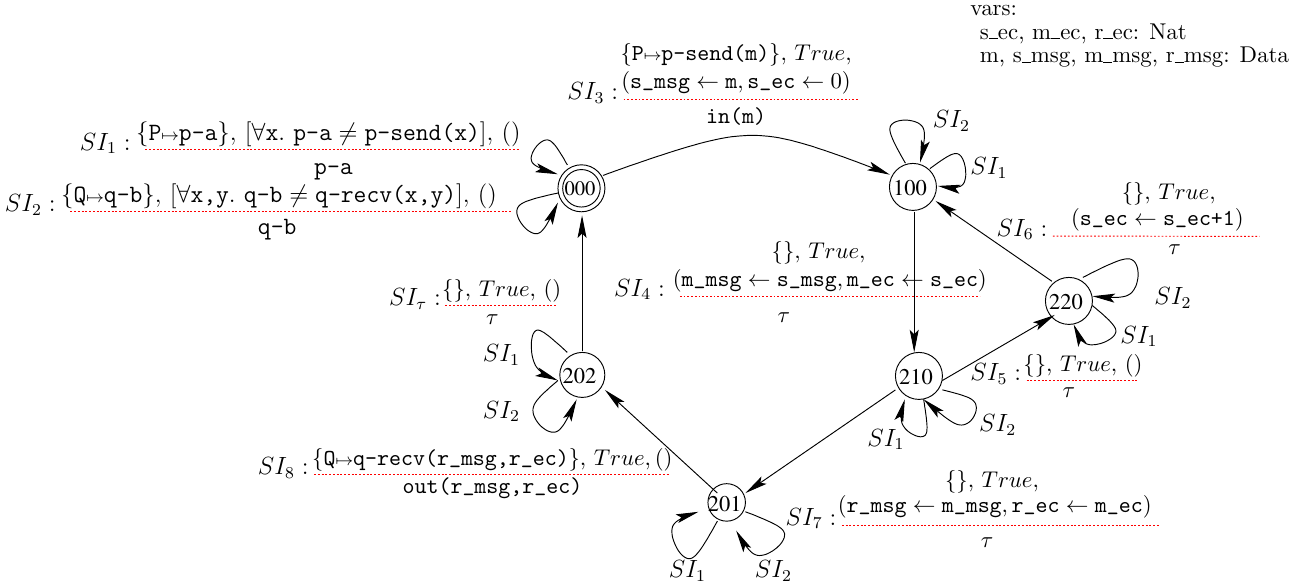}}
  \caption{Open automaton for $\symb{SimpleProtocolImpl}$}  \label{SimpleProtCounter:ImplOA}
\end{figure}

    Figure \ref{SimpleProtCounter:ImplOA} shows the open automaton of  \symb{SimpleProtocolImpl} from Figure \ref{SimpleProt:Impl}. In this drawing, we have short labels for states, representing $\ostate{s_0,m_0,r_0}$ by \texttt{000}. Note that open transitions are denoted $\texttt{SI}_i$ and tau open transition by  $\texttt{SI}_{\tau}$. The resulting behaviour is quite simple:  we have a main loop including receiving a message from $P$ and transmitting the same message to $Q$, with some intermediate $\tau$ actions from the internal communications between the protocol processes. In most of the transitions, you can observe that data is propagated between the successive pLTS variables (holding the message, and the error counter value). On the right of the figure, there is a loop of $\tau$ actions ($\texttt{SI}_4$, $\texttt{SI}_5$ and $\texttt{SI}_6$)  showing the handling of errors and the incrementation of the error counter.

\subsection{pNet Composition Properties: composition of open transitions}
The semantics of open pNets allows us to prove two crucial properties relating pNet composition with pNet semantics: open transition of a composed pNet can be decomposed into open transitions of its composing sub-pNets, and conversely, from the open transitions of sub-pNets,  an open transition of the composed pNet can be built.

We start with a decomposition property: from one open transition of $P[Q]_{j_0}$, we exhibit 
corresponding behaviours of $P$ and $Q$, and determine the relation between their 
predicates.
\begin{lem}[Open transition decomposition\label{lem-decompose}] Consider two pNets $P$ and $Q$ that are not pLTSs\footnote{A similar lemma can be proven for a pLTS $Q$}.
	Let $\Leaves(Q)=p_l^{l\in L_Q}$ and suppose:
	\[ P[Q]_{j_0}  
		\models
		{\openrule
			{
				\beta_j^{j\in J}, \Pred,  
				\Post}
			{\ostate{s_i^{i\in L}} \OTarrow {\alpha}
				\ostate{s_i'^{\, i\in L}}}
		}
	\]
		with  $J\cap\Holes(Q)\neq\emptyset$ or $\exists i\in L_Q.\,s_i\neq s'_i$, i.e. $Q$ takes part in the reduction.
		 Then there exist $\alpha_Q$, $\Pred\,'$, $\Pred\,''$, 
		$\Post\,'$, $\Post\,''$ s.t.:\\[-2ex]
		\begin{mathpar}
		P\models{\openrule
			{
				\beta_j^{j\in (J\setminus \Holes(Q)) \cup \{j_0\}}, 
				\Pred\,',  
				\Post\,'}
			{\ostate{s_i^{i\in L\setminus L_Q}} \OTarrow {\alpha}
				\ostate{s_i'^{\,i\in L\setminus L_Q}}}
		}
	\vspace{-2.2ex}\\\text{and~\qquad\qquad~}
		Q\models{\openrule
			{
				\beta_j^{j\in J\cap\Holes(Q)}, \Pred\,'',  
				\Post\,''}
			{\ostate{s_i^{i\in L_Q}} \OTarrow {\alpha_Q}
				\ostate{s_i'^{\,i\in L_Q}}}
		}
		\end{mathpar}
		and  $\Pred \iff \Pred\,'
		\land \Pred\,''\land \alpha_Q=\beta_{j_0}$, $\Post=\Post\,'\uplus 
		\Post\,''$ where $\Post\,''$ is the restriction of $\Post$ over variables of 
		$Q$.
\end{lem}

Lemma \ref{lem-compose} is combining an open transition of $P$ with
an open transition of $Q$, and building a corresponding transition of
$P[Q]_{j_0}$  by assembling their elements.

\begin{lem}[Open transition composition]\label{lem-compose} 
	Suppose $j_0\in J$ and:\\[-1ex]
\begin{mathpar}
P\models{\openrule
	{
		\beta_j^{j\in J}, 
		\Pred,  
		\Post}
	{\ostate{s_i^{i\in L}} \OTarrow {\alpha}
		\ostate{s_i'^{\, i\in L}}}
}
\quad\text{~~and~~}\quad
Q\models{\openrule
	{
		\beta_j^{j\in J_Q},
		 \Pred\,',  
		\Post\,'}
	{\ostate{s_i^{i\in L_Q}} \OTarrow {\alpha_Q}
		\ostate{s_i'^{\, i\in L_Q}}}
}
\end{mathpar}
Then, we have:\\[-1ex]
	\[ P[Q]_{j_0}  
	\models
	{\openrule
		{
			\beta_j^{(j\in J\setminus\{j_0\}) \uplus J_Q}, 
			\Pred\land\Pred\,'\land \alpha_Q=\beta_{j_0},  
			\Post\uplus \Post\,'}
		{\ostate{s_i^{i\in L\uplus L_Q}} \OTarrow {\alpha}
			\ostate{s_i'^{\, i\in L\uplus L_Q}}}
	}
	\]
\end{lem}

Note that this does not mean that any two pNets can be composed and produce an open 
transition. Indeed, the predicate $\Pred\land\Pred\,'\land \alpha_Q=\beta_{j_0}$ is often not  satisfiable, in particular if the action  $\alpha_Q$ cannot be matched with $\beta_{j_0}$.
Note also that $\beta_{j_0}$ is  only used as an intermediate term inside formulas in the composed open transition: it 
does not appear as global action, and will not appear as an action of a hole.

\subsection{Bisimulation for open pNets -- a composable bisimulation theory}
\label{section:bisimulation-PN}
As  our symbolic operational semantics provides an open automaton, we can apply the notion of
	strong (symbolic) bisimulation on automata to open pNets.
\begin{defi}[FH-bisimulation for open pNets]\label{def:bisim-pnets}
Two pNets are FH-bisimilar if there exists a relation between their associated 
automata that is an FH-bisimulation and their initial states are in the relation (i.e. the predicate associated with the initial states is verifiable).
\end{defi}

We can now prove that pNet composition  preserves
FH-bisimulation. More precisely, one can define two preservation
properties, namely 1) when one hole of a pNet is filled by two bisimilar other (open) pNets; and 2) when the same hole in two bisimilar pNets are
filled by the same pNet, in other words, composing a pNet with two
bisimilar contexts. The general case will be obtained by
transitivity of the bisimulation relation (Theorem~\ref{thm-equiv}). 

\begin{thm}[Congruence]\label{thm-congr-eq}
	Consider an open pNet:
	$\pNet = \mylangle \pNet_i^{i\in I}, \Sort_j^{j\in J}, 
	\set{\symb{SV}}\myrangle$.
	Let $j_0\in J$ be a hole. Let $\pNetQ$ and $\pNetQ'$ be two FH-bisimilar pNets such that\footnote{Note that $\Sortop(\pNetQ)=\Sortop(\pNetQ')$ is 
	ensured by 
	strong bisimilarity.} 
	$\Sortop(\pNetQ)=\Sortop(\pNetQ')=\Sort_{j_0}$. Then 
	$\pNet[\pNetQ]_{j_0}$ and 
	$\pNet[\pNetQ']_{j_0}$ are FH-bisimilar.
\end{thm}

\begin{thm}[Context equivalence]\label{thm-ctxt-eq}
	Consider two  open pNets
	$\pNet = \mylangle \pNet_i^{i\in I}, \Sort_j^{j\in J}, 
	\set{\symb{SV}}\myrangle$ and 	$\pNet' = \mylangle {\pNet'}_i^{i\in I}, 
	\Sort_j^{j\in 
	J}, 	\set{\symb{SV'}}\myrangle$ that are FH-bisimilar
	(recall they must have the same holes to be bisimilar).
	Let $j_0\in J$ be a hole, and $Q$ be a pNet such that $\Sortop(Q)=\Sort_{j_0}$. Then 
	$\pNet[Q]_{j_0}$ and 
	$\pNet'[Q]_{j_0}$ are FH-bisimilar.
\end{thm}

Finally, the previous theorems can be composed to state a general theorem about 
composability and FH-bisimilarity.
\begin{thm}[Composability] \label{thm-composability}
	Consider two FH-bisimilar pNets with an arbitrary number of holes, when replacing, 
	inside those two original pNets, a subset of the holes by FH-bisimilar pNets, we 
	obtain two FH-bisimilar pNets.
\end{thm}
This theorem is quite powerful. It somehow implies that the theory of open pNets is convenient to study properties of process composition. Open pNets can indeed be used to study process operators and process algebras, as shown in~\cite{henrio:Forte2016} where compositional properties are extremely useful. In the case of interaction protocols~\cite{BHHM:FACS11}, composition of bisimulation can justify abstractions used in some parts of the application.

\section{Weak bisimulation}\label{sec:weak}

Weak symbolic bisimulation was introduced to relate transition systems
that have indistinguishable behaviour, with respect to some definition
of \emph{internal actions} that are considered local to some
subsystem, and consequently cannot be observed, nor used for
synchronisation with their context.
The notion of non-observable actions varies in different contexts,
e.g. $tau$ in CCS, and $i$ in Lotos, we could define classically a set of
\emph{internal/non-observable actions} depending on a specific action
algebra. In this paper, to simplify the notations, we will simply use $\tau$ as the single non-observable action; the generalisation of our results to a set of non-observable actions is trivial. 
Naturally, a non-observable action cannot be synchronised with
actions of other systems in its environment. 
We show here that under such assumption of non-observability of $\tau$ actions, see Definition~\ref{def:Non-ObsTau}, we can define a weak bisimulation relation that is compositional, in the sense of open pNet composition. In this section we will first define a notion of weak open transition similar to open transition. In fact a weak open transition is made of several open transitions labelled as non-observable transitions, plus potentially one observable open transition. This allows us to define weak open automata, and a weak bisimulation relation based on these weak open automata. Finally, we apply this weak bisimulation to open pNets, obtain a weak bisimilarity relationship for open pNets, and prove that this relation has compositional properties.



\subsection{Preliminary definitions and notations}

We first specify in terms of open transition, what it means for an action to be non-observable. Namely, we constraint ourselves to system where the emission of a $\tau$ action by a sub-pNet cannot be observed by the surrounding pNets. In other words, a pNet cannot change its state, or emit a specific observable action when one of its holes emits a $\tau$ action.

More precisely, we state that $\tau$ is not observable if the automaton always allows any $\tau$ transition from holes, and additionally the global transition resulting from a $\tau$ action of a hole is a $\tau$ transition not changing the pNet's state.
We define $\Id(V)$ as the identity function on the set of variables $V$.
\begin{defi}[Non-observability of $\tau$ actions for open automata]\label{def:Non-ObsTau}
An open automaton $A = \mylangle J,\mathcal{S},s_0,V,\mathcal{T}\myrangle$ \emph{cannot observe $\tau$ actions} if and only if for all $j$ in $J$ and $s$ in $\mathcal{S}$ we have:
\begin{enumerate}
\item
\[ \openrule
         {
           (j\mapsto\tau),\True,\Id(V)}
         {s \OTarrow {\tau} s}
         \in \mathcal{T}
\]
and 
\item for all $\beta_j$, $J$,  $\alpha$,  $s$, $s'$, $\Pred$, $\Post$  such that
\[ \openrule
         {
           \beta_j^{j\in J},\Pred,\Post}
         {s \OTarrow {\alpha} s'}
         \in \mathcal{T} \] If there exists $j$ such that $\beta_j=\tau$ then we have: \[ \alpha=\tau\land s=s'\land \Pred=\True\land\Post=\Id(V) \land J=\{j\}
\]
\end{enumerate}
\end{defi}
The first statement of the definition states that the open automaton must allow a hole to do a silent action at any time, and must not observe it, i.e. it cannot change its internal state because a hole did a $\tau$ transition. The second statement ensures that there cannot be in the open automaton other transitions that would be able to observe a $\tau$ action from a hole: statement (2) states that all the open transitions where a hole does  a $\tau$ action must be of the shape given in statement (1). The condition $J=\{j\}$ is a bit restrictive, it could safely be replaced by $\forall j\in J.\, \beta_j=\tau$, allowing the other holes to perform $\tau$ transitions too (because these $\tau$ actions cannot be observed).

By definition, one weak open transition contains  several open transitions, where  each open transition can require an observable action from a given hole, the same hole might have to emit several observable actions for a single weak open transition to occur. 
Consequently, for a weak open transition to trigger, a sequence of actions from a given hole may be required.

Thus, we let $\gamma$ range over sequences of action terms and use $\dotcup$ as the concatenation operator that appends sequences of action terms: given two sequences of action terms  $\gamma\dotcup\gamma '$ concatenates the two sequences. The operation is lifted to indexed sets of sequences:   at each index $i$, $\set {\gamma_1}\dotcup \set {\gamma_2}$ concatenates the sequences of actions at index $i$ of $\set{\gamma_1}$ and the one at index $i$ of $\set {\gamma_2}$\footnote{One of the two sequences is empty when $i\not \in \dom(\set{\gamma_1})$ or $i\not \in \dom(\set{\gamma_2})$ .}. $[a]$ denotes a sequence with a single element.

As required actions are now sequences of observable actions, we need an operator to build them from set of actions that occur in open transitions, i.e. an operator that takes a set of actions performed by one hole and produces a sequence of observable actions.

Thus we define $\vis{\set\beta}$ as the mapping $\set\beta$  with only observable actions of the holes in $I$, but where each element is either empty or a list of length 1:
 \[\vis{\beta_i^{i\in I}} = [\beta_i]^{i\in I'}\text{ where }I'=\left\{i| i\in I \land \beta_i\neq \tau\right\}\]

As an example the $\vis{\set\beta}$ built from the transition $OT_1$ in Example~\ref{OT:SimpleProt}, page \pageref{OT:SimpleProt} is $\texttt{P}\mapsto [\texttt{p-send(m)}]$. Remark that in our simple example no $\tau$ transition involves any visible action from a hole, so we have no $\beta$ sequences of length longer than 1 in the weak automaton.

\subsection{Weak open transition definition}

Because of the non-observability property (Definition~\ref{def:Non-ObsTau}), it is possible to add any number of $\tau$ transitions of the holes before or after any open transition freely. This property justifies the fact that we can abstract away $\tau$ transitions from holes in the definition of a weak open transition.
We define weak open transitions similarly to open transitions except that holes can perform \emph{sequences of observable actions} instead of single actions (observable or not). Compared to the definition of open transition, this small change has a significant impact as a single weak transition is the composition of several transitions of the holes.

%

\def\InvAct{\mathcal{Inv}}
%

\begin{defi}[Weak open transition (WOT)]\label{def:weakOT}
A weak open transition over a
	set $J$ of holes with sorts $\Sort_j^{j\in J}$ and a set of states $\mathcal{S}$ is 
	a structure of the form:	
\begin{mathpar}
 \openrule
         {
           \gamma_j^{j\in J'},\Pred,\Post}
         {s \OTWeakarrow {\alpha} s'}
 \end{mathpar}
	Where $J'\subseteq J$, $s, s'\in\mathcal{S}$ and $\gamma_j$
        is a list of transitions of the hole $j$, with each element of the list in $\Sort_j$. $\alpha$ is an action 
        label denoting the resulting action
        of this open transition. \Pred\ and \Post\ are defined similarly to Definition~\ref{def:OT}. We use $\WT$ to range over sets of weak open transitions.

A weak open automaton $\mylangle J,\mathcal{S},s_0,V,\WT\myrangle$ is similar to an open automaton  except that $\WT$ is a set of weak open transitions over $J$ and $\mathcal{S}$.
\end{defi}

A weak open transition labelled $\alpha$ can be seen as a sequence of open transitions that are all labelled $\tau$ except one that is labelled $\alpha$; however conditions on predicates, effects, and states must be verified for this sequence to be fired.

We are now able to build a weak open automaton from an open automaton. This is done in a way that resembles the process of $\tau$ saturation: we add  $\tau$ open transitions before or after another (observable or not) open transition.
\begin{defi}[Building a weak open automaton]\label{def:buildweakOT}
  Let $A = \mylangle J,\mathcal{S},s_0,V,\mathcal{T}\myrangle$ be an open automaton. 
The weak open automaton \emph{derived} from $A$ is an open automaton  $\mylangle J,\mathcal{S},s_0,V,\WT\myrangle$ where $\WT$ is derived from $\mathcal{T}$ by saturation, applying the following rules:
\begin{mathpar}
 \openrule
         {
           \emptyset,\symb{True},\Id(V)}
         {s \OTWeakarrow {\tau} s} \in \WT  \qquad \WTUn
 \end{mathpar}
and
\begin{mathpar}
\inferrule{
 \openrule
         {
           \set{\beta},\Pred,\Post}
         {s \OTarrow {\alpha} s'} \in \mathcal{T}
} 
{ \openrule
         {
           \vis{\set{\beta}}\!,\Pred,\Post
				 } {s \OTWeakarrow {\alpha} s'} \in \WT
}\qquad \WTDeux
 \end{mathpar}
 and
\begin{mathpar}
\inferrule {\openrule
         {
           \set{\gamma_1},\Pred_1,\Post_1   }
         {s \OTWeakarrow {\tau} s_1} \in \WT
\qquad
\openrule
         {
           \set{\gamma_2},\Pred_2,\Post_2  }
         {s_1 \OTWeakarrow {\alpha} s_2} \in \WT
\qquad
\openrule
         {
           \set{\gamma_3},\Pred_3,\Post_3  }
         {s_2 \OTWeakarrow {\tau} s'} \in\WT
\\
\Pred=\Pred_1\land\Pred_2\subst{\Post_1}\land \Pred_3\subst{\Post_2\shortodot\Post_1}
\\
\set{\gamma}=\set{\gamma_1}\dotcup \set{\gamma_2}\subst{\Post_1}\dotcup\set{\gamma_3}\subst{\Post_2\shortodot \Post_1}\\
\alpha'=\alpha\subst{\Post_1}
}
{
\openrule
         {\set{\gamma}
           ,
		\Pred,
				\Post_3\shortodot\Post_2\shortodot\Post_1} 
         {s \OTWeakarrow {\alpha'} s'} \in\WT
} \WTTrois
\end{mathpar}
 
\end{defi}
Rule $\WTUn$ states that it is always possible to do a non-observable transition, where the state is unchanged and the holes perform no action. Rule~$\WTDeux$ states that each open transition can be considered as a weak open transition. The last rule is the most interesting:  Rule~$\WTTrois$ allows any number of $\tau$ transitions before or after any weak open transition. This rules carefully composes predicates, effects, and actions of the holes, indeed in the rule, predicate $\Pred_2$ manipulates variables of $s_1$ that result from the first weak open transition. Their values thus depend on the initial state but also on the effect (as a substitution function $\Post_1$) of the first weak open transition. In the same manner, $\Pred_3$ must be applied the substitution defined by the composition $\Post_2\shortodot\Post_1$. Similarly, effects on variables must be applied to obtain the global effect of the composed weak open transition, it must also be applied to observable actions of the holes, and to the global action of the weak open transition.



\begin{figure}[ht]
   \centerline{\includegraphics[width=13cm]{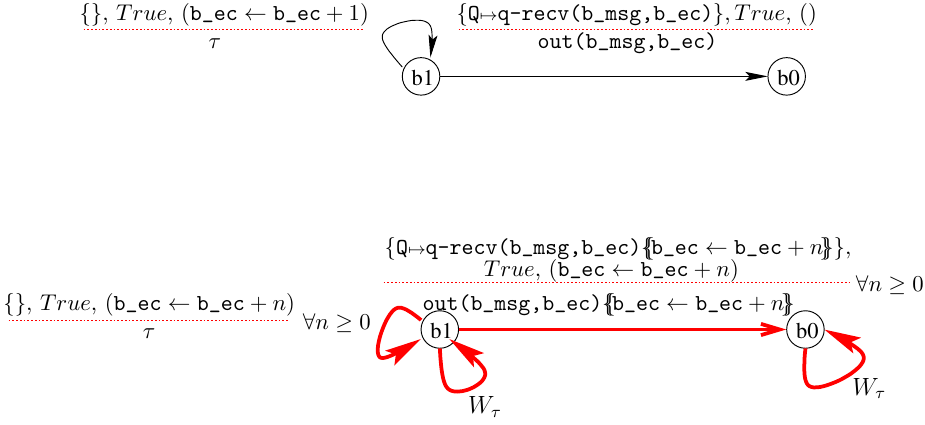}}
  \caption{Construction of an example of weak open transition}
   \label{WOT2}
\end{figure}

\begin{example} [A weak open-transition]
Figure \ref{WOT2} shows the construction of one of the weak transitions of the open automaton of \symb{SimpleProtocolSpec}. On the top we show the subset of the original open automaton (from Figure \ref{SimpleProtCounter:SpecOA}) considered here, and at the bottom the generated weak transition.  For readability, we abbreviate the weak open transitions encoded by $\openrule   {\{\}, True,	() } {s \OTWeakarrow \tau s'}$  as $W_\tau$. The weak open transition shown here is the transition delivering the result of the algorithm to hole $Q$ by applying rules: \WTUn,\WTDeux, and \WTTrois. First rule \WTUn~ adds a $WT_\tau$ loop on each state. Rule \WTDeux~ transforms each 3 OTs into WOTs.   Then consider application of Rule \WTTrois~ on a sequence 3  WOTs.   $\openrule
         {\{\}, True,
			(\texttt{b\_ec}\gets \texttt{b\_ec}+1) }
         {\texttt{b1}\OTWeakarrow \tau \texttt{b1}}$; $\openrule
         {\{\}, True,
			(\texttt{b\_ec}\gets \texttt{b\_ec}+1) }
         {\texttt{b1} \OTWeakarrow \tau \texttt{b1}}$;  $\openrule
         {\{\}, True,
			() }
         {\texttt{b1} \OTWeakarrow \tau \texttt{b1}}$. The result will be:   $\openrule
         {\{\}, True,
			(\texttt{b\_ec}\gets \texttt{b\_ec}+2) }
         {\texttt{b1} \OTWeakarrow \tau \texttt{b1}}$. We can iterate this construction an arbitrary number of times, getting for any natural number $n$ a weak open transition:
  $\openrule
         {\emptyset, True,
			(\texttt{ec}\gets \texttt{ec}+n) }
        {\texttt{b1} \OTWeakarrow \tau \texttt{b1}}\, \forall n \geq 0$.  Finally,  applying again \WTTrois, and using the central open transition having \texttt{\nounderline{out(b\_msg,b\_ec)}}  as $\alpha$, we get the resulting weak open transition between b1 and b0 (as shown in Figure \ref{WOT2}).  Applying the substitutions finally yields the weak transitions family $WS_7$ in Figure  \ref{SimpleProtCounter:WeakSpecOA}.

\end{example}

\begin{figure}[h]
   \centerline{\includegraphics[width=11cm]{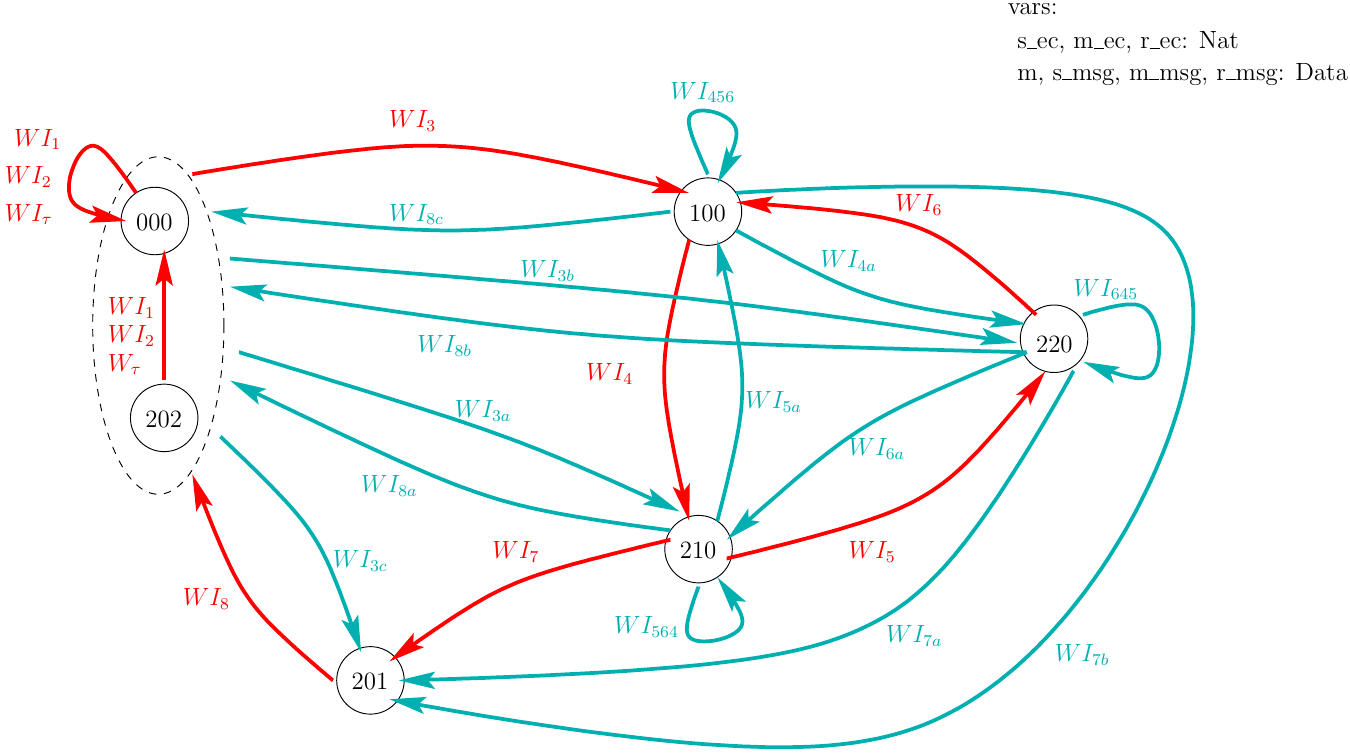}}
  \caption{Weak Open Automaton of $\symb{SimpleProtocolImpl}$}
   \label{SimpleProtCounter:ImplWOA2}
\end{figure}

  \begin{example}[Weak open automata]
    Figures \ref{SimpleProtCounter:WeakSpecOA} and \ref{SimpleProtCounter:ImplWOA2} respectively show the weak automata of \symb{SimpleProtocolSpec} and \symb{SimpleProtocolImpl}. We encode weak open transitions  by $WS$ on the specification model and by $WI$ on the implementation model.

\begin{figure}[h]
   \centerline{\includegraphics[width=15cm]{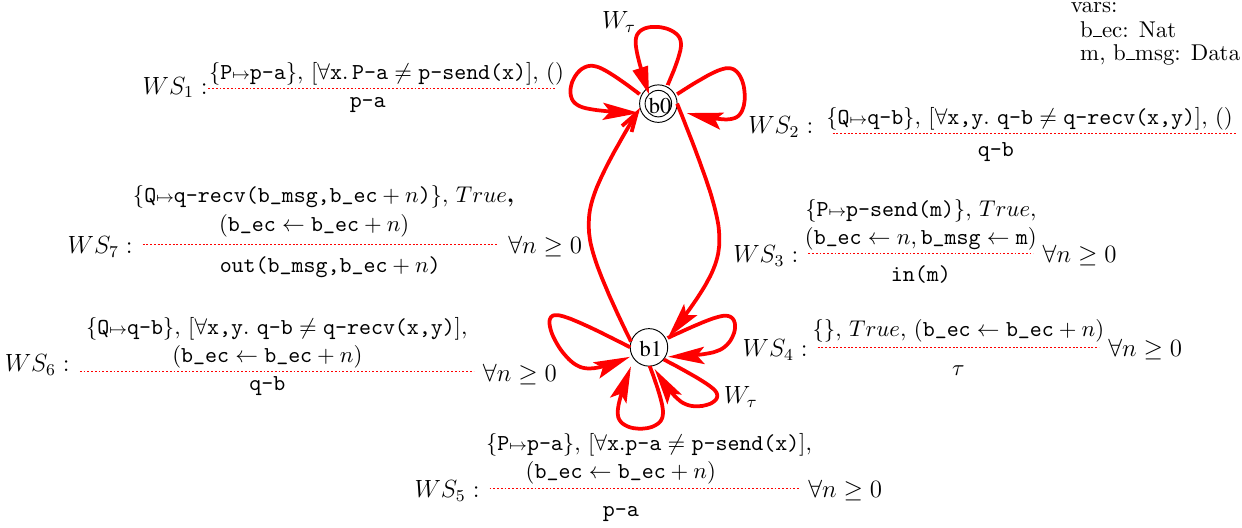}}
  \caption{Weak Open Automaton of  $\symb{SimpleProtocolSpec}$}
   \label{SimpleProtCounter:WeakSpecOA}
\end{figure}

 For readability, we only give names to the  weak open transitions of \symb{SimpleProtocolImpl} in Figure \ref{SimpleProtCounter:ImplWOA2}; we detail some of these transitions below and the full list is included in Appendix \ref{Appendix:FullExample}.
Let us point out that the weak OT loops ($WI_1$,$WI_2$ and $W_\tau$) on state ${000}$ are also present in all other states, we did not repeat them. Additionally, many WOTs are similar, and numbered accordingly as 3, 3a, 3b, 3c and 8, 8a, 8b, 8c respectively: they only differ by their respective source or target states; the "variant" WOTs appear in blue in   Figure~\ref{SimpleProtCounter:ImplWOA2}.
\end{example}


Now let us give some details about the construction of the weak automaton of the \symb{SimpleProtocolImpl} pNet, obtained by application of the weak rules as explained above. We concentrate on weak open transitions $WI_3$ and $WI_4$. Let us denote as $post_n$ the effect (as a substitution function) of the strong open transitions $SI_n$ from Figure \ref{SimpleProtCounter:ImplOA}:
\smallskip

$post_3 = (\texttt{s\_msg}\gets \texttt{m}, \texttt{s\_ec}\gets 0)$

$post_4 = (\texttt{m\_msg}\gets \texttt{s\_msg}, \texttt{m\_ec}\gets \texttt{s\_ec})$

$post_5 = ()$

$post_6 = (\texttt{s\_ec}\gets \texttt{s\_ec+1})$

\medskip

Then the effect of one single $100 \xrightarrow{OT_4} 210 \xrightarrow{OT_5} 220 \xrightarrow{OT_6} 100$ loop is\footnote{when showing  the result of $Posts$ composition, we will omit the identity substitution functions introduced by the $\shortodot$ definition in page \pageref{def:substitutions}}:
\[post_{456} = post_6 \shortodot\ post_5 \shortodot\ post_4
= (\texttt{s\_ec}\gets \texttt{s\_ec}+1)\]

So if we denote ${post_{456*}}$ any iteration of this loop, we get ${post_{456*}} = (\texttt{s\_ec}\gets \texttt{s\_ec}+n)$ for any $n\ge 0$, and the \Post~ of the weak OT  $WI_{3}$ is:\\
 $Post_3 = post_{456*}\shortodot\ post_3 = (\texttt{s\_msg}\gets m, \texttt{s\_ec}\gets n), \forall n\ge 0$ and \Post~ of  $WI_{3a}$  is:\\ $post_4\shortodot {post_{456*}}\shortodot\ post_3 = (\texttt{m\_msg}\gets m, \texttt{m\_ec}\gets n), \forall n\ge 0$.
\medskip

We can now show some of the weak OTs of Figure \ref{SimpleProtCounter:ImplWOA2} (the full table is included in Appendix \ref{Appendix:FullExample}).
As we have seen above, the effect of rule $WT_3$ when a silent action have an effect on the variable $ec$ will generate an infinite family of WOTs, depending on the number of iterations through the loops. We denote these families using a "meta-variable" $n$, ranging over $\texttt{Nat}$.
\smallskip

$ WI_1 = \openrule
{\{\texttt{P}\mapsto \texttt{p-a}\}, [\forall \texttt{x}. \texttt{p-a} \neq \texttt{p-send(x)} ], ()}
{s \OTWeakarrow {\texttt{p-a}} s}
$ (for any  $s \in S$)

$ \forall n\ge 0.\, WI_3(n) = \openrule
  {\{\texttt{P}\mapsto \texttt{p-send(m)}\}, True,
    (\texttt{s\_msg}\gets \texttt{m}, \texttt{s\_ec}\gets n)}
  {000 \OTWeakarrow {\nounderline{\texttt{in(m)}}} 100}
        $

$  \forall n\ge 0.\, WI_4(n) = \openrule
         {\{\}, True, 
   (\texttt{m\_msg}\!\gets\! \texttt{s\_msg}, \texttt{m\_ec}\!\gets\! \texttt{s\_ec}+n, \texttt{s\_ec}\!\gets\! \texttt{s\_ec}+n)}
         {100 \OTWeakarrow {\tau} 210}
        $

$  \forall n\ge 0.\, WI_{456}(n) = \openrule
         {\{\}, True, 
    (\texttt{s\_ec}\gets \texttt{s\_ec}+n)}
  {100 \OTWeakarrow {\tau} 100}
        $
        
   \medskip     
The  \Post~of the weak OT $ WI_{6a}$ is:
$$
\begin{array}{l@{}l}
Post_{6a}&= post_{4}\shortodot\ post_{456*}\shortodot\ post_{6}\\ 
&=  (\texttt{m\_msg}\gets \texttt{s\_msg}, \texttt{m\_ec}\gets \texttt{s\_ec})\,
\shortodot (\texttt{s\_ec}\gets \texttt{s\_ec}+n) 
\shortodot (\texttt{s\_ec}\gets \texttt{s\_ec}+1) \\
&= (\texttt{m\_msg}\!\gets \texttt{s\_msg}, \texttt{m\_ec}\!\gets \texttt{s\_ec}+1\!+\!n, \texttt{s\_ec}\!\gets \texttt{s\_ec}+1\!+\!n)
\end{array}
$$

So we get:

$ \forall n\ge 0.\, WI_{6a}(n) = \openrule
         {\{\}, True, 
         (\texttt{m\_ec}\gets \texttt{s\_ec}+1+n, \texttt{s\_ec}\gets \texttt{s\_ec}+1+n)}
         {220 \OTWeakarrow {\tau} 210}$ 


\subsection{Composition properties: composition of weak open transitions}
We now have two different semantics for open pNets: a strong semantics, defined  as an open automaton, and as a weak semantics, defined as a weak open automaton. Like the open automaton, the weak open automaton features valuable composition properties. We can exhibit  a composition property and a decomposition property that relate open pNet composition with their semantics, defined as weak open automata. These are however technically more complex than the ones for open automata because each hole performs a set of actions, and thus a composed transition is the composition of one transition of the top-level pNet and a sequence of transitions of the sub-pNet that fills its hole. They can be found as Lemmas~\ref{lem-decomposeWOT}, Lemma~\ref{lem-Weakcompose1}, and Lemma~\ref{lem-Weakcompose} in Appendix~\ref{sec:app-composition}.

\subsection{Weak FH-bisimulation}
For defining a bisimulation relation between weak open automata, two options are possible. Either we define a simulation similar to the strong simulation but based on open automata, this would look like the FH-simulation but would need to be adapted to weak open transitions. Or we define directly and classically a weak FH-simulation as a relation between two open automata, relating the open transition of the first one with the transition of the weak open automaton derived from the second one. 

The definition below specifies how a set of weak open transitions can simulate an open transition, and under which condition; this is used to relate, by weak FH-bisimulation, two open automata by reasoning on the weak open automata that can be derived from the strong ones.
This is defined formally as follows.

\begin{defi}[Weak FH-bisimulation]\label{def-Weak-bisim} ~\\
\noindent
Let $A_1 = \mylangle J,\mathcal{S}_1, s_0,V_1,
    \mathcal{T}_1\myrangle$ and $A_2 = \mylangle J,\mathcal{S}_2,t_0, V_2, \mathcal{T}_2\myrangle$ be open automata with disjoint sets of variables.
Let $\mylangle J,\mathcal{S}_1, s_0,V_1,
    \WT_1\myrangle$ and $\mylangle J,\mathcal{S}_2,t_0, V_2, \WT_2\myrangle$ be the
weak open automata derived from $A_1$ and $A_2$ respectively.
Let $\mathcal{R}$ a relation over
$\mathcal{S}_1$ and $\mathcal{S}_2$, as in Definition~\ref{def-FH-bisim}.

Then 
   $\mathcal{R}$ is a weak FH-bisimulation iff for any  states
$s\in\mathcal{S}_1$ and
$t\in\mathcal{S}_2$ such that $(s,t|\Pred_{s,t})\in\mathcal{R}$, we 
   have the following:

 \begin{itemize}
 \item  For any open transition $OT$ in $\mathcal{T}_1$:
 \begin{mathpar}
     \openrule
         {
           \beta_j^{j\in J'},\Pred_{OT},\Post_{OT}}
         {s \OTarrow {\alpha} s'}

\end{mathpar}
 there exists an indexed set of weak open transitions $\symb{WOT}_x^{\;x\in X} \subseteq \WT_2$:
 \begin{mathpar}
    \openrule
         {
           \gamma_{j x}^{j\in J_{x}}, \Pred_{OT_x},\Post_{OT_x}}
         {t \OTWeakarrow {\alpha_x} t_x}
\end{mathpar}
 such that  $\forall x.\, \{j\in J'|\beta_j\neq\tau\}=J_{x}, (s',t_x|\Pred_{s',t_x})\in \mathcal{R}$; 
 and  
\begin{multline*}
\Pred_{s,t} \land \Pred_{OT}\\
\hspace{1cm} \implies\!\!\! \displaystyle{\bigvee_{x\in X}\!
   \left( \forall j\in J_x. \vis{\beta_j}\!=\!\gamma_{jx}\! \land\! \Pred_{OT_x}
     \!\land\! \alpha\!=\!\alpha_x\! \land\!  
     \Pred_{s',t_x}\subst{\Post_{OT}\uplus\Post_{OT_x}}\right)}
\end{multline*}
    
 \item  and symmetrically any open transition from $t$ in $\mathcal{T}_2$ can be 
      covered by a set of weak transitions from $s$ in $\WT_1$.
 \end{itemize}

Two pNets are weak FH-bisimilar if there exists a relation between their associated 
automata that is a weak FH-bisimulation and their initial states are in the relation, i.e. 
the predicate associated to the relation between the initial states is \True.
 \end{defi}

Compared to strong bisimulation, except the obvious use of weak open transitions to simulate an open transition, the condition on predicate is slightly changed concerning actions of the holes. Indeed only the visible actions of the holes must be compared and they form a list of actions, but of length at most one.

Our first important result is that weak FH-bisimilarity is an equivalence in the same way as strong FH-bisimilarity:

\begin{thm}[Weak FH-Bisimulation is an equivalence]\label{thm-weak-equiv} Suppose $\mathcal{R}$ 
is a weak FH-bisimulation. Then $\mathcal{R}$ is an equivalence, that is, $\mathcal{R}$ is 
reflexive, symmetric and transitive.
\end{thm}
The proof  is detailed in Appendix \ref{app-WFH-equiv}, it follows a similar pattern as the proof that strong FH-bisimulation is an equivalence, but technical details are different, and in practice we rely on a variant of the definition of weak FH-bisimilarity; this equivalent version simulates a \emph{weak} open transition with a set of weak open transition. The careful use of the best definition of weak FH-bisimilarity makes the proof similar to the strong FH-bisimulation case.

\subsubsection*{Proving bisimulation in practice}

In practice, we are dealing with finite representations of the (infinite) open automata. In \cite{hou:hal-02406098}, we defined a slightly modified definition of the ``coverage'' proof obligation, in the case of strong FH-Bisimulation. This modification is required to manage in a finite way all possible instantiations of an OT. In the case of weak FH-Bisimulation, the proof obligation from Definition \ref{def-Weak-bisim} becomes:
      \begin{multline*}
\forall fv_{OT}.\, \Big\{ \Pred_{s,t} \land \Pred_{OT} \implies\\
 \bigvee_{x\in X}\!
  \left[\exists fv_{OT_x}.
    \left( \forall j\in J_x. \vis{\beta_j}\!=\!\gamma_{jx}\! \land\! \Pred_{OT_x}
     \!\land\! \alpha\!=\!\alpha_x\! \land\!  
     \Pred_{s',t_x}\subst{\Post_{OT}\uplus\Post_{OT_x}}\right)\right]\Big\}
\end{multline*}

where $fv_{OT}$ denotes the set of free variables of all expressions in $OT$.
\subsection{Weak bisimulation for open pNets}

Before defining a weak open automaton for the semantics of open pNets,
it is necessary to state under which condition a pNet is unable to
observe silent actions of its holes. In the setting of pNets this can
simply be expressed as a condition on the synchronisation
vectors. Precisely, the set of synchronisation vectors must contain
vectors that let silent actions go through the pNet,
i.e. synchronisation vectors where one hole does a $\tau$ transition,
and the global visible action is a $\tau$. Additionally, no other
synchronisation vector must be able to react on a silent action from a
hole, i.e. if a synchronisation vector observes a $\tau$ from a hole
it cannot synchronise it with another action nor emit an action that
is not $\tau$. This is formalised as follows:

\begin{defi}[Non-observability of silent actions for pNets]\label{def:non-obspNet}~\\
A pNet $\mylangle \pNet_i^{i\in I} , \Sort_j^{j\in J}, \set{\symb{SV}}\myrangle$
 \emph{cannot observe silent actions} if it verifies:\\ $\forall i\in I\uplus J.\, \SV{(i\mapsto \tau)}{\tau}{\True}\in \set{\symb{SV}}$ and 
\begin{equation*}
\forall \left(\SV{{(\alpha_i)}^{i\in I'}} 
{\alpha'} 
{e_b}\in \set{\symb{SV}}\right), 
\forall i\in I'\cap J.\, \alpha_i=\tau \implies 
\alpha'=\tau ~\land~ 
I'=\{i\}
\end{equation*}
\end{defi}

\begin{example}[CCS choice (counter-example)]\label{ex:CCSplus}
Here is the encoding of a choice operator.\\[1.5ex]
\begin{tabular}{cp{.53\linewidth}}
 \begin{minipage}{.43\linewidth}
 \includegraphics[width=\linewidth]{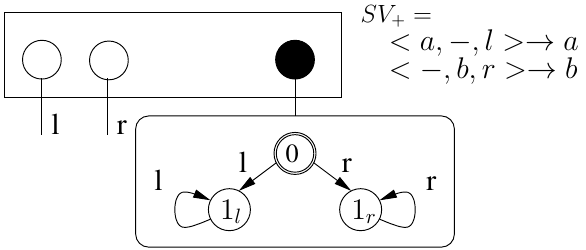}\vspace{-18ex}
\end{minipage} 
&
The left hole is
  indexed $l$ the right hole $r$. The third subnet, contains an LTS encoding
  the control part. We obtain the specific behaviour with the 
  synchronisation vector. The first action  of one of
  the holes decides which branch of the LTS is activated;
   all subsequent actions will be from the same side. 
\\
\end{tabular}

This pNet does not satisfy Definition~\ref{def:non-obspNet}. Indeed, if $a$ or $b$ is $\tau$ then the $+$ operator can indeed observe the $\tau$ transition.
On the other side, the parallel operator defined similarly \emph{satisfies} Definition~\ref{def:non-obspNet}.

\end{example}

With this definition, it is easy to check that the open automaton that gives the semantics of such an open pNet cannot observe silent actions in the sense of Definition~\ref{def:Non-ObsTau}.

\begin{property}[Non-observability of silent actions]
The  semantics of a pNet, as provided in Definition~\ref{def:operationalSemantics}, that cannot observe silent actions is an open automaton that  cannot observe silent actions.
\end{property}


Under this condition, it is safe to define the weak open automaton that provides a weak semantics to a given pNet. This is simply obtained by applying Definition~\ref{def:buildweakOT} to generate a weak open automaton from the open automaton that is the strong semantics of the open pNet, as provided by Definition~\ref{def:operationalSemantics}.

\begin{defi}[Semantics of pNets as a weak open automaton]
Let $A$ be the open automaton expressing the semantics of an open pNet $\pNet$; let $\mylangle J,\mathcal{S},s_0, V, \WT\myrangle$ be the weak open automaton derived from $A$; we call this weak open automaton the weak semantics of the pNet $\pNet$. Then, we denote $\pNet \models \WOT$ whenever $\WOT\in\WT$.
\end{defi}

From the definition of the weak open automata of pNets, we can now study the properties of weak bisimulation concerning open pNets.

\subsection{Properties of weak bisimulation for open pNets}

When silent actions cannot be observed, weak bisimulation is a congruence for open pNets: if $P$ and $Q$ are weakly bisimilar to $P'$ and $Q'$ then the composition of $P$ and $Q$ is weakly bisimilar to the composition of $P'$ and $Q'$, where composition is the hole replacement operator: 	$\pNet[\pNetQ]_{j}$ and 
	$\pNet'[\pNetQ']_{j}$ are weak FH-bisimilar. This can be shown by proving the two following theorems.
The detailed proof of these theorem can be found in Appendix~\ref{sec:app-composition}. The proof strongly relies on the fact that weak FH-bisimulation is an equivalence, but also on the composition properties for open automata.

\begin{thm}[Congruence for  weak bisimulation]\label{weak-thm-congr-eq}
	Consider an open pNet $P$
 that cannot observe silent actions, of the form 	$\pNet = \mylangle \pNet_i^{i\in I}, \Sort_j^{j\in J}, 
	\set{\symb{SV}}\myrangle$.
	Let $j_0\in J$ be a hole. Let $\pNetQ$ and $Q'$ be two weak FH-bisimilar pNets such that\footnote{Note that $\Sortop(\pNetQ)=\Sortop(\pNetQ')$ is 
	ensured by 
	weak bisimilarity.} 
	$\Sortop(\pNetQ)=\Sortop(\pNetQ')\subseteq\Sort_{j_0}$. Then 
	$\pNet[\pNetQ]_{j_0}$ and 
	$\pNet[\pNetQ']_{j_0}$ are weak FH-bisimilar.
\end{thm}

\begin{thm}[Context equivalence for  weak bisimulation]\label{weak-thm-ctxt-eq}
	Consider two  open pNets
	$\pNet = \mylangle \pNet_i^{i\in I}, \Sort_j^{j\in J}, 
	\set{\symb{SV}}\myrangle$ and 	$\pNet' = \mylangle {\pNet'}_i^{i\in I}, 
	\Sort_j^{j\in 
	J}, 	\set{\symb{SV'}}\myrangle$ that are weak FH-bisimilar
	(recall they must have the same holes to be bisimilar) and that cannot observe silent actions.
	Let $j_0\in J$ be a hole, and $\pNetQ$ be a pNet such that $\Sortop(\pNetQ)\subseteq\Sort_{j_0}$. Then 
	$\pNet[\pNetQ]_{j_0}$ and 
	$\pNet'[\pNetQ]_{j_0}$ are weak FH-bisimilar.
\end{thm}

Finally, the previous theorems can be composed to state a general theorem about 
composability and weak FH-bisimilarity.

\begin{thm}[Composability of weak bisimulation]\label{weak-compos}
	Consider two weak FH-bisimilar pNets with an arbitrary number of holes, such that the two pNets cannot observe silent actions. When replacing, 
	inside those two original pNets, a subset of the holes by weak FH-bisimilar pNets, we 
	obtain two weak FH-bisimilar pNets.
\end{thm}

\begin{example}[CCS Choice] Consider the $+$ operator of CCS, shown in  Example~\ref{ex:CCSplus}.  It is well-known that weak bisimulation is not a congruence in CCS, and this is reflected here because we have shown that the $+$ operator can observe the $\tau$ transitions. Thus, even if we can define a weak bisimulation for CCS with $+$ it does not verify the necessary requirements for being a congruence.
\end{example}

\subsubsection*{Running example}
\label{subsubsection:runnig example}
In Section~\ref{sec:weak} we have shown the full saturated weak automaton for both \symb{SimpleProtocolSpec} and \symb{SimpleProtocolImpl}. We will show here how we can check if some given relation between these two automata is a weak FH-Bisimulation. 

\medskip
\noindent
Preliminary remarks:
  \begin{itemize}
    \item Both pNets trivially verify the ``non-observability''
      condition: the only vectors having $\tau$ as an action of a
      sub-net are of the form ``$< -, \tau, -> \rightarrow \tau$''.
    \item We must take care of variable name conflicts: in our example, the variables of the 2 systems already have different names, but the action parameters occurring in the transitions (\texttt{m}, \texttt{msg}, \texttt{ec}) are the same, that is not correct. In the tools, this is managed by the static semantic layer; in the  example, we rename the only conflicting variables $m$ into $m1$ for \symb{SimpleProtocolSpec}, and $m2$ for \symb{SimpleProtocolImpl}.
    \end{itemize}

  \medskip
  Now consider the relation $\mathcal{R}$ defined by the following triples:

\noindent
  \begin{tabular}{|c|c|l|}
\hline
     \symb{SimpleProtocolSpec} states&   \symb{SimpleProtocolImpl} states& Predicate\\
    \hline
    b0 & $000$ & True\\
    b0 & $202$ & True\\
    b1 & $100$ & $\texttt{b\_msg = s\_msg} \land \texttt{b\_ec = s\_ec}$\\
    b1 & $210$ & $\texttt{b\_msg = m\_msg} \land \texttt{b\_ec = m\_ec}$\\
    b1 & $220$ & $\texttt{b\_msg = s\_msg} \land \texttt{b\_ec = s\_ec}$\\
    b1 & $201$ & $\texttt{b\_msg = r\_msg} \land \texttt{b\_ec = r\_ec}$\\
    \hline
    \end{tabular}

  \medskip
  Checking that $\mathcal{R}$ is a weak FH-Bisimulation means checking, for each of these triples, that each (strong) OT of one the states corresponds to a set of WOTs of the other, using the conditions from Definition \ref{def-Weak-bisim}.
  We give here one  example: consider the second triple from the table, and transition $SS_3$ from state \texttt{b0}. Its easy to guess that it will correspond to $WI_3(0)$ of state $202$.

  $ SS_3 = \openrule
  {\{\texttt{P}\mapsto \texttt{p-send(m1)}\}, True,
    (\texttt{b\_msg}\gets \texttt{m1}, \texttt{b\_ec}\gets 0)}
  {\text{b0} \OTarrow {\nounderline{\texttt{in(m1)}}} \text{b1}}
  $
  
$ WI_3(0) = \openrule
  {\{\texttt{P}\mapsto \texttt{p-send(m2)}\}, True,
    (\texttt{s\_msg}\gets \texttt{m2}, \texttt{s\_ec}\gets 0)}
  {000 \OTWeakarrow {\nounderline{\texttt{in(m2)}}} 100}
  $
  
  Let us check formally the conditions:
  
\begin{itemize}
  \item Their sets of active (non-silent) holes is the same: $J' = J_x = \{\texttt{P}\}$.
  \item Triple ($\text{b1},  100,  \texttt{b\_msg = s\_msg} \land \texttt{b\_ec = s\_ec}$) is in $\mathcal{R}$.
  \item The verification condition \\
    $\forall fv_{OT}.\, \{ \Pred \land \Pred_{OT}\\
\hspace{1cm} \implies\!\!\! \displaystyle{\bigvee_{x\in X}\!
   \left[\exists fv_{OT_x}.
  \left( \forall j\in J_x. \vis{\beta_j}\!=\!\gamma_{jx}\! \land\! \Pred_{OT_x}
     \!\land\! \alpha\!=\!\alpha_x\! \land\!  
     \Pred_{s',t_x}\subst{\Post_{OT}\uplus\!\Post_{OT_x}}\right)\right]\}}$

\medskip Gives us:

$\forall \texttt{m1}.\, \{True \land True\hspace{3mm} \implies
\exists \texttt{m2}. \\
(\texttt{[p-send(m1)]} = \texttt{[p-send(m2)]}
\land True
\land \nounderline{\texttt{in(m1)}} = \nounderline{\texttt{in(m2)}} \land 
\\
(\texttt{b\_msg\!\! = \!\!s\_msg} \land \texttt{b\_ec\!\! =\!\! s\_ec}) \subst{(\texttt{b\_msg}\!\gets\! \texttt{m1}, \texttt{b\_ec}\!\gets\! 0)\uplus (\texttt{s\_msg}\!\gets\! \texttt{m2},\texttt{s\_ec}\!\gets\! 0)})\}$

\medskip That is reduced to:

$\forall \texttt{m1}. \exists \texttt{m2}.\,
(\texttt{p-send(m1)} = \texttt{p-send(m2)}
\land \texttt{in(m1)} = \texttt{in(m2)}
\land \texttt{m1}=\texttt{m2} \land 0=0)$

\medskip That is a tautology.

\end{itemize}

\section{Related Works}\label{sec:RW}

 To the best of our knowledge, there are not many research works on Weak Bisimulation Equivalences between such complicate system models (open, symbolic, data-aware, with loops and assignments).
We give a brief overview of other related publications, focussing first on Open and Compositional approaches, then on Symbolic Bisimulation for data-sensitive systems.


\subsubsection*{Open and Compositional systems}
In
\cite{Johnson:2013:CBSE,Johnson:2014:qosa}, the authors investigate several methodologies for the compositional verification of software systems, with the aim to verify reconfigurable component systems. To improve scaling and compositionality, the authors  decouple the verification problem that is to be resolved by a SMT (satisfiability modulo theory) solver into independent sub-problems on independent sets of variables. These works clearly highlight the interest of incremental and compositional verification in a very general setting. In our own work on open pNets, adding more structure to the composition model, we show how to enforce a compositional proof system that is more powerful than independent sets of variables. Our theory has also been encoded into an SMT solver and it would be interesting to investigate how the examples of evolving systems studied by the authors could be encoded into pNet and verified by our framework. However, the models of Johnson et al. are quite different from ours, in particular they are much less structured, and translating them is clearly outside the scope of this article.
In previous work \cite{gaspar:hal-00916115}, we also have shown how (closed) pNet models could be used to encode and verify finite instances of reconfigurable component systems. 


Methodologies for reasoning about abstract semantics of open systems can be found in \cite{BaldanBB:2002, BaldanBB:07,Dubut:20}, authors introduce  behavioural equivalences for open systems from various symbolic approaches. Working in the setting of process calculi, some close relations  exist with the work  of the authors of \cite{BaldanBB:2002,BaldanBB:07}, where both approaches are based
on some kinds of labelled transition systems. The distinguishing feature of their approach is the transitions systems are labelled with logical formulae that provides  an abstract characterization of the structure that a hole must possess and of the actions it can perform in order to allow a transition to fire. Logical formulae are suitable formats that capture the general class of components that can act as the placeholders of the system during its evolution. In our approach we purposely leave the algebra of action terms undefined but the only operation we allow on action of holes is the comparison with other actions. Defining properly the interaction between a logical formulae in the action and the logics of the pNet composition seems very difficult.

Among the approaches for modelling open systems, one can cite  \cite{Beohar:2020} that uses transition conditions depending on an external environment, and introduce bisimulation relations based on this approach. The approach of  \cite{Beohar:2020} is 
  highly based on logics and their bisimulation theory richer in this aspect, while our theory is highly structural and focuses on relation between structure and equivalence. Also, we see composition as a structural operation putting systems together, and do not focus on the modelisation of an unknown outside world. Overall we believe that the two approaches are complementary but checking the compatibility of the two different bisimulation theories is not  trivial.


There is also a clear relation with the seminal works on rule formats for Structured Operational Semantics, e.g. DeSimone format, GSOS, and conditional rules with or without negative premisses \cite{deSimone85,Bloom88bisimulationcant,GROOTE1992202,VANGLABBEEK2004229}.  The Open pNets model provides a way to define operators similar to these rules formats, but with quite different aim and approach. A formal comparison would be interesting, though not trivial. What we can say easily is that: the pNet format syntactically encompasses both DeSimone, GSOS, and conditional premisses rules. Then our compositionality result is more powerful than their classical results, but this is not a surprise, as we rely on a (sufficient) syntactic hypothesis on a particular system, rather than the general rules defining an operator.  Last, we intentionally do not accept negative premisses, that would be more  to put into practice in our implementation. an extension could be studied in future work.

\subsubsection*{Symbolic and data-sensitive systems}
 As mentioned in the \emph{Introduction}, the work that brought us a lot of inspirations are those of Lin et al. \cite{IngolfsdottirL:2001,HennessyLin:TCS95,Linconcur96}. They developed the theory of symbolic transition graphs (STG), and the associated symbolic (early and late, strong and weak) bisimulations, they also study STGs with assignments as a model for message-passing processes. Our work extends these in several ways: first our models are compositional, and our bisimulations come with effective conditions for being preserved by pNet composition (i.e. congruent), even for the weak version. This result is more general than the bisimulation congruences for value-passing CCS in \cite{IngolfsdottirL:2001}. Then our settings for management of data types are much less restrictive, thanks to our use of satisfiability engines, while Lin's algorithms were limited to data-independent systems. 

In a similar way, \cite{Abriola:2018} presents a notion of ''data-aware'' bisimulations on data graphs, in which computation of such bisimulations is studied based on XPath logical language extended with tests for data equality.


Research related to the keyword "Symbolic Bisimulation" refer to two very different domains, namely BDD-like techniques for modelling and computing finite-state bisimulations, that are not related to our topic; and symbolic semantics for data-dependant or high-order systems, that are very close in spirit to our approach. In this last area, we can mention
Calder's work \cite{calder2001symbolic}, that  defines a symbolic semantic for full Lotos, with a symbolic bisimulation over it;
Borgstrom et al., Liu et al, Delaune et al. and Buscemi et al. providing symbolic semantic and equivalence for different variants of pi calculus respectively \cite{borgstrom2004symbolic,delaune2007symbolic,liu2010complete,buscemi2008open}; and more recently
 Feng et al. provide a symbolic bisimulation for quantum processes \cite{feng2014symbolic}.
All the above works, did not give a complete approach for verification, and the models on which these works based are definitely different from ours.


\section{Conclusion and Discussion}
\label{section:conclusion}
 pNets (Parameterised Networks of Automata) is a formalism adapted to the representation of the behaviour of a parallel or distributed systems. One  strength of pNets is their parameterised nature, making them adapted to the representation of systems of arbitrary size, and making the modelling of parameterised system possible. Parameters are also crucial to reason on interaction protocols that can address one entity inside an indexed set of processes. pNets have been successfully used to represent behavioural specification of parallel and distributed components and verify their correctness~\cite{AmeurBoulifa2017,HKM-FASE16}. VCE is the specification and verification platform that uses pNets as an intermediate representation.

Open pNets are pNets with holes; they are adapted to represent processes parameterised by the behaviour of other processes, like composition operators or interaction 
protocols that synchronise the actions of processes that can be plugged afterwards. Open
pNets are hierarchical composition of automata with holes and parameters. We
defined here a semantics for open pNets and a complete bisimulation theory for them. 
The semantics of open pNets relies on the definition of open automata that are automata with holes and parameters, but no hierarchy. Open automata are somehow labelled transition systems with parameters and holes, a notion that is useful to define semantics, but makes less sense when modelling a system, compared to pNets. To be precise, it is on open automata that we define our bisimulation relations.

This article defines a strong and a weak bisimulation relation that are adapted to parameterised systems and hierarchical composition. Our bisimulation principle handles pNet parameters in the sense that two states might be or not in relation depending on the value of parameters. Our strong  bisimulation is compositional by nature in the sense that bisimulation is maintained when composing processes. We also identified a simple and realistic condition on the semantics of non-observable actions that allows weak bisimulation to be also compositional. Overall we believe that this article paved the way for a solid theoretical foundation for compositional verification of parallel and distributed systems.

pNets support the refinement checking at the automata level through a simulation approach, with symbolic evaluation of
the guards and transitions. 
 The definition of simulation on open automata should be  stronger than a strict simulation since it matches a transition with a family of transitions. Such a relation should be able to check the refinement between two open automata with the same level of abstraction but specified differently, for example, by duplicating states, removing transitions, reinforcing guards, modifying variables. Additionally, composition of 
pNets gives the possibility to either add new holes to a system or  fill holes. A useful simulation relation  should thus support the  comparison of automata that do not have the same number of holes. Designing such a simulation relation is a non-trivial extension of this work that we are investigating.

We are currently extending this work,  looking at further properties of FH-bisimulation, but also
the relations with existing equivalences on both closed and open systems. In particular, our model being significantly different from those considered in \cite{IngolfsdottirL:2001}, it
would be interesting to compare our "FH" family of bisimulations with the hierarchy of symbolic bisimulations from these authors.
We also plan to apply open pNets to the study of complex composition
operators in a symbolic way, for example in the area of parallel
skeletons, or distributed algorithms.
We have developed tool support for computing the
symbolic semantics in term of open automata \cite{QBMZ-AVOCS18}, and have
developed algorithms to check strong FH-bisimulation \cite{hou:hal-02406098}.
More recently we published preliminary work for the case of weak FH-Bisimulation \cite{}. 
The challenges here, in the context of our symbolic systems, is not so much algorithmic complexity,
as was the case with classical weak bisimulation on finite models, but decidability and termination.
The naive approach using an explicit construction of the weak transition, may in itself introduce non-termination,
so we prefer a direct implementation of the weak bisimulation definition, without
constructing the weak automata, but searching \emph{on demand} to construct the required weak transitions.
Beside, we explore in \cite{wang:hal-03126313} more pragmatic approaches using weak bisimulation preserving (pattern-based) 
reduction rules.

\bibliographystyle{alpha}

\bibliography{biblio}

\newpage
\appendix    
\section{Proof on FH-bisimulation}
\subsection{Bisimulation is an equivalence: Proof of Theorem~\ref{thm-equiv}}\label{thm-equiv-proof}~\\
        \emph{Suppose $\mathcal{R}$ 
       	is an FH-bisimulation. Then $\mathcal{R}$ is an equivalence, that is, 
       	$\mathcal{R}$ is 
       	reflexive, symmetric and transitive.
       	}
       	\proof
       	It is trivial to check reflexivity and symmetry. Here we focus on the
       	transitivity. 
       	To prove transitivity of strong FH-bisimulation on pNets it is sufficient to 
       	prove 
       	transitivity of the strong FH-bisimulation on states. Consider 3 open automata 
       	$\mathcal{T}_1$, $\mathcal{T}_2$, $\mathcal{T}_3$ and states $s$, $t$, $u$ 
       	in those 
       	automata\footnote{We omit the constraints stating that each $s_x,\,t_x,\,u_x$ is 
       	in the 
       	states of 
       		$\mathcal{T}_1,\,\mathcal{T}_2,\,\mathcal{T}_3$ for the sake of readability}.
       	Suppose we have $\mathcal{R}$ an FH-bisimulation relation between states of 
       	$\mathcal{T}_1$ and of  $\mathcal{T}_2$; members of $\mathcal{R}$ are of the form 
       	$(s,t|\Pred_{s,t})$.
       	Suppose we also  have $\mathcal{R}'$ an FH-bisimulation relation between states 
       	of 
       	$\mathcal{T}_2$ and of  $\mathcal{T}_3$; members of $\mathcal{R}'$ are of the 
       	form 
       	$(t,u|\Pred_{t,u})$.
       	
       	Let $\mathcal{R}''$ be the relation: 
       	\[\mathcal{R}'' = 
       	\{(s,u|\Pred_{s,u})\,\,\Big|\,\Pred_{s,u}=\hspace{-0.7cm}\bigvee_{\begin{array}{c}       		
       		(s,t|\Pred_{s,t})\in\mathcal{R}\\ (t,u|\Pred_{t,u})\in\mathcal{R}' 	
       		\end{array}
       }\hspace{-0.5cm}\Pred_{s,t}\land\Pred_{t,u}\}\]

This relation is the adaptation of the transitivity to the conditional relationship that 
defines a bisimulation. Indeed the global disjunction together with the conjunction of 
predicates plays exactly the role of the intermediate element in a transitivity rule: 
``there exists an intermediate state'' corresponds to the global disjunction, and the 
conjunction of states expresses the intermediate predicate is used to ensure 
satisfiability of the predicate relating the first state to the last one.
       	
       	The relation is built as follows: for each pair of states $s$, $u$, for each 
       	state 
       	$t$ such that $\mathcal{R}$ relates $s$ and $t$, and $\mathcal{R}'$ relates 
       	$t$ 
       	and $u$, we take the conjunction of the two predicates. The predicates for 
       	different 
       	values of $t$ are collected by a disjunction. 
       	
       	We will show
       	that $\mathcal{R}''$ is an FH-bisimulation. Consider 
       	$(s,u|\Pred_{s,u})\in\mathcal{R}''$. Then there is a set of states of 
       	$\mathcal{T}_2$ relating $s$ and $u$, let $(t_p)^{p\in P}$ be this family.  
       	       	We have $\Pred_{s,u} = \displaystyle{\bigvee_{p\in P} \Pred_{s,p}\land\Pred_{p,u}}$.

\medskip

       	For any $p\in P$ by definition of $\mathcal{R}''$,
       	$(s,t_p|\Pred_{s,p})\in\mathcal{R}$,  and 
       	$(t_p,u|\Pred\,'_{p,u})\in\mathcal{R}'$. 
       	We have 
       	the 
       	following by definition of bisimulation:
       	For any open transition $OT$ in $\mathcal{T}_1$ originating from $s$.
       	\begin{mathpar}
       	\openrule
       	{
       		\beta_j^{j\in J_1},\Pred_{OT},\Post_{OT}}
       	{s \OTarrow {\alpha} {s}'}     	
       	\end{mathpar}
       	
       	There exists an indexed set of open transitions $OT_{p x}^{x\in X} \subseteq \mathcal{T}_2$:
       	
       	\begin{mathpar} 
       	\openrule
       	{
       		\beta_{j p x}^{j\in J_{px}}, \Pred_{OT_{px}},\Post_{OT_{px}}}
       	{t_p \OTarrow {\alpha_{px}} t_{p x}}\qquad (*)
       	\end{mathpar}
       	
       	such that  $\forall x, J_1=J_{px}, (s',t_{px}|\Pred_{{px}})\in 
       	\mathcal{R}$;
       	and  
\begin{multline*}
       	\Pred_{s,p} \land \Pred_{OT}\\
 \implies \bigvee_{x\in X}
       	\Big( \forall j. \beta_j=\beta_{jpx}  \land \Pred_{OT_{px}}
       	\land \alpha\!=\!\alpha_{px} \land
       	\Pred_{{px}}\subst{\Post_{OT}\uplus\Post_{OT_{px}}}\Big)
\end{multline*}

For any open transition $OT_{px}$, since
       	$(t_p,u|\Pred_{p,u})\in\mathcal{R}'$ there exists an indexed set of open transitions
       	$OT_{pxy}^{y\in Y} \subseteq \mathcal{T}_3$: 
       	
       	\begin{mathpar}  	
       	\openrule
       	{
       		\beta_{j p x y}^{j\in J_{pxy}}, 
       		\Pred_{OT_{pxy}},\Post_{OT_{pxy}}}
       	{u \OTarrow {\alpha_{pxy}} u_{pxy}}\qquad (**)
       	\end{mathpar}
       	such that  $\forall y, J_{px}=J_{pxy}, 
       	(t_{px},u_{pxy}|\Pred_{{pxy}})\in \mathcal{R}'$; and  
\begin{multline*}
\Pred\,'_{p,u} \land \Pred_{OT_{p x}}\\
       	\hspace{1cm} \implies\!\!\! \displaystyle{\bigvee_{y\in Y}\!
       	\Big( \forall j.\beta_{jpx}\!=\!\beta_{jpxy} \land\! \Pred_{OT_{pxy}}\!
       	\land\! \alpha_{px}\!=\!\alpha_{pxy} \!\land\!
       	\Pred_{{pxy}}\subst{\Post_{OT_{px}}\!\uplus\!\Post_{OT_{pxy}}}\Big)}
\end{multline*}
       	
       	This is verified for each $p\in P$.\\ Overall,  we have a family of open 
       	transitions 
       	$OT_{pxy}^{p\in 
       		P, x\in X, 
       		y\in Y} \subseteq \mathcal{T}_3$ that should simulate $OT$.

       	First, we have $\forall y, \forall x, \forall p,  J_1=J_{px}=J_{pxy}, 
       	({s}',u_{pxy}|\Pred\,'_{{pxy}})\in \mathcal{R}''$ for some $\Pred\,'_{{pxy}}$. 
       	Indeed for any 
       	$p$, 
       	$x$, and 
       	$y$, $t_{px}$
       	relates ${s}'$ and $u_{pxy}$, we have
       	$(s',t_{px}|\Pred_{{px}})\in \mathcal{R}$
       	and $(t_{px},u_{pxy}|\Pred_{{pxy}})\in \mathcal{R}'$. 
       	More precisely,  $t_{px} \in ({t'_p})^{p\in P'}$ where $({t'_p})^{p\in 
       		P'}$ and $P'\subseteq P$  is 
       	the set of states relating ${s}'$ and $u_{pxy}$ (the states used in the open transition must belong to the set of states ensuring the transitive relation).
       	Additionally, for all $p$, $x$, $y$, $\Pred_{px}\land 
       	\Pred_{{pxy}}\implies 
       	\Pred\,'_{{pxy}}$ (this is one element of the  disjunction defining the 
       	predicate $\Pred\,'_{{pxy}}$
       	relating ${s}'$ and $u_{pxy}$ in the definition of $\mathcal{R}''$).

One can notice that, as bisimulation predicates are used to relate states that 
belong to two different open automata, the free variables of these predicates 
that do not belong to the two related automata can safely be renamed to avoid any 
name clash. In practice,
we can suppose that $\Pred\,'_{{pxy}}$ does not 
contain the variables of $\mathcal{T}_2$ because it is used to relate states of $\mathcal{T}_1$ and $\mathcal{T}_3$ . Indeed if $\Pred\,'_{{pxy}}$ uses variables of 	$\mathcal{T}_2$, we can consider instead another predicate that is equivalent to $\Pred\,'_{{pxy}}$ and does not 
contain the variables of 	$\mathcal{T}_2$ (this is safe according to the semantic interpretation of open automata and relations). 
Similarly, we can suppose that $\Pred_{{px}}$ contains no 
variable in $\mathcal{T}_3$, and $\Pred_{{pxy}}$ contains no 
variable in $\mathcal{T}_1$.
       	
       	Second, by definition of bisimulation we need (recall that $\Pred_{s,u}$ is the 
       	original predicate relating $s$ and $u$ by definition of the transitive 
       	closure):\\
       	\begin{small}
       		$\Pred_{s,u} \land \Pred_{OT}\\
       		\hspace{1cm} \implies\!\!\! \displaystyle{\bigvee_{x\in X}\bigvee_{y\in Y}\bigvee_{p\in P}
       		\Big( \forall j. \beta_j=\beta_{jpxy}  \land \Pred_{OT_{pxy}}
       		\land \alpha\!=\!\alpha_{pxy} \land
       		\Pred\,'_{{pxy}}\subst{\Post_{OT}\uplus\Post_{OT_{pxy}}}\Big)}$.
       	\end{small}

\medskip
\noindent From (*) and (**) we have:\\ 
for all $p$,
       	\begin{small}
       		$\Pred_{s,p} \land \Pred_{OT}\land\Pred_{p,u}\\
       		\hspace{1cm} \implies\!\!\! \displaystyle{\bigvee_{x\in X}
       		\Big( \forall j. \beta_j=\beta_{jpx}  \land \Pred_{OT_{px}}
       		\land \alpha\!=\!\alpha_{px} \land
       		\Pred_{{px}}\subst{\Post_{OT}\uplus\Post_{OT_{px}}}\Big)\land\Pred_{p,u}}\\
       		\hspace{1cm} \implies\!\!\! \bigvee_{x\in X}
       		\Big( \forall j. \beta_j=\beta_{jpx}  \land 
       		(\Pred_{OT_{px}}\land\Pred_{p,u})
       		\land \alpha\!=\!\alpha_{px} \land
       		\Pred_{{px}}\subst{\Post_{OT}\uplus\Post_{OT_{px}}}\Big)\\
       		\hspace{1cm} \implies\!\!\! \bigvee_{x\in X}
       		\Big( \forall j. \beta_j=\beta_{jpx}  \land \bigvee_{y\in Y} 
       		\Big( \forall j'. \beta_{j'px}=\beta_{j'pxy}  \land \Pred_{OT_{pxy}}
       		\land \alpha_{px}\!=\!\alpha_{pxy}\\~\hspace*{2em}~ \land
       		\Pred_{{pxy}}\subst{\Post_{OT_{px}}\uplus\Post_{OT_{pxy}}}\Big)
       		\land \alpha\!=\!\alpha_{px} \land
       		\Pred_{{px}}\subst{\Post_{OT}\uplus\Post_{OT_{px}}}\Big)\\
       		\hspace{1cm} \implies\!\!\! \bigvee_{x\in X} \bigvee_{y\in Y}
       		\Big( \forall j, j'. \beta_j=\beta_{jpx} \land \beta_{j'px}=\beta_{j'pxy}
       		\land \Big( 
       		\Pred_{OT_{pxy}}
       		\land \alpha\!=\alpha_{px}\!=\!\alpha_{pxy}\\~\hspace*{2em}~ \land
       		\Pred_{{pxy}}\subst{\Post_{OT_{px}}\uplus\Post_{OT_{pxy}}}
       		\land
       		\Pred_{{px}}\subst{\Post_{OT}\uplus\Post_{OT_{px}}}\Big)\Big)
       		$
       		
       	\end{small}
       	
By construction, four substitution functions $\subst{~}$ only have an effect on the  
variables of the open automaton they belong to, they also produce terms containing only 
variables of the open automaton they belong to. Finally, because of the domain of the 
substitution functions of the predicates, we have:\\
       	$\Pred_{{px}}\subst{\Post_{OT}\uplus\Post_{OT_{px}}}\land 
       	\Pred_{{pxy}}\subst{\Post_{OT_{px}}\uplus
       	\Post_{OT_{pxy}}}\Leftrightarrow\\ 
       	\Pred_{{px}}\subst{\Post_{OT}\uplus\Post_{OT_{px}}\uplus\Post_{OT_{pxy}}}\land
       	\Pred_{{pxy}}\subst{\Post_{OT}\uplus\Post_{OT_{px}}\uplus\Post_{OT_{pxy}}}\\
       	\implies\Pred\,'_{{pxy}}\subst{\Post_{OT}\uplus
       	\Post_{OT_{px}}\uplus\Post_{OT_{pxy}}}
       	\Leftrightarrow\\ 
       	 \Pred\,'_{{pxy}}\subst{\Post_{OT}\uplus\Post_{OT_{pxy}}} $\\
       	
       	This allows us to conclude, with $\Pred_{s,u} = \bigvee_{p\in P} 
       	\Pred_{s,p}\land\Pred_{p,u}$:

      	\begin{small}     	
$\displaystyle{\Pred_{s,u} \land \Pred_{OT}}\\
{~}\hspace{1cm}~ \implies\!\!\! \bigvee_{p\in P} (
	\Pred_{s,p}\land\Pred_{p,u} \land \Pred_{OT})\\
{~}\hspace{1cm}~ \implies\!\!\!\bigvee_{p\in P}
 \bigvee_{x\in X} \bigvee_{y\in Y}
\Big( \forall j, j'.\beta_j=\beta_{jpx} \land \beta_{j'px}=\beta_{j'pxy}
\land \Pred_{OT_{pxy}}\\
~\hspace*{8.5em}~\land \alpha\!=\!\alpha_{px}\!=\!\alpha_{pxy}\land\Pred\,'_{{pxy}}\subst{\Post_{OT}\uplus \Post_{OT_{pxy}}}\Big)
\\
{~}\hspace{1cm}~ \implies\!\!\! \bigvee_{x\in X}\bigvee_{y\in Y}\bigvee_{p\in P}
\Big( \forall j. \beta_j\!=\!\beta_{jpxy}  \land \Pred_{OT_{pxy}}
\land \alpha\!=\!\alpha_{pxy} \land
\Pred\,'_{{pxy}}\subst{\Post_{OT}\uplus\Post_{OT_{pxy}}}\Big)$
  \end{small}
       	
       	\smallskip
       	Concerning the other direction of bisimulation, it is sufficient to notice that 
       	the role 
       	of $s$ and $u$ in the definition of $\mathcal{R}''$ is symmetrical, and thus 
       	the 
       	proof is similar.
       	
\qed

\subsection{Composition Lemmas}       
       The proofs of the composition theorems for FH-bisimulation rely on two main lemmas,
dealing respectively with the decomposition of a composed behaviour
between the context and the internal pNet, and with their recomposition. 

\subsubsection*{{\bf Lemma~\ref{lem-decompose}}: Open transition decomposition} ~\\
 Consider two pNets $P$ and $Q$ that are not pLTSs\footnote{A similar lemma can be proven for a pLTS $Q$}.
	Let $\Leaves(Q)=p_l^{l\in L_Q}$; suppose:
	\[ P[Q]_{j_0}  
		\models
		{\openrule
			{
				\beta_j^{j\in J}, \Pred,  
				\Post}
			{\ostate{s_i^{i\in L}} \OTarrow {\alpha}
				\ostate{s_i'^{\, i\in L}}}
		}
	\]
		with  $J\cap\Holes(Q)\neq\emptyset$ or $\exists i\in L_Q.\,s_i\neq s'_i$, i.e. $Q$ takes part in the reduction.
		 Then, there exist $\alpha_Q$, $\Pred\,'$, $\Pred\,''$, 
		$\Post\,'$, $\Post\,''$ s.t.:\\[-2ex]
		\begin{mathpar}
		P\models{\openrule
			{
				\beta_j^{j\in (J\setminus \Holes(Q)) \cup \{j_0\}}, 
				\Pred\,',  
				\Post\,'}
			{\ostate{s_i^{i\in L\setminus L_Q}} \OTarrow {\alpha}
				\ostate{s_i'^{\,i\in L\setminus L_Q}}}
		}
	\vspace{-2.2ex}\\\text{and~~}
		Q\models{\openrule
			{
				\beta_j^{j\in J\cap\Holes(Q)}, \Pred\,'',  
				\Post\,''}
			{\ostate{s_i^{i\in L_Q}} \OTarrow {\alpha_Q}
				\ostate{s_i'^{\,i\in L_Q}}}
		}
		\end{mathpar}
		and  $\Pred \iff \Pred\,'
		\land \Pred\,''\land \alpha_Q=\beta_{j_0}$, $\Post=\Post\,'\uplus 
		\Post\,''$ where $\Post\,''$ is the restriction of $\Post$ over variables  
		$\vars(Q)$.

\textbf{Preliminary note:}
The introduction of fresh variables introduce alpha-conversion at many points 
of the proof; we 
	only 
	give major arguments concerning alpha-conversion to make the proof readable; in 
	general, fresh variables appear in each transition inside
        terms $\beta_j$, $v$, and 
	$\Pred$.

\proof
Consider rule \TrDeux\ in 
       Definition~\ref{def:operationalSemantics}, applied to the pNet $P[Q]_{j_0}$. 	
	\noindent
\begin{small}	
\begin{mathpar}
    \mprset {vskip=.8ex}
\inferrule
    {
\Leaves(\mylangle {\pNet}_m^{m\in I}, \set{\Sort}, \symb{SV}_k^{k\in 
    	K}\myrangle) \!=\! \pLTS_l^{l\in L} \qquad  	
k\!\in\! K \qquad SV_k \!=\! \SV{(\alpha'_m)^{m \in I_1\uplus I_2\uplus J}}{\alpha'}{e_b} 
\\
\\     	
	\forall m\!\in\! I_1. {\pNet_m 
	\models\openrule
    	{
    	\beta_{j}^{j\in J_m}, \Pred_m, \Post_m}
    	{\ostate{s_{i}^{i \in L_m}} \OTarrow {\alpha_m}
    		\ostate{(s_i^\prime)^{i\in L_m}}} }	
  \qquad
\forall m\!\in\! I_2.		{ \pNet_m 
    	 \models
    	\openrule
    	{\emptyset, \Pred_m, \Post_m}
    	{\ostate{s_m} \OTarrow {\alpha_m}
    		\ostate{s_m'}} }\\\\
     J' = \biguplus_{m\in I_1}\!\! J_m \uplus J 	\\
    	\Pred = \bigwedge_{m\in I_1\uplus I_2}\!\! \Pred_m \land
    	\Predsv(SV_k,\alpha_m^{m\in I_1\uplus I_2},\beta_j^{j\in J},\alpha)\\ 
    	\forall i\in	L\backslash \left(\biguplus_{m\in I_1}\!\! L_m \uplus I_2\right).\,s'_i=s_i \\
    \fresh(\alpha'_m,\alpha',\beta_j,\alpha) 
    }
    {\mylangle {\pNet}_m^{m\in I}, \set{\Sort}, \symb{SV}_k^{k\in K}\myrangle
    	\models
    	{\openrule
    		{
    		{\beta_j}^{j\in J^\prime}, \Pred,  \biguplus_{m\in I_1\uplus I_2} 
    		\Post_m}
    		{\ostate{s_i^{i\in L}} \OTarrow {\alpha}
    			\ostate{(s_i^\prime)^{i\in L}}}
    	}
    }\quad\TrDeux
\end{mathpar} 

\end{small}

We know each premise is \True\ for $P[Q]_{j_0}$. 
  $j_0\in I_1$ because $Q$ is not a pLTS. We try to prove the equivalent premise for 
$P$.

First, $K$ and the synchronisation vector $SV_k$ are unchanged (however 
$j_0$ passes from 
the set of sub-pNets to the set of holes). 
We have $\Leaves(P[Q]_{j_0})=\Leaves(P)\uplus \Leaves(Q)$. 

Now focus on the OTs of the sub-pNets. For each $m\in I_1\uplus I_2$ we have one of the two 
following OT:\\
either $m$ in $I_1$
\[
\pNet_m \models\openrule
    	{
    	\beta_{j}^{j\in J_m}, \Pred_m, \Post_m}
    	{\ostate{s_{i}^{i \in L_m}} \OTarrow {\alpha_m}
    		\ostate{(s_i^\prime)^{i\in L_m}}}\]
or, $m$ in $I_2$
\[{ \pNet_m 
    	 \models
    	\openrule
    	{\emptyset, \Pred_m, \Post_m}
    	{\ostate{s_m} \OTarrow {\alpha_m}
    		\ostate{s_m'}}}\]

Only elements of $(I_1\uplus I_2)\backslash\{j_0\}$   are useful to assert the premise for reduction of $P$; the last 
one ensures the open transition for the pNet $Q$ (note that $Q$ is at place $j_0$, and by 
definition of the open transition 
for $P[Q]_{j_0}$, 
$L_{j_0}=L_Q$, and $J_{j_0}=	J\cap\Holes(Q)$):\\[-2ex]
	\[Q\models{\openrule
		{
			\beta_j^{j\in J\cap\Holes(Q)}, \Pred_{j_0},  
			\Post\,''}
		{\ostate{s_i^{i\in L_Q}} \OTarrow {\alpha_{j_0}}
			\ostate{(s_i')^{\, i\in L_Q}}}
	}\]

This already ensures the second part of the conclusion of the lemma, i.e. the OT for $Q$ 
if we 
choose  $\alpha_Q=\alpha_{j_0}$ and $\Pred\,''= \Pred_{j_0}$. 
Considering 
the OT of $P$ we have another  $J'$ that is $J'_p=J'\setminus\Holes(Q)\uplus 
\{j_0\}$; we denote $I'_1=I_1\setminus \{j_0\}$ the predicate is 
$\displaystyle{\Pred\,'=\!\!\!\!\!\bigwedge_{m\in I_1'\uplus I_2}\!\!\!\!\!\Pred_m  \land \Predsv(SV_k,\alpha_m^{m \in I'_1\uplus I_2},\beta_j^{j\in J\cup \{j_0\}},\alpha)}$\\
where
\begin{multline*}
\Predsv(SV_k,\alpha_m^{m \in I'_1\uplus I_2},\beta_j^{j\in 
J\cup 
\{j_0\}},\alpha)\Leftrightarrow \\
\forall i\in I'_1\uplus I_2.\, \alpha_i=\alpha'_i\land \forall j \in J\cup \{j_0\}.\, 
\beta_j=\alpha'_j 
\land 
\alpha=\alpha'
\land e_b
\end{multline*}

Modulo renaming of fresh variables, this is identical to the predicate that 
occurs in the source open transition except $\alpha_{j_0}=\alpha'_{j_0}$ has been replaced by  $\beta_{j_0}=\alpha'_{j_0}$. As $\alpha_{j_0}=\alpha_Q$ and $\beta_{j_0}$ is free, we have $\beta_{j_0}=\alpha'_{j_0}\land \beta_{j_0}=\alpha_Q \iff \alpha_{j_0}=\alpha'_{j_0}$.
Thus, $\Pred \iff (\Pred\,'
		\land \Pred\,'')\land \alpha_Q=\beta_{j_0}$. 
Finally, Post 
into conditions of the context $P$ and the pNet $Q$ (they are
built similarly as they only deal with  
leaves): $\Post=\Post\,'\uplus \Post\,''$. This concludes the 
proof as we checked all the premises of the open transition for both $P$ and $Q$. We obtain the following reduction by the rule \TrDeux:
\medskip
	
\noindent
	\begin{small}
\begin{mathpar}
    \mprset {vskip=.8ex}
\inferrule
    {
\Leaves(\mylangle {\pNet}_m^{m\in I\setminus\{j_0\}}, \set{\Sort}, \symb{SV}_k^{k\in 
    	K}\myrangle) \!=\! \pLTS_l^{l\in L} \qquad  	
k\!\in\! K \qquad SV_k \!=\! \SV{(\alpha'_m)^{m \in I_1\uplus I_2\uplus J}}{\alpha'}{e_b} 
\\
\\     	
	\forall m\!\!\in\!\! I_1\setminus\{j_0\}. {\pNet_m 
	\models\openrule
    	{
    	\beta_{j}^{j\in J_m}, \Pred_m, \Post_m}
    	{\ostate{s_{i}^{i \in L_m}} \OTarrow {\alpha_m}
    		\ostate{(s_i^\prime)^{i\in L_m}}} }	
  \qquad
\forall m\!\!\in\!\! I_2.		{ \pNet_m 
    	 \models
    	\openrule
    	{\emptyset, \Pred_m, \Post_m}
    	{\ostate{s_m} \OTarrow {\alpha_m}
    		\ostate{s_m'}} }\\\\
     J' = \biguplus_{m\in I_1 \setminus \{j_0\}}\!\! J_m \uplus J 	\\
    	\Pred\,' = \bigwedge_{m\in I_1\uplus I_2\setminus\{j_0\}}\!\! \Pred_m \land
    	\Predsv(SV_k,\alpha_m^{m\in I_1\uplus I_2 \setminus \{j_0\}},\beta_j^{j\in J\cup\{j_0\}},\alpha)\\ 
    	\forall i\in	L\backslash \left(\biguplus_{m\in I_1\setminus\{j_0\}}\!\! L_m \uplus I_2\right).\,s'_i=s_i \\
    \fresh(\alpha'_m,\alpha',\beta_j,\alpha) 
    }
    {\mylangle {\pNet}_m^{m\in I\setminus\{j_0\}}, \set{\Sort}, \symb{SV}_k^{k\in K}\myrangle
    	\models
    	{\openrule
    		{
    		{\beta_j}^{j\in J\setminus \Holes(Q)\uplus \{j_0\}}, \Pred\,',  \biguplus_{m\in I_1\setminus\{j_0\}\uplus I_2} 
    		\Post_m}
    		{\ostate{s_i^{i\in L\setminus L_Q}} \OTarrow {\alpha}
    			\ostate{(s_i^\prime)^{i\in L\setminus L_Q}}}
    	}
    }
\end{mathpar}
 \end{small}
  \qedhere

In general, the actions that can be emitted by $Q$ is  a subset of the possible 
actions of the holes, and the predicate involving $v_Q$ and the synchronisation vector is 
 more restrictive than the one involving only the variable $\beta_{j_0}$. This has no 
 impact 
 on the previous proof and this restriction  results from the composition of predicates.

\subsubsection*{ \bf{Lemma~\ref{lem-compose}}:  Open transition composition}~\\
	Consider two pNets $P$ and $Q$ where $P$ is not a pLTS. Suppose $j_0\in J$ and:\\[-1ex]
\begin{mathpar}
P\models{\openrule
	{
		\beta_j^{j\in J}, 
		\Pred,  
		\Post}
	{\ostate{s_i^{i\in L}} \OTarrow {\alpha}
		\ostate{s_i'^{\, i\in L}}}
}
\text{~~\qquad\qquad~and~\qquad\qquad~~}
Q\models{\openrule
	{
		\beta_j^{j\in J_Q},
		 \Pred\,',  
		\Post\,'}
	{\ostate{s_i^{i\in L_Q}} \OTarrow {\alpha_Q}
		\ostate{s_i'^{\, i\in L_Q}}}
}
\end{mathpar}
Then, we have:\\[-2ex]
	\[ P[Q]_{j_0}  
	\models
	{\openrule
		{
			\beta_j^{(j\in J\setminus\{j_0\}) \uplus J_Q}, 
			\Pred\land\Pred\,'\land \alpha_Q=\beta_{j_0},  
			\Post\uplus \Post\,'}
		{\ostate{s_i^{i\in L\uplus L_Q}} \OTarrow {\alpha}
			\ostate{s_i'^{\, i\in L\uplus L_Q}}}
	}
	\]

Note that this does not mean that any two pNets can be composed and produce an open 
transition. Indeed, the predicate $\Pred\land\Pred\,'\land \alpha_Q=\beta_{j_0}$ will not 
be satisfiable if the action of $\alpha_Q$ cannot be matched with $\beta_{j_0}$.
Note also that $\beta_{j_0}$ is now only used as an intermediate term inside formulas: it 
does not appear neither as global action nor as an action of a hole.

\proof
Let $P=\mylangle {\pNet}_m^{m\in I}, \set{\Sort}, \symb{SV}_k^{k\in K}\myrangle$.
Consider first the open transition derived from $P$.
       Consider each premise of the open transition (constructed by \TrDeux\ rule in 
Definition~\ref{def:operationalSemantics}). 
	\begin{small}
\begin{mathpar}
    \mprset {vskip=.8ex}
\inferrule
    {
\Leaves(\mylangle {\pNet}_m^{m\in I}, \set{\Sort}, \symb{SV}_k^{k\in 
    	K}\myrangle) \!=\! \pLTS_l^{l\in L} \qquad  	
k\!\in\! K \qquad SV_k \!=\! \SV{(\alpha'_m)^{m \in I_1\uplus I_2\uplus J}}{\alpha'}{e_b} 
\\
\\     	
	\forall m\!\in\! I_1. {\pNet_m 
	\models\openrule
    	{
    	\beta_{j}^{j\in J_m}, \Pred_m, \Post_m}
    	{\ostate{s_{i}^{i \in L_m}} \OTarrow {\alpha_m}
    		\ostate{(s_i^\prime)^{i\in L_m}}} }	
  \qquad
\forall m\!\in\! I_2.		{ \pNet_m 
    	 \models
    	\openrule
    	{\emptyset, \Pred_m, \Post_m}
    	{\ostate{s_m} \OTarrow {\alpha_m}
    		\ostate{s_m'}} }\\\\
     J' = \biguplus_{m\in I_1}\!\! J_m \uplus J 	\\
    	\Pred = \bigwedge_{m\in I_1\uplus I_2}\!\! \Pred_m \land
    	\Predsv(SV_k,\alpha_m^{m\in I_1\uplus I_2},\beta_j^{j\in J},\alpha)\\ 
    	\forall i\in	L\backslash \left(\biguplus_{m\in I_1}\!\! L_m \uplus I_2\right).\,s'_i=s_i \\
    \fresh(\alpha'_m,\alpha',\beta_j,\alpha) 
    }
    {\mylangle {\pNet}_m^{m\in I}, \set{\Sort}, \symb{SV}_k^{k\in K}\myrangle
    	\models
    	{\openrule
    		{
    		{\beta_j}^{j\in J^\prime}, \Pred,  \biguplus_{m\in I_1\uplus I_2} 
    		\Post_m}
    		{\ostate{s_i^{i\in L}} \OTarrow {\alpha}
    			\ostate{(s_i^\prime)^{i\in L}}}
    	}
    }\quad \TrDeux
\end{mathpar}  
\end{small}
We know each premise is \True\ for $P$ and try to prove the equivalent premise for 
$P[Q]_{j_0}$ (using the open transition of $Q$). $P[Q]_{j_0}$ exhibits a similar \TrDeux~rule where
$K$ and the synchronisation vector are unchanged ($j_0$ is now in the set of sub-pNets); 
$SV_k=\SV{(\alpha'_j)^{j\in I\uplus\{j_0\}\uplus 
	J}}{\alpha'}{e_b}$. $\Leaves(P[Q]_{j_0})=\Leaves(P)\uplus \Leaves(Q)$. $I$ and $J$ 
	are the 
    set of leaves and holes of $P$, $I_1\uplus I_2$ and $J'$ are the sets of moving 
    leaves and holes 
    in the reduction of $P$. All sub-pNets of 
    must
    be 
reduced, we need:
\[
    	\forall m\!\in\! I_1\uplus\{j_0\}. {\pNet_m 
    	\models\openrule
    	{
    	\beta_{j}^{j\in J_m}, \Pred_m, \Post_m}
    	{\ostate{s_{i}^{i \in L_m}} \OTarrow {\alpha_m}
    		\ostate{(s_i^\prime)^{i\in L_m}}} }	
  \]\[
\forall m\!\in\! I_2.		{ \pNet_m 
    	 \models
    	\openrule
    	{\emptyset, \Pred_m, \Post_m}
    	{\ostate{s_m} \OTarrow {\alpha_m}
    		\ostate{s_m'}} }\]
the sub-pNet at position $j_0$ is the one filled by $Q$ (we define $P_{j_0}=Q$ and similarly $J_m=J_Q$, $\Pred_m=\Pred\,'$, $\Post_m=\Post\,'$,\ldots are the elements of the OT of Q) which offers an open transition 
by hypothesis, the other open transitions are immediate consequence of the open 
transition that can be performed by $P$ (premises of \TrDeux).
The set of moving leaves is the union of the moving leaves in the open transition for $P$ 
and the ones for $Q$; similarly the moving holes are the union of the moving 
holes, minus $j_0$: $J'_{PQ}=  J\setminus\{j_0\}\uplus J_Q$. The 
predicate for the open 
transition is:\\
 $\displaystyle{\Pred\,''=\!\!\!\!\bigwedge_{m\in I _1\uplus I_2}\!\!\!\Pred_m \land \Pred\,'
\land \Pred(SV_k,v_i^{i\in I_1\uplus I_2}\uplus(j_0\mapsto v_Q),\beta_j^{j\in 
J},\alpha)}$. \\
By definition we have:\\
$\Pred(SV_k,\alpha_i^{i\in I}\uplus(j_0\mapsto \alpha_Q),\beta_j^{j\in 
J},v)\Leftrightarrow
	\forall i\in I.\, \alpha_i=\alpha'_i\land \forall j \in J.\, \beta_j=\alpha'_j \land 
	\alpha=\alpha'\land \alpha_Q=\alpha_{j_0}\land e_b$,
this is equivalent to $\forall i\in I.\, \alpha_i=\alpha'_i\land \forall j \in J.\, 
\beta_j=\alpha'_j \land 
	\alpha=\alpha'\land \alpha_Q=\beta_{j_0}\land \beta_{j_0}=\alpha_{j_0}\land e_b$
 and by definition of $\Pred$ (as obtained by applying \TrDeux~rule),
	$\Pred\,''\iff \Pred\land\Pred\,'\land \alpha_Q=\beta_{j_0}$. The post-condition 
	gathers the post-conditions related to all 
	the leaves: $\displaystyle{\biguplus_{m\in I_1\cup\{j_0\}\uplus I_2} 
    		\!\!\!\!\!\Post_m = 	\Post\uplus \Post\,'}$.\\ Finally, the composed open transition can be
        built by \TrDeux\ rule as follows:
	\noindent
	\begin{small}
\begin{mathpar}
    \mprset {vskip=.8ex}
\inferrule
    {
\Leaves(\mylangle {\pNet}_m^{m\in I\cup\{j_0\}}, \set{\Sort}, \symb{SV}_k^{k\in 
    	K}\myrangle) \!=\! \pLTS_l^{l\in L} \qquad  	
k\!\in\! K \qquad SV_k \!=\! \SV{(\alpha'_m)^{m \in I_1\uplus I_2\uplus J}}{\alpha'}{e_b} 
\\
\\     	
	\forall m\!\!\in\!\! I_1\cup\{j_0\}. {\pNet_m 
	\models\openrule
    	{
    	\beta_{j}^{j\in J_m}, \Pred_m, \Post_m}
    	{\ostate{s_{i}^{i \in L_m}} \OTarrow {\alpha_m}
    		\ostate{(s_i^\prime)^{i\in L_m}}} }	
  \qquad
\forall m\!\!\in\!\! I_2.		{ \pNet_m 
    	 \models
    	\openrule
    	{\emptyset, \Pred_m, \Post_m}
    	{\ostate{s_m} \OTarrow {\alpha_m}
    		\ostate{s_m'}} }\\\\
     J'_{PQ}=  J\setminus\{j_0\}\uplus J_Q	\\
    	\Pred\,'' =\bigwedge_{m\in I _1\uplus I_2}\Pred_m \land \Pred\,'
\land \Pred(SV_k,v_i^{i\in I_1\uplus I_2}\uplus(j_0\mapsto v_Q),\beta_j^{j\in 
J},\alpha)\\ 
    	\forall i\in	L\backslash \left(\biguplus_{m\in I_1\cup\{j_0\}}\!\! L_m \uplus I_2\right).\,s'_i=s_i \\
    \fresh(\alpha'_m,\alpha',\beta_j,\alpha) 
    }
    {\mylangle {\pNet}_m^{m\in I}, \set{\Sort}, \symb{SV}_k^{k\in K}\myrangle
    	\models
    	{\openrule
    		{
    		{\beta_j}^{j\in  J'_{PQ}}, \Pred\,'',  	\Post\uplus \Post\,'}
    		{\ostate{s_i^{i\in L\uplus L_Q}} \OTarrow {\alpha}
    			\ostate{(s_i^\prime)^{i\in L\uplus L_Q}}}
    	}
    }
\end{mathpar}  
\end{small}
   This provides the desired conclusion.   
\qed

Note that we also have the following lemma (trivial):

\begin{lem}[Open transition composition -- inactive]\label{lem-compose-2} ~\\	This lemma is the simple case where the pNet filling the hole is not involved in the transition. Suppose $j_0\not\in J$ and $L_Q=\Leaves({Q})$:\\[-2ex]
\begin{mathpar}
P\models{\openrule
	{
		\beta_j^{j\in J}, 
		\Pred,  
		\Post}
	{\ostate{s_i^{i\in L}} \OTarrow {\alpha}
		\ostate{s_i'^{\, i\in L}}}
}
\end{mathpar}
Then, for any state $\ostate{s_i^{i\in L_Q}}$ of $Q$, we have:\\[-1ex]
	\[ P[Q]_{j_0}  
	\models
	{\openrule
		{
			\beta_j^{j\in J}, 
			\Pred,  
			\Post}
		{\ostate{s_i^{i\in L}\uplus s_i^{i\in L_Q}} \OTarrow {\alpha}
			\ostate{s_i'^{\, i\in L}\uplus s_i^{i\in L_Q}}}}
	\]
\end{lem}
The proof is trivial.

 \subsection{Proof of Theorem ~\ref{thm-congr-eq}}~\\
\noindent{\bf  Congruence}:
\textit{	Consider an open pNet:
	$\pNet = \mylangle \pNet_i^{i\in I}, \Sort_j^{j\in J}, 
	\set{\symb{SV}}\myrangle$.
	Let $j_0\in J$ be a hole. Let $\pNetQ$ and $\pNetQ'$ be two FH-bisimilar pNets such that 
	$\Sortop(\pNetQ)=\Sortop(\pNetQ')\subseteq\Sort_{j_0}$\footnote{Note that $\Sortop(\pNetQ)=\Sortop(\pNetQ')$ is 
	ensured by 
	strong bisimilarity.}. Then 
	$\pNet[\pNetQ]_{j_0}$ and 
	$\pNet[\pNetQ']_{j_0}$ are FH-bisimilar.
}
\medskip
\proof
 The proof of Theorem~\ref{thm-congr-eq} exhibits classically a bisimulation relation for 
 a 
 composed system.  It considers then an open transition of $P[Q]_{j_0}$ that should be 
 simulated. It then uses  Lemma~\ref{lem-decompose} to decompose the open transition 
 of $P[Q]_{j_0}$ and obtain an open transition of $P$ and $Q$; the FH-bisimulation 
 property can 
 be applied  to $Q$ to obtain an equivalent family of open transitions of $Q'$; this 
 family is 
 then recomposed by Lemma~\ref{lem-compose} to build a set of open transitions of 
 $P[Q']_{j_0}$ 
 that will simulate the original one.

 Let $\Leaves(Q)=p_l^{l\in L}$, $\Leaves(Q')={p'}_l^{l\in L'}$, 
 $\Leaves(P)=p_l^{l\in L_P}$.
 Consider $Q$ FH-bisimilar to $Q'$. It means that there is a relation 
 $\mathcal{R}$ that is an FH-bisimulation between the open automata of the two pNets. 
 We will consider the relation
$\mathcal{R}'=\{(s,t|\Pred_{s,t})|s=s'\uplus s'' \land 
 t=t'\uplus s'' \land s''\in \mathcal{S}_P \land (s',t'|\Pred_{s,t})\in\mathcal{R}\}$ 
   where $\mathcal{S}_P$ is the set of states of the open automaton of $P$.	We will prove 
 that $\mathcal{R}'$ is an open FH-bisimulation. Consider a pair of FH-bisimilar 
 states: $(\ostate{s_{i}^{i \in L_P\uplus L}},\ostate{{t}_{i}^{i \in L'}\uplus 
 	{s}_{i}^{i \in L_P}}|\Pred_{s,t})\in\mathcal{R}'$. 
 Consider an 
 open transition $OT$ of $P[Q]_{j_0}$. 
 \\[-2ex]     
 \[P[Q]_{j_0}\models\openrule
 {
 	\beta_j^{j\in J},\Pred_{OT},\Post_{OT}}
 { \ostate{s_{i}^{i \in L_P\uplus L}}\OTarrow {\alpha} \ostate{{s'}_{i}^{~i \in 
 			L_P\uplus 
 			L}}}\]
 Let $J'=J\setminus \Holes(Q) \cup \{j_0\}$.	 By 
 Lemma~\ref{lem-decompose} we have:\\[-2ex]
 	\begin{mathpar}
 P\models{\openrule
 	{
 		\beta_j^{j\in J'}, 
 		\Pred\,',  
 		\Post\,'}
 	{\ostate{s_{i}^{i\in L_P}} \OTarrow {\alpha}
 		\ostate{s_{i}'^{\,i\in L_P}}}
 }\\
 Q\models{\openrule
 	{
 		\beta_j^{j\in J\cap\Holes(Q)}, \Pred\,'',  
 		\Post\,''}
 	{\ostate{s_{i}^{i\in L}} \OTarrow {\alpha_Q}
 		\ostate{s_{i}'^{\,i\in L}}}
 }\end{mathpar}
 and  $\Pred_{OT}\iff \Pred\,'
 \land \Pred\,''\land \alpha_Q= \beta_{j_0}$, $\Post_{OT}=\Post\,'\uplus 
 \Post\,''$ ($\Post\,''$ is the restriction of $\Post$ over $\vars(Q)$). As $Q$ is FH-bisimilar to $Q'$ and $(\ostate{s_{i}^{i \in 
 		L}},\ostate{{t}_{i}^{i \in L'}}|\Pred_{s,t})\in\mathcal{R}$ there is a family 
 $OT'_x$ 
 of 	open transitions of the automaton of $Q'$ such that\\[-2ex] 
 \begin{mathpar}
 Q'\models\openrule
 {
 	\beta_{j x}^{j\in J\cap\Holes(Q)}, 
 	\Pred_{OT_x},\Post_{OT_x}}
 {\ostate{t_{i}^{i\in L'}} \OTarrow {\alpha_x} \ostate{t_{i x}^{i\in L'}}}
 \end{mathpar}
 and  $\forall x, (\ostate{s_{i}^{i\in L}},\ostate{t_{i x}^{i\in 
 		L'}}|\Pred_{s x})\in 
 \mathcal{R}$; 
 and  \\
 $\Pred_{s,t} \land \Pred\,''
 \implies$\\ $\displaystyle{\bigvee_{x\in X}
 \Big( \forall j\in J\cap \Holes(Q). \beta_j\!=\!\beta_{jx}\!  \land\! 
 \Pred_{OT_x}
 \land \alpha_Q\!=\!\alpha_x \land  
 \Pred_{s, x}\subst{\Post\,''\uplus\Post_{OT_x}}\Big)}$

 We can now apply Lemma~\ref{lem-compose} on each of the $OT'_x$ together with 
 the transition of $P$ and obtain a new family $OT_x$ of open transitions (where for 
 $i\in L_P$, $t_{i}=s_{i}$ and $t_{i x}=s'_{i}$, and for $j\in Holes(P)$, 
 $\beta_{j x}=\beta_j$):\\[-2ex]
 \[ P[Q']_{j_0}  
 \models
 {\openrule
 	{
 		\beta_{j x}^{j\in J}, 
 		\Pred\,'\land\Pred_{OT_x}\land \alpha_x=\beta_{j_0 x} ,  
 		\Post\,'\uplus \Post_{OT_x} }
 	{\ostate{t_{i }^{i\in L'\uplus L_Q}} \OTarrow {\alpha_x}
 		\ostate{{t}_{i x}^{i\in L'\uplus L_Q}}}
 }
 \]

 Observe that we used the fact that $J=(J\setminus\Holes(Q)\cup 
 \{j_0\})\setminus\{j_0\}\cup 
 (J\cap\Holes(Q))$. Now we have to verify the conditions for the 
 FH-bisimulation between $OT$ and $OT_x$.
 $\forall x, (\ostate{{s'}_{i}^{~i\in L_P\uplus L}},\ostate{t_{i 	x}^{i\in L_P\uplus L'}}|\Pred_{s, x})\in 
 \mathcal{R}'$ (by definition of
 $\mathcal{R}'$) and in three steps we get:
 
 \noindent                        
 \begin{small} $\Pred_{s,t} \land \Pred_{OT} 
 \implies 	\Pred_{s,t}\land\Pred\,'\land \Pred\,''\land\alpha_Q=\beta_{j_0}\\ %
 	\implies  \hspace{-2ex}
 	{\displaystyle\bigvee_{x\in X}
 	\Big( \forall j\in J\cap \Holes(Q). \beta_j=\beta_{jx}  \land 
 	\Pred_{OT_x}
 	\land \alpha_Q\!=\!\alpha_x \land  
 	\Pred_{s,x}\subst{\Post\,''\uplus\Post_{OT_x}}\Big)\land
 }
 \\~~\hspace*{3.5em}~\Pred\,'\land\alpha_Q=\beta_{j_0} \\
 	\implies
 	 \hspace{-2ex}
 	{\displaystyle\bigvee_{x\in X}
 		\Big( \forall j\in J\cap \Holes(Q). \beta_j=\beta_{jx}  \land 
 		\Pred\,'\land \Pred_{OT_x}
 		\land \alpha_Q\!=\!\alpha_x \land \alpha_Q\!=\!\beta_{j_0 x}\, \land} \\ 
 		~\hspace*{3.5em}~ \Pred_{s, x}\subst{\Post\,''\uplus\Post_{OT_x}}\Big)
 	 $\end{small}
 	 
 Note that, $\beta_{j_0}$ can be transformed into  $\beta_{j_0 x}$ because of the 
 implication hypothesis.
%
 The obtained formula reaches the goal except for 
 two points:
 \begin{itemize}
 	\item We need $\forall j\!\in\! J$ instead of $\forall j\!\in\! J\cap\Holes(Q)$  but  
 	adding prerequisite on more variables 
 	does not   	change the validity of the formula (those variables are not used).
\fussy
 	\item Concerning the last term, we need 
 	$\Pred_{s x}\subst{\Post_{OT}\uplus(\Post\,'\uplus \Post_{OT_x})}$, i.e.
 	$\Pred_{s, x}\subst{(\Post\,'\uplus 	\Post\,'')\uplus(\Post\,'\uplus \Post_{OT_x})}$. We 
 	can conclude by observing that	$\Pred_{s,x}$ does not use any variable of $P$ 
 	and thus the substitution $\subst{Post\,'}$ has no effect on it.
\sloppy
 \end{itemize}	
Finally: \\
\begin{small} $\Pred_{s,t} \land \Pred_{OT} \implies\\
{\displaystyle\bigvee_{x\in X}
 		\!\Big( \forall j\in J. \beta_j=\beta_{jx} \! \land\! 
 		\left(\Pred\,'\land\! \Pred_{OT_x}
 		 \land \alpha_Q\!=\!\beta_{j_0 x}\right)\! \land\! \alpha_Q\!=\!\alpha_x\land\!  \Pred_{s, x}\subst{\Post\,''\!\uplus\!\Post_{OT_x}}\Big)
 		}$
 		\end{small}\\
This proves the  condition of the FH-simulation, the other direction is 
 similar.
\qed
        \subsection{Proof of Theorem ~\ref{thm-ctxt-eq}: Context equivalence}
\textit{	Consider two FH-bisimilar open pNets:
	$\pNet = \mylangle \pNet_i^{i\in I}, \Sort_j^{j\in J}, 
	\set{\symb{SV}}\myrangle$ and 	$\pNet' = \mylangle {\pNet'}_i^{i\in I}, 
	\Sort_j^{j\in 
	J}, 	\set{\symb{SV'}}\myrangle$ 
	(recall they must have the same holes to be bisimilar).
	Let $j_0\in J$ be a hole, and $Q$ be a pNet such that $\Sortop(Q)\subseteq\Sort_{j_0}$. Then 
	$\pNet[Q]_{j_0}$ and 
	$\pNet'[Q]_{j_0}$ are FH-bisimilar.
}

\medskip
\proof
The proof of Theorem~\ref{thm-ctxt-eq} exhibits  a bisimulation relation for a 
composed system. It then uses  Lemma~\ref{lem-decompose} to decompose the open transition 
of $P[Q]_{j_0}$ and obtain an open transition of $P$ on which the FH-bisimulation 
property can 
be applied  to obtain an equivalent family of open transitions of $P'$; this family is 
then recomposed by Lemma~\ref{lem-compose} to build a set of open transitions of 
$P'[Q]_{j_0}$ 
that will simulate the original one.

        Let $\Leaves(Q)=p_l^{l\in L_Q}$, 
$\Leaves(P)=p_l^{l\in L}$, $\Leaves(P')={p'}_l^{l\in L'}$.
	Consider $P$ FH-bisimilar to $P'$. It means that there is a relation 
	$\mathcal{R}$ that is an FH-bisimulation between the open automata of the two pNets. 
	We will consider the relation $\mathcal{R}'=\{(s,t|\Pred_{s,t})|s=s'\uplus s'' \land 
	t=t'\uplus s'' \land s\in \mathcal{S}_Q \land (s',t'|\Pred_{s,t})\in\mathcal{R}\}$ 
	where $\mathcal{S}_Q$ is the set of states of the open automaton of $Q$.	We will prove 
	that $\mathcal{R}'$ is an open FH-bisimulation. Consider a pair of FH-bisimilar 
	states: $(\ostate{s_{1 i}^{i \in L\uplus L_Q}},\ostate{{s}_{2 i}^{i \in L'}\uplus 
	{s}_{1 i}^{i \in L_Q}}|\Pred)\in\mathcal{R}'$. 
Consider an 
	open transition $OT$ of $P[Q]_{j_0}$. 
\\[-2ex]     
	\[P[Q]_{j_0}\models\openrule
	{
		\beta_j^{j\in J},\Pred_{OT},\Post_{OT}}
	{ \ostate{s_{i}^{i \in L\uplus L_Q}}\OTarrow {\alpha} \ostate{{s'}_{i}^{~i \in 
	L\uplus 
	L_Q}}}\]
Let $J'=J\setminus \Holes(Q) \cup \{j_0\}$.	 By 
	Lemma~\ref{lem-decompose} we have:\\[-2ex]
			\begin{mathpar}
				P\models{\openrule
				{
					\beta_j^{j\in J'}, 
					\Pred\,',  
					\Post\,'}
				{\ostate{s_{1}^{i\in L}} \OTarrow {\alpha}
					\ostate{s_{i}'^{~ i\in L}}}
			}\\
			Q\models{\openrule
				{
					\beta_j^{j\in J\cap\Holes(Q)}, \Pred\,'',  
					\Post\,''}
				{\ostate{s_{i}^{i\in L_Q}} \OTarrow {\alpha_Q}
					\ostate{s_{i}'^{~ i\in L_Q}}}
			}\end{mathpar}
			and  $\Pred_{OT} \iff \Pred\,'
		\land \Pred\,''\land \alpha_Q=\beta_{j_0}$, $\Post_{OT}=\Post\,'\uplus 
			\Post\,''$ ($\Post\,''$ is the restriction of $\Post$ over $\vars(Q)$). As $P$ is FH-bisimilar to $P'$ and $(\ostate{s_{i}^{i \in 
			L}},\ostate{{t}_{i}^{i \in L'}}|\Pred_{s,t})\in\mathcal{R}$ there is a family 
			$OT'_x$ 
			of 	open transitions of the automaton of $P'$ such that\\[-2ex] 
			\begin{mathpar}
			P'\models\openrule
			{
				\beta_{j x}^{j\in J'}, 
				\Pred_{OT_x},\Post_{OT_x}}
			{\ostate{t_{i}^{i\in L'}} \OTarrow {\alpha_x} \ostate{t_{i x}^{i\in L'}}}
			\end{mathpar}
			and  $\forall x, (\ostate{s_{i}^{i\in L}},\ostate{t_{i x}^{i\in 
			L'}}|\Pred_{s x})\in 
			\mathcal{R}$; 
			and  \\
			$\displaystyle{\Pred_{s,t} \land \Pred\,'
		 \!\!\implies\!\!\! \bigvee_{x\in X}\!
			\Big( \forall j\in J'. \beta_j=\beta_{jx}\!  \land\! 
			\Pred_{OT_x}\!
			\land\! \alpha\!=\!\alpha_x\! \land\!  
			\Pred_{s,x}\subst{\Post\,'\uplus\Post_{OT_x}}\Big)}$
			
			We can now apply Lemma~\ref{lem-compose} on each of the $OT'_x$ together with 
			the transition of $Q$ and obtain a new family $OT_x$ of open transitions (where for 
			$i\in L_Q$, $t_{i}=s_{i}$ and $t_{i x}=s'_{i}$, and for $j\in Holes(Q)$, 
			$b_{j x}=b_j$):\\[-2ex]
				\[ P'[Q]_{j_0}   
				\models
				{\openrule
					{
						\beta_{j x}^{j\in J}, 
						\Pred_{OT_x} \land\Pred\,''\land\alpha_Q=\beta_{j_0 x},  
						\Post_{OT_x}\uplus \Post\,''}
					{\ostate{t_{ i }^{i\in L'\uplus L_Q}} \OTarrow {\alpha_x}
						\ostate{{t}_{ i x}^{i\in L'\uplus L_Q}}}
				}
				\]
			Observe that $J=(J\setminus\Holes(Q)\cup \{j_0\})\setminus\{j_0\}\cup 
			(J\cap\Holes(Q))$. Now we have to verify the conditions for the 
			FH-bisimulation between $OT$ and $OT_x$.
			 $\forall x, (\ostate{{s'}_{i}^{~i\in L\uplus L_Q}},\ostate{t_{i 
			x}^{~i\in L'\uplus L_Q}}|\Pred_{s, x})\in 
			\mathcal{R}'$ (by definition of
                        $\mathcal{R}'$) and in four steps we get:

\noindent                        
\begin{small} 
$\Pred_{s,t} \land \Pred_{OT} \implies
 \Pred_{s,t}\land\Pred\,'\land \Pred\,''\land \alpha_Q=\beta_{j_0}\\  
{~}\hspace{1cm}~ \implies  \hspace{-2ex}
{\displaystyle{\bigvee_{x\in X}\!\Big(\forall j\!\in\! J'. \beta_j\!=\!\beta_{jx}  
\!\land \!
\Pred_{OT_x}\!\land\! \alpha_Q\!=\!\beta_{j_0}
\land\! \alpha\!=\!\alpha_x}} 
\land \Pred_{s, x}\subst{\Post\,'\uplus\Post_{OT_x}}) \land
\Pred\,''\Big)\\ %
{~}\hspace{1cm}~ \implies  \hspace{-2ex}
	{\displaystyle{\bigvee_{x\in X}\!(\forall j\!\in\! J'. \beta_j\!=\!\beta_{jx}  
\land \!
\left(\Pred_{OT_x}\!\land\!
\Pred\,''\!
\land \alpha_Q\!=\!\beta_{j_0 x}\right)
\land \alpha\!=\!\alpha_x}} 
\land \Pred_{s,x}\subst{\Post\,'\uplus\Post_{OT_x}}) 	
$\end{small}


 The obtained formula reaches the goal except for two points:
\begin{itemize}
	\item We need $\forall j\!\in\! J$ instead of $\forall j\!\in\! J'$ with 
	$J'\!=\!J\!\setminus\! \Holes(Q) \cup \{j_0\}$ but the formula under the quantifier 
	does not depend on 
	$b_{j_0}$ now (thanks to 
	the substitution). Concerning $\Holes(Q)$, adding prerequisite on more variables 
	does not 
	change the validity of the formula (those variables are not used).
	\item We need $\Pred_{s, x}\subst{\Post_{OT}\uplus(\Post_{OT_x}\uplus \Post\,'')}$, i.e.,
	$\Pred_{s,x}\subst{(\Post\,'\uplus \Post\,'')\uplus(\Post_{OT_x}\uplus \Post\,'')}$. We 
	can conclude by observing that	$\Pred_{s, x}$ does not use any variable of $Q$ 
	and thus the substitution involving $\Post\,''$ has no effect.
\end{itemize}	
This proves the  condition of the FH-simulation, the other direction is 
similar.
\qed

       \section{Weak FH-bisimulation lemmas and proofs}

We  define a quantified composition operator for effects, i.e. $\Post$ elements of the open transitions.
We use ${\displaystyle \bigodot_{i=n}^{0} \Post_i}$ to denote  $\Post_n\otimes .. \otimes \Post_0$. By convention ${\displaystyle \bigodot_{i=-1}^{0} \Post_i}$ is the identity.

\subsection{Weak bisimulation is an equivalence}\label{app-WFH-equiv}

In this section, we first define two alternative definitions, one for weak open transition, one for  weak bisimulation. We use these two alternative definitions to show that weak bisimulation is an equivalence, we will also re-use these alternative definitions in the  proofs of the theorems in next sections.

\begin{lem}[Alternative definition of weak open transitions]
\label{lem-rel-OT-WOT} Let $A = \mylangle J,\mathcal{S}, s_0,V_1,
    \mathcal{T}\myrangle$ be an open automaton and $\mylangle J,\mathcal{S}, s_0,V_2,
    \WT\myrangle$ be the
weak open automaton derived from $A$.  The two following statements are equivalent
\begin{enumerate}
\item Either $   
\alpha = \tau \wedge  \set{\gamma}=\emptyset \wedge \Pred =\True \wedge \Post =\Id(s)\wedge s=s'$; or \\ 
there exist   $\set{\beta_{1i}}$, $\set{\beta_{2i}}$, and $\set{\beta_{3i}}$, $\Pred_{1i}$, $\Pred_{3i}$, $\Post_{1i}$, and  $\Pred_2$, $\Post_2$, $n\geq -1$, $m\geq -1$ s.t.\footnote{$n=-1$ (resp. $m=-1$) corresponds to the case where there is no $\tau$ transition before (resp. after) the transition $\alpha$.}:

\begin{mathpar}
\forall i \in [0..n].\openrule
    {
       \set{\beta_{1i}},\Pred_{1i},\Post_{1i}   }
         {s_{1i} \OTarrow {\tau} s_{1(i+1)}} \in \mathcal{T} \quad \wedge
\quad
\openrule
         {
           \set{\beta_2},\Pred_2,\Post_2 }
         {s_2 \OTarrow {\alpha} s'_2} \in \mathcal{T}
\quad \wedge\\
\forall i \in [0..m].\openrule
         {
           \set{\beta_{3i}},\Pred_{3i},\Post_{3i}    }
         {s_{3i} \OTarrow {\tau} s_{3(i+1)}} \in \mathcal{T}
 \end{mathpar}
\item  there exist $\set{\gamma}$, $\Pred$, $\Post$ s.t.
 \begin{mathpar}
\openrule
         {
           \set{\gamma},
		\Pred, \Post
				 } {s \OTWeakarrow {\alpha'} s'} \in\WT
\end{mathpar}
where\\
$
\alpha'=\alpha\displaystyle{\subst{\bigodot_{j=n}^{0}\Post_{1j}}}\\
s=s_{10} \wedge s_{1(n+1)}=s_2 \wedge s'_2 = s_{30} \wedge s_{3(m+1)}=s'\\
\displaystyle{
\set{\gamma}=\mybigdotcup_{i=0}^n \vis{\set{\beta_{1i}}\subst{\bigodot_{j=i-1}^{0} \Post_{1j} } }  \dotcup  \vis{\set{\beta_2}\subst{\bigodot_{j=n}^{0}\Post_{1j}}} \dotcup}\\
{}~\qquad \displaystyle{\mybigdotcup_{i=0}^m \vis{\set{\beta_{3i}} \subst{\bigodot_{j=i-1}^{0}\Post_{3j}\shortodot\Post_2\shortodot\bigodot_{j=n}^{0}\Post_{1j}} }}
\\
\Pred=\bigwedge_{i=0}^n\Pred_{1i}\subst{\bigodot_{j=i-1}^{0}\Post_{1j}}\land\Pred_2 \subst{\bigodot_{j=n}^{0}\Post_{1j}}\land\\ 
{}~\qquad\quad\Big(\bigwedge_{i=0}^m\Pred_{3i}\subst{\bigodot_{j=i-1}^{0}\Post_{3j}\shortodot\Post_2\shortodot\bigodot_{j=n}^{0}\Post_{1j}}\Big)\\
\Post=\bigodot_{j=m}^{0}\Post_{3j}\shortodot\Post_2\shortodot\bigodot_{j=n}^{0}\Post_{1j}
$

\end{enumerate}
\end{lem}
\proof 
 ($\Rightarrow$) We present an induction on $n$ and $m$, focusing on the incrementation on $n$: we prove that the property is valid for $m=-1$, $n=-1$ apply a first induction proof for going from $n$ to $n+1$, a similar induction can be applied to go from $m$ to $m+1$ (omitted).
\begin{itemize}
\item The base case there is one transition, so $n=-1$ and $m=-1$, we have:
\begin{mathpar}
 \openrule
         {
           \set{\beta},\Pred,\Post}
         {s \OTarrow {\alpha} s'} \in \mathcal{T}
\end{mathpar} 
by rule \WTDeux\ we can directly conclude the implication: 
 \begin{mathpar}

\openrule
         {
           \set{\beta},\Pred,\Post}
         {s \OTarrow {\alpha} s'} \in \mathcal{T} \Rightarrow 
{ \openrule
         {
           \vis{\set{\beta}}\!,\Pred,\Post
				 } {s \OTWeakarrow {\alpha} s'} \in \WT
}
\end{mathpar} 
\item For the inductive step, first we have by induction hypothesis that the formula holds for some lengths $m$ and $n$. Induction step is to infer that formula holds for transitions of length  $n+1$. 
We consider the case $n'=n+1$.  We want to prove (1) $\Rightarrow$ (2) in the lemma, and in (1) we  focus on the case where there is a set of open transitions (this is the case: $s\neq s'$). In other words, we consider the sequence of ($n+m+4$) open transitions: 
\begin{mathpar}
\Big(\forall i \in [0..n+1].\openrule
    {
       \set{\beta_{1i}},\Pred_{1i},\Post_{1i}   }
         {s_{1i} \OTarrow {\tau} s_{1(i+1)}} \in \mathcal{T} \quad \wedge
\quad
\openrule
         {
           \set{\beta_2},\Pred_2,\Post_2 }
         {s_2 \OTarrow {\alpha} s'_2} \in \mathcal{T}
\quad \wedge\\
~~\qquad\qquad\forall i \in [0..m].\openrule
         {
           \set{\beta_{3i}},\Pred_{3i},\Post_{3i}   }
         {s_{3i} \OTarrow {\tau} s_{3(i+1)}} \in \mathcal{T}
\Big) 
\end{mathpar}

By recurrence hypothesis we suppose that  (1) $\Rightarrow$ (2) holds for $n$ and $m$ (compared to the line above, we remove the first $\tau$ transition). We have:\\

$
\Big(\forall i \in [1..n+1].\openrule
    {
       \set{\beta_{1i}},\Pred_{1i},\Post_{1i}   }
         {s_{1i} \OTarrow {\tau} s_{1(i+1)}} \in \mathcal{T} \,\wedge\,
\openrule
         {
           \set{\beta_2},\Pred_2,\Post_2 }
         {s_2 \OTarrow {\alpha} s'_2} \in \mathcal{T}
\, \wedge\\
{}~~\hspace{2.25cm}\forall i \in [0..m].\openrule
         {
           \set{\beta_{3i}},\Pred_{3i},\Post_{3i}   }
         {s_{3i} \OTarrow {\tau} s_{3(i+1)}}\! \in\! \mathcal{T}
\Big)  \Rightarrow\!  {
\openrule
         {
           \set{\gamma},
		\Pred, \Post
				 } {s'' \OTWeakarrow {\alpha'} s'} \in\WT
}
$
where\\
\begin{align*}
s''=s_{10} &\wedge s_{1(n+2)}=s_2 \wedge s'_2 = s_{30} \wedge s_{3(m+1)}=s'\\
\alpha'=&\alpha\displaystyle{\subst{\bigodot_{j=n+1}^{1}\!\!\Post_{1j}}}\\
\set{\gamma}=&\mybigdotcup_{i=1}^{n+1} \vis{\set{\beta_{1i}}\subst{\bigodot_{j=i-1}^{1} \Post_{1j} } }  \dotcup  \vis{\set{\beta_2}\subst{\bigodot_{j=n+1}^{1}\Post_{1j}}} \dotcup\\
& \mybigdotcup_{i=0}^m \vis{\set{\beta_{3i}} \subst{\bigodot_{j=i-1}^{0}\Post_{3j}\shortodot\Post_2\shortodot\bigodot_{j=n}^{0}\Post_{1j}} }\\
\Pred=&\bigwedge_{i=1}^{n+1}\Pred_{1i}\subst{\bigodot_{j=i-1}^{1}\Post_{1j}}\land\Pred_2 \subst{\bigodot_{j={n+1}}^{1}\Post_{1j}}\land\\ 
&\quad\Big(\bigwedge_{i=0}^m\Pred_{3i}\subst{\bigodot_{j=i-1}^{0}\Post_{3j}\shortodot\Post_2\shortodot\bigodot_{j={n+1}}^{1}\Post_{1j}}\Big)\\
\Post=&\bigodot_{j=m}^{0}\Post_{3j}\shortodot\Post_2\shortodot\bigodot_{j={n+1}}^{1}\Post_{1j}
\end{align*}

We need to prove that   by adding the following open transition the implication remains true:
\begin{mathpar}
 \openrule
         {
           \set{\beta_{10}},\Pred_{10},\Post_{10}}
         {s_{10} \OTarrow {\tau} s_{11}} \in \mathcal{T}
\end{mathpar}
First by using rule \WTDeux~ we have:
\begin{mathpar}

 \openrule
         {
           \set{\beta_{10}},\Pred_{10},\Post_{10}}
         {s_{10} \OTarrow {\tau} s_{11}} \in \mathcal{T}
\Rightarrow 
{ \openrule
       {
           \vis{\set{\beta_{10}}}\!,\Pred_{10},\Post_{10}
				 } {s_{10} \OTWeakarrow {\tau}s_{11}} \in \WT
}
\end{mathpar} 

On the other hand, by rule \WTUn~ we have the following weak open transition:
\begin{mathpar}
 \openrule
         {
          \emptyset,\True,\Id(s')
				 } {s' \OTWeakarrow {\tau} s'} \in \WT
\end{mathpar}

Finally by applying rule \WTTrois~on  the above weak open transitions:

\begin{mathpar}
\inferrule {\openrule
       {
           \vis{\set{\beta_{10}}}\!,\Pred_{10},\Post_{10}
				 } {s_{10} \OTWeakarrow {\tau}s_{11}}\! \in\! \WT
\\
\openrule
         {
           \set{\gamma},
		\Pred, \Post
				 } {s'' \OTWeakarrow {\alpha'} s'}\! \in\!\WT
\\
\openrule
         {
          \emptyset,\True,\Id(s'')
				 } {s' \OTWeakarrow {\tau} s'}\! \in\! \WT
\\\\
\set{\gamma''}= \vis{\set{\beta_{10}}} \dotcup \set{\gamma}\subst{\Post_{10}}\\
\Pred\,''\!=\!\Pred_{10}\land\Pred\subst{\Post_{10}}\\\\
\Post\,''=\Id(s'')\shortodot\Post\shortodot\Post_{10} \\ \alpha''=\alpha'\subst{\Post_{10}}
}
{
\openrule
         {
          \set{\gamma''}, \Pred\,'',\Post\,''
				 } 
         {s_{10} \OTWeakarrow {\alpha''} s'} \in\!\WT
} \WTTrois
\end{mathpar}
where we obtain the conclusion of the lemma, as required with the following assertions (derived from previous assertions):
\begin{align*}
s_{10}=s_{10} &\wedge s_{1(n+2)}=s_{2} \wedge s'_2 = s_{30} \wedge s_{3(m+1)}=s'\\
\alpha''=&\alpha\subst{\bigodot_{j=n+1}^{1}\Post_{1j}}\subst{\Post_{10}} = \alpha\subst{\bigodot_{j=n+1}^{0}\Post_{1j}}\\
\set{\gamma''}=&\mybigdotcup_{i=0}^{n+1} \vis{\set{\beta_{1i}}\subst{\bigodot_{j=i-1}^{1} \Post_{1j} } }  \dotcup  \vis{\set{\beta_2}\subst{\bigodot_{j=n+1}^{1}\Post_{1j}}} \dotcup\\
&\mybigdotcup_{i=0}^m \vis{\set{\beta_{3i}} \subst{\bigodot_{j=i-1}^{0}\Post_{3j}\shortodot\Post_2\shortodot\bigodot_{j=n}^{0}\Post_{1j}} }
\\
\Pred=&\bigwedge_{i=0}^{n+1}\Pred_{1i}\subst{\bigodot_{j=i-1}^{0}\Post_{1j}}\land\Pred_2 \subst{\bigodot_{j={n+1}}^{0}\Post_{1j}}\land\\ 
&\Big(\bigwedge_{i=0}^m\Pred_{3i}\subst{\bigodot_{j=i-1}^{0}\Post_{3j}\shortodot\Post_2\shortodot\bigodot_{j={n+1}}^{0}\Post_{1j}}\Big)
\\
\Post=&\bigodot_{j=m}^{0}\Post_{3j}\shortodot\Post_2\shortodot\bigodot_{j={n+1}}^{0}\Post_{1j}
\end{align*}

The right part of the disjunction, i.e.
\begin{mathpar}
\Big(\alpha = \tau \wedge  \set{\gamma}=\emptyset \wedge \Pred =\True \wedge \Post =\Id(s)\wedge s=s'\Big)
\end{mathpar}
is handled trivially by rule \WTUn. 
\end{itemize}

\noindent ($\Leftarrow$) We proceed by structural induction on the rules  building  the weak transition (as described in the original definition). The recurrence hypothesis being that the  original definition implies the characterization (1), with the conditions stated at the bottom of the theorem. We consider the different rules:

\begin{itemize}
\item Case rule \WTUn. We have:
\begin{mathpar}
{\openrule
         {
           \emptyset,\True,\Id(s)
				 } {s \OTWeakarrow {\tau} s} \in \WT
}
\end{mathpar}
We can directly conclude by the right part of the disjunction the following:\\

$
{ \openrule
         {
           \emptyset,\True,\Id(s)
				 } {s \OTWeakarrow {\tau} s} \in\! \WT
}\! \Rightarrow\! \Big(\alpha = \tau \wedge \set{\gamma}=\emptyset \wedge \Pred \!=\!\True \wedge \Post =\Id(s)\wedge s\!=\!s'\Big)
$
\item Case rule \WTDeux. We have:
\begin{mathpar}
\openrule
         {
           \set{\gamma},\Pred,\Post}
         {s \OTWeakarrow {\alpha} s'} \in \WT \Rightarrow
         \openrule
         {
           \set{\beta},\Pred,\Post}
         {s \OTarrow {\alpha} s'} \in \mathcal{T}         
\end{mathpar}
where $\set{\gamma} = \vis{\set{\beta}}$.\\

These two cases above prove the implication with $n=-1$ and $m=-1$.

\item Case rule \WTTrois. We have:
\begin{mathpar}
\inferrule {\openrule
         {
           \set{\gamma_1},\Pred_1,\Post_1   }
         {s \OTWeakarrow {\tau} s'} \in \WT
\\
\openrule
         {
           \set{\gamma_2},\Pred_2,\Post_2  }
         {s' \OTWeakarrow {\alpha} s''} \in \WT
\\
\openrule
         {
           \set{\gamma_3},\Pred_3,\Post_3    }
         {s'' \OTWeakarrow {\tau} s'''} \in\WT
\\
\Pred=\Pred_1\land\Pred_2\subst{\Post_1}\land \Pred_3\subst{\Post_2\shortodot\Post_1}
\\
\set{\gamma}=\set{\gamma_1}\dotcup \set{\gamma_2}\subst{\Post_1}\dotcup\set{\gamma_3}\subst{\Post_2\shortodot \Post_1}\\
\alpha'=\alpha\subst{\Post_1}
}
{
\openrule
         {
           \set{\gamma},
		\Pred,
				\Post_3\shortodot{\Post_2}\shortodot{\Post_1} }
         {s \OTWeakarrow {\alpha'} s'''} \in\WT
}
\end{mathpar}


\begin{enumerate}

\item By induction hypothesis this means each tau weak open transition can be written as a series of $n_1$ tau open transitions such $n_1=n+m+3$, hence by simplification we have (strictly speaking, by induction we might also have the case $\alpha=\tau\land\set \gamma=\emptyset\land \ldots$ but in this case, rule \WTUn~ allows us to obtain a similar reduction with $n_1=1$): 
\begin{mathpar}
\openrule
         {
           \set{\gamma_1},\Pred_1,\Post_1   }
         {s \OTWeakarrow {\tau} s'} \in\! \WT 
\Rightarrow \,
 \forall i \in [0..n_1].\openrule
    {
       \set{\beta_{1i}},\Pred_{1i},\Post_{1i}   }
         {s_{1i} \OTarrow {\tau} s_{1(i+1)}} \in \mathcal{T}
\end{mathpar}
where \\
$ 
s=s_{10} \wedge s_{1{(n_1+1)}}\!=s',\quad
\displaystyle{\set{\gamma_1}= \mybigdotcup_{i=0}^{n_1} \vis{\set{\beta_{1 i}}\subst{\bigodot_{j=i-1}^{0} \Post_{1 j} } }} \\
\Pred_1\!=\!\bigwedge_{i=0}^{n_1}(\Pred_{1i} \subst{\bigodot_{j=i-1}^{0}\Post_{i j}} ), \quad \Post_1= \bigodot_{i=n_1}^{0} \subst{\Post_{1i}} 
$

\item Similarly, a series of $m_1$ open transitions such that $m_1=n+m+3$   can be simplified as follows: 
\begin{mathpar}
\openrule
         {
           \set{\gamma_3},\Pred_3,\Post_3   }
         {s'' \OTWeakarrow {\tau} s'''} \in\! \WT
\Rightarrow\, 
\forall i \in [0..m_1].\openrule
    {
       \set{\beta_{3i}},\Pred_{3i},\Post_{3i}   }
         {s_{3i} \OTarrow {\tau} s_{3(i+1)}} \in \mathcal{T}        
\end{mathpar}
where
\begin{mathpar}
s''=s_{30} 

\land

 s_{3{(m_1+1)}}\!=s'''

\land

\displaystyle{\set{\gamma_3}= \mybigdotcup_{i=0}^{m_1} \vis{\set{\beta_{3 i}}\subst{\bigodot_{j=i-1}^{0} \Post_{3 j} } }} 

\land 

\Pred_3\!=\!\bigwedge_{i=0}^{m_1}(\Pred_{3i} \subst{\bigodot_{j=i-1}^{0}\Post_{3 j}} )

\land 

 \Post_3= \bigodot_{i=m_1}^{0} \subst{\Post_{3i}} 
\end{mathpar}

\item Concerning the middle reduction, by induction hypothesis there exists a set of open transitions in $\mathcal{T}$ such that:
\begin{mathpar}
\openrule
         {
           \set{\gamma_2},\Pred_2,\Post_2  }
         {s' \OTWeakarrow {\alpha'} s''} \in\! \WT
\Rightarrow 
\bigg(\forall i \in [0..n_2].\openrule
    {
       \set{\beta_{2i}},\Pred_{2i},\Post_{2i}}
         {s_{2i} \OTarrow {\tau} s_{2(i+1)}} \in \mathcal{T}  \wedge

\openrule
         {
           \set{\beta'},\Pred\,',\Post\,' }
         {s_2 \OTarrow {\alpha''} s_2'} \in \mathcal{T}
 \wedge
\forall i \in [0..m_2].\openrule
         {
           \set{\beta_{2i}'},\Pred_{2i}^{\,\prime},\Post'_{2i}    }
         {s_{2i}' \OTarrow {\tau} s_{2(i+1)}'} \in \mathcal{T}
\bigg)         
\end{mathpar}

where
\begin{align*}
s'=s_{20} &\wedge s_{2(n_2+1)}=s_2 \wedge  s_2' = s_{20}' \wedge s_{2(m_2+1)}'=s''\\
\alpha''=&\alpha'\subst{\bigodot_{j=n_2}^{0}\Post_{2j}}\\
\set{\gamma_2}=&\mybigdotcup_{i=0}^{n_2} \vis{\set{\beta_{2i}}\subst{\bigodot_{j=i-1}^{0} \Post_{2 j} } }  \dotcup  \vis{\set{\beta'}\subst{\bigodot_{j=n_2}^{0}\Post_{2j}}} \dotcup\\
&
 \mybigdotcup_{i=0}^{m_2} \vis{\set{\beta'_{2i}} \subst{\bigodot_{j=i-1}^{0}\Post\,'_{2j}\shortodot\Post\,'\shortodot\bigodot_{j=n_2}^{0}\Post_{2j}}}
\\
\Pred_2=&\bigwedge_{i=0}^{n_2}\Pred_{2i}\subst{\bigodot_{j=i-1}^{0}\Post_{2j}}\land\Pred\,' \subst{\bigodot_{j={n_2}}^{0}\Post_{2j}}\land\\ 
&\Big(\bigwedge_{i=0}^{m_2}\Pred\,'_{2i}\subst{\bigodot_{j=i-1}^{0}\Post\,'_{2j}\shortodot\Post\,'\shortodot\bigodot_{j={n_2}}^{0}\Post_{2j}}\Big)\\
\Post_2=&\bigodot_{j=m_2}^{0}\Post\,'_{2j}\shortodot\Post\,'\shortodot\bigodot_{j={n_2}}^{0}\Post_{2j}
\end{align*}

\end{enumerate}
~~\\
Therefore, we can deduce that we have:
\begin{mathpar}
{ \openrule
         {
           \set{\gamma}\!,\Pred,\Post
				 } {s \OTWeakarrow {\alpha} s'} \in \WT
}
 \Rightarrow
\bigg(\forall i\in [0..(n_1\!+\!n_2)].\openrule
    {
       \set{\beta_{4i}},\Pred_{4i},\Post_{4i}   }
         {s_{4i} \OTarrow {\tau} s_{4(i+1)}} \in\mathcal{T} \wedge
         
\openrule
         {
           \set{\beta'},\Pred\,',\Post\,' }
         {s_2 \OTarrow {\alpha''} s_2'} \in \mathcal{T}
\wedge\,
\forall i \in [0..(m_1\!+\!m_2)].\openrule
         {\set{\beta_{5i}},\Pred_{5 i},\Post_{5i}    }
         {s_{5i}\OTarrow {\tau} s_{5({i+1})}}\in\mathcal{T}\bigg)  
\end{mathpar}

such that
\begin{align*}
s_{4i}=\begin{cases}
			s_{1i} & \mbox{if }i<n_1\\
			s_{2i-n_1} & \mbox{if }i\geq n_1
	   \end{cases}
&\qquad\qquad\qquad&
s_{5i}=\begin{cases}
			s_{3i} & \mbox{if }i<m_1\\
s_{2i-m_1}' & \mbox{if }i\geq m_1
	\end{cases}\\
\end{align*}

 and similarly for $\Pred_{4i}$, $\Pred_{5i}$, $\Post_{4i}$, and $\Post_{5i}$.  

Also, we have the following assertions:
\begin{align*}
s\!=\!s_{40}& ~\wedge~s_{4(n_1+n_2+1)}\!=s_2 ~\wedge~s_2' = s_{50} ~\wedge~ s_{5(m_1+m_2+1)}\!=\!s'\\
\alpha''=&\alpha\subst{\!\!\bigodot_{j=n_1+n_2}^{0}\!\!\!\Post_{4j}}\\
\set{\gamma}=&\mybigdotcup_{i=0}^{n_1+n_2} \vis{\set{\beta_{4i}}\subst{\bigodot_{j=i-1}^{0} \Post_{4 j} } }  \dotcup  \vis{\set{\beta'}\subst{\!\!\bigodot_{j=n_1+n_2}^{0}\!\!\!\Post_{4j}}}\dotcup \\
&\mybigdotcup_{i=0}^{m_1+m_2}\!\!\vis{\set{\beta_{5i}}\subst{\bigodot_{j=i-1}^{0}\Post\,'_{5i}\shortodot\Post\,'\shortodot\!\!\bigodot_{j=n_1+n_2}^{0}\!\!\!\Post_{4i}} }\\
\Pred=&
\bigwedge_{i=0}^{n_1+n_2}\!\!\Pred_{4i}\subst{\bigodot_{j=i-1}^{0}\Post_{4j}}\land\Pred\,' \subst{\!\!\bigodot_{j={n_1+n_2}}^{0}\!\!\!\Post_{4j}}\land\\ 
&\quad\bigwedge_{i=0}^{m_1+m_2}\Pred_{5i}\subst{\bigodot_{j=i-1}^{0}\!\!\Post_{5j}\shortodot\Post\,'\shortodot\!\!\!\bigodot_{j={n_1+n_2}}^{0}\!\!\!\Post_{4j}}\\
\Post=&\bigodot_{j=m_1+m_2}^{0}\!\!\!\Post_{5j}\shortodot\Post\,'\shortodot\!\!\bigodot_{j={n_1+n_2}}^{0}\!\!\Post_{4j}
\end{align*}

This concludes the inductive step, showing  that the decomposition expressed by the $\Leftarrow$ direction of the lemma is always possible with the right side conditions. 
\qedhere
\end{itemize}

\begin{lem}[Alternative definition of weak bisimulation]\label{lem-alternative-weak-bisim} The definition of weak bisimulation given in Definition \ref{def-Weak-bisim} is equivalent to the following one:

\fussy
 Let $A_1 = \mylangle J,\mathcal{S}_1, s_0, V_1,
    \mathcal{T}_1\myrangle$ and $A_2 = \mylangle J,\mathcal{S}_2,t_0, V_2, \mathcal{T}_2 \myrangle$ be open automata; $\mylangle J,\mathcal{S}_1, s_0,V_1,
    \WT_1\myrangle$ and $\mylangle J,\mathcal{S}_2,t_0, V_2, \WT_2\myrangle$ be the
weak open automaton derived from $A_1$ and $A_2$ respectively.
For any  states
$s\in\mathcal{S}_1$ and
$t\in\mathcal{S}_2$ such that $(s,t|\Pred_{s,t})\in\mathcal{R}$, we 
   have:
\sloppy
\begin{itemize}
 \item  For any open transition $\symb{WOT}$ in $\WT_1$:
 \begin{mathpar}
     \openrule
         {
           \gamma_j^{j\in J'},\Pred_{OT},\Post_{OT}}
         {s \OTWeakarrow {\alpha} s'}

\end{mathpar}
 there exists an indexed set of weak open transitions $\symb{WOT}_x^{x\in X} \subseteq \WT_2$:
 \begin{mathpar}
    \openrule
         {
           \gamma_{j x}^{j\in J_{x}}, \Pred_{OT_x},\Post_{OT_x}}
         {t \OTWeakarrow {\alpha_x} t_x}
\end{mathpar}
 such that  $\forall x, J'=J_{x}, (s',t_x|\Pred_{s',x})\in \mathcal{R}$; 
 and  \\
$\Pred_{s,t} \land \Pred_{OT}\!\!\implies\!\!\!
\displaystyle{\bigvee_{x\in X}
   \left( \forall j\in J_x.\gamma_j\!=\!\gamma_{jx}\!\land\! \Pred_{OT_x}
     \land \alpha\!=\!\alpha_x \!\land\!  
     \Pred_{s',x}\subst{\Post_{OT}\!\uplus\!\Post_{OT_x}}\right)}$
    
 \item  and symmetrically any open transition from $\symb{WOT}$ in $\WT_2$ can be 
      covered by a set of weak transitions from $s$ in $\WT_1$.
 \end{itemize}

\end{lem}

\proof Note that Definition \ref{def-Weak-bisim} is a particular case of the definition above, thus we only need to prove one direction of the equivalence between the two definitions, namely: \\
($\Rightarrow$) We prove that Definition \ref{def-Weak-bisim} implies the definition above. In other words,
suppose that $\Pred_{s,t}\in \mathcal{R}$ 
and  suppose that the following statement holds:
\begin{mathpar}
     \openrule
         {
           \gamma_j^{j\in J'},\Pred_{OT},\Post_{OT}}
         {s \OTWeakarrow {\alpha} s'} \in \WT_1

\end{mathpar}
Moreover, by using Lemma \ref{lem-rel-OT-WOT} we know that:
\begin{mathpar}
    \openrule
         {
           \gamma_j^{j\in J'},\Pred_{OT},\Post_{OT}}
         {s \OTWeakarrow {\alpha} s'} \in \WT_1
         \Rightarrow
\bigg(\forall i \in [0..n].\openrule
    {
       \beta_{1ij}^{j\in J_1'},\Pred_{1i},\Post_{1i}   }
         {s_{1i} \OTarrow {\tau} s_{1(i+1)}} \in \mathcal{T}_1  \wedge
\\{\qquad\qquad\qquad\qquad\qquad}
\openrule
         {
           \beta_{2j}^{j\in J_2'},\Pred_2,\Post_2 }
         {s_{20} \OTarrow {\alpha'} s_{21}} \in \mathcal{T}_1
 \wedge
\forall i \in [0..m].\openrule
         {
           \beta_{3ij}^{j\in J_3'},\Pred_{3i},  \Post_{3i}  }
         {s_{3i} \OTarrow {\tau} s_{3(i+1)}} \in \mathcal{T}_1
\bigg)
\end{mathpar}
where

\begin{align*}
s=s_{10} &\wedge s_{1(n+1)}=s_{20} \wedge s_{21} = s_{30} \wedge s_{3(m+1)}=s'\\
\alpha=&\alpha'\subst{\bigodot_{j=n}^{0}\Post_{1j}}\\
\gamma_j^{j\in J'}=&\mybigdotcup_{i=0}^n \vis{\set{\beta_{1i}}\subst{\bigodot_{j=i-1}^{0} \Post_{1j} } }  \dotcup  \vis{\set{\beta_2}\subst{\bigodot_{j=n}^{0}\Post_{1j}}} \dotcup\\
&\mybigdotcup_{i=0}^m \vis{\set{\beta_{3i}} \subst{\bigodot_{j=i-1}^{0}\Post_{3j}\shortodot\Post_2\shortodot\bigodot_{j=n}^{0}\Post_{1j}} }
\\
\Pred_{OT}=&\bigwedge_{i=0}^n\Pred_{1i}\subst{\bigodot_{j=i-1}^{0}\Post_{1j}}\land\Pred_2 \subst{\bigodot_{j=n}^{0}\Post_{1j}}\land\\ 
&\qquad\Big(\bigwedge_{i=0}^m\Pred_{3i}\subst{\bigodot_{j=i-1}^{0}\Post_{3j}\shortodot\Post_2\shortodot\bigodot_{j=n}^{0}\Post_{1j}}\Big)\\
\Post_{OT}=&\bigodot_{j=m}^{0}\Post_{3j}\shortodot\Post_2\shortodot\bigodot_{j=n}^{0}\Post_{1j}
\end{align*}

For the sake of simplicity, we prove the rule in the  restricted case where $n$ and $m$ are equal to $0$, hence a single tau open transition will be considered on each side of the potentially visible one. The proof may be easily generalized to the multiple tau open transitions by using the same reasoning and \WTTrois~ rule. Consider each open transition separately:
\begin{enumerate}
\item For the first open transition in $\mathcal{T}_1$:
\begin{mathpar}
 \openrule
    {
       \beta_{1j}^{j\in J_1'},\Pred_1,\Post_1   }
         {s_{10} \OTarrow {\tau} s_{11}}          
\end{mathpar}
by hypothesis we have $(s,t|\Pred_{s,t})\in\mathcal{R}$  and $s=s_{10}$. Thus, by  Definition \ref{def-Weak-bisim} we  can deduce there exists an indexed set of weak open transitions $\symb{WOT}_a^{a\in A} \subseteq \WT_2$:
 \begin{mathpar}
    \openrule
         {
           \gamma_{j a}^{j\in J_{a}}, \Pred_{OT_a},\Post_{OT_a}}
         {t \OTWeakarrow {\alpha_{1 a}} u_a}
\end{mathpar}
 such that  $\forall a, J_{a}=\{j\in J_1'|\beta_{1j}\neq\tau\},   (s_{11},u_a|\Pred_{s_{11},a})\in \mathcal{R}$ and \\
 $\Pred_{s,t} \land \Pred_1\implies\\
 \displaystyle{\bigvee_{a\in A}}
  \!\! \left( \forall j\in J_a.\vis{\beta_{1j}}\!=\!\gamma_{ja} \!\land\! \Pred_{OT_a}\!
     \land\! \alpha_{1 a}\!=\!\tau \land  
     \Pred_{s_{11},a}\subst{\Post_1\uplus\Post_{OT_a}}\right)$
   
 Note that, because $\AlgE\cap\AlgA=\emptyset$  (actions and expressions are disjoint) and $\alpha_{1 a}\!=\!\tau$ we have 
directly ($\alpha_{1 a}$ cannot be a variable, and cannot contain expressions/variables because $\tau$ has no parameter):
\begin{mathpar}
    \openrule
         {
           \gamma_{j a}^{j\in J_{a}}, \Pred_{OT_a},\Post_{OT_a}}
         {t \OTWeakarrow {\tau} u_a}
\end{mathpar}
%

\item Concerning the middle open transition in $\mathcal{T}_1$:
 \begin{mathpar}
     \openrule
         {
           \beta_{2j}^{j \in J_2'},\Pred_2,\Post_2 }
         {s_{20} \OTarrow {\alpha'} s_{21}} 
\end{mathpar}
we have  $(s_{11},u_a|\Pred_{s_{11},a})\in \mathcal{R}$  and $s_{11}=s_{20}$. Again 
 by Definition \ref{def-Weak-bisim} we can deduce there exists an indexed set of weak open transitions $\symb{WOT}_{b}^{b\in B} \subseteq \WT_2$:
 \begin{mathpar}
    \openrule
         {
           \gamma_{j b}^{j\in J_{b}}, \Pred_{OT_{b}},\Post_{OT_{b}}}
         {u_a \OTWeakarrow {\alpha_{2 b}} v_{b}}
\end{mathpar}
such that  $\forall b, J_{b}=\{j\in J_2'|\beta_{2j}\neq\tau\}, 
(s_{21},v_{b}|\Pred_{s_{21},b})\in \mathcal{R}$; \\
 $\Pred_{s_{11},a} \land \Pred_2\implies\\
\displaystyle{\bigvee_{b\in B}}
 \!\!  \left( \forall j\in J_{b}. \vis{\beta_{2j}}\!=\!\gamma_{jb}\land \Pred_{OT_{b}}\!\land \alpha'\!=\!\alpha_{2 b}\! \land\!  
     \Pred_{s_{21},b}\subst{\Post_2\uplus\Post_{OT_{b}}}\right)$

\item Similarly to the case 1, we consider the third  open transition in $\mathcal{T}_1$:
     \begin{mathpar}
 \openrule
    {
       \beta_{3j}^{j \in J_3'},\Pred_3,\Post_3   }
         {s_{30} \OTarrow {\tau} s_{31}} \in \mathcal{T}          
\end{mathpar}
From previous case, we have $(s_{21},v_{b}|\Pred_{s_{21},b})\in \mathcal{R}$, and we have $s_{21}=s_{30}$. Then, by
 Definition \ref{def-Weak-bisim} there exists an indexed set of weak open transitions $\symb{WOT}_{a}^{c\in C} \subseteq \WT_2$:
 \begin{mathpar}
    \openrule
         {
           \gamma_{jc}^{j\in J_{c}}, \Pred_{OT_{c}},\Post_{OT_{c}}}
         {v_{b} \OTWeakarrow {\tau} w_{c}}
\end{mathpar}
 such that  $\forall c, J_{c}= \{j\in J_3'|\beta_{3j}\neq\tau\}, 
(s_{31},w_{c}|\Pred_{s_{31},c})\in \mathcal{R}$ and \\
 $\Pred_{s_{21},b} \land \Pred_3\implies 
\displaystyle{\bigvee_{c\in C}}
   \!\left( \forall j\in J_{c}. \vis{\beta_{3j}}\!=\!\gamma_{jc} \! \land\! \Pred_{OT_{c}}\!
     \land\!  
     \Pred_{s_{31},c}\subst{\Post_3\uplus\Post_{OT_{c}}}\right)$
         
\end{enumerate}
\medskip     
Based on cases described above by applying \WTTrois~rule on the resulting $\symb{WOT}$s we have:

\begin{mathpar}
  \mprset {vskip=.5ex}
\inferrule {
\openrule
         {
           \gamma_{j a}^{j\in J_{a}}, \Pred_{OT_x},\Post_{OT_a}}
         {t \OTWeakarrow {\tau} u_a}
\qquad
\openrule
         {
           \gamma_{j b}^{j\in J_{b}}, \Pred_{OT_{ y}},\Post_{OT_{b}}}
         {u_a \OTWeakarrow {\alpha_2} v_{b}}
\qquad
\openrule
         {
           \gamma_{jc}^{j\in J_{c}}, \Pred_{OT_{c}}, \Post_{OT_{c}}}
         {v_{b} \OTWeakarrow {\tau} w_{c}}
\\ 
\set{\gamma'}=\gamma_{j a}^{j\in J_{a}} \dotcup \gamma_{jb}^{j\in J_{b}} \subst{\Post_{OT_a}} \dotcup\gamma_{jc}^{j\in J_{c}} \subst{\Post_{OT_{ b}}\shortodot\Post_{OT_a}}\\
\Pred=\Pred_{OT_a}\land\Pred_{OT_{b}}\subst{\Post_{OT_a}}\land \Pred_{OT_{c}}\subst{\Post_{OT_{b}}\shortodot\Post_{OT_a}}\\
\Post= \Post_{OT_{c}} \shortodot \Post_{OT_{b}}\shortodot\Post_{OT_{a}}\quad \alpha''=\alpha_{2}\subst{\Post_{OT_a}}\\
}
{
\openrule
         {
           \set{\gamma},
		\Pred, \Post}  
         {t \OTWeakarrow {\alpha''} w_{c}} 
} 
 \end{mathpar}
It remains to be proven that the following statement holds:\\
$\Pred_{s,t} \land \Pred\implies
 \displaystyle{\bigvee_{x\in X}} 
  \left( \forall j\in J. \gamma_j'=\gamma_{j}  \land \Pred
     \land \alpha \!=\! \alpha'' \land  
     \Pred_{s',x}\subst{\Post_{OT}\uplus\Post\,}\right)$\\
We have:\\
$\Pred_{OT}= \Pred_1 \wedge \Pred_2\subst{\Post_1} \wedge \Pred_3\subst{\Post_2\shortodot\Post_1}$\\
$\Post_{OT}=\Post_3\shortodot \Post_2\shortodot\Post_1$\\
Moreover, we have the following  statement:\\
$\Pred_{s,t} \land \Pred_1\implies
 \displaystyle{\bigvee_{a\in A}}
 \! \! \left( \forall j\in J_a. \vis{\beta_{1j}}=\gamma_{ja}  \land \Pred_{OT_a}
      \land  
     \Pred_{s_{11},a}\subst{\Post_1\uplus\Post_{OT_a}}\right)$\\
With the conjunction of the predicate $\Pred_2\subst{\Post_1}$ on both sides of the implication, we get:
\begin{multline*}
\Pred_{s,t} \land \Pred_1)\wedge\Pred_2\subst{\Post_1}\implies \\
 \bigvee_{a\in A}
  \!\! \left( \forall j\in J_a. \vis{\beta_{1j}}\!=\!\gamma_{ja}\!  \land\! \Pred_{OT_a}\! \land\!
     \Pred_{s_{11},a}\subst{\Post_1\uplus\Post_{OT_a}} \wedge \Pred_2\subst{\Post_1\uplus\Post_{OT_a}} \right) \hspace{-2ex}
\end{multline*}
Note that  on the right side of the implication we added  the substitution  of $\Post_{OT_x}$ without affecting the validity of the statement, because  the domain  of the substitution function $\Post_{OT_x}$ is disjoint from the others. Hence a little rewriting gives:
\begin{multline*}
\Pred_{s,t} \land \Pred_1)\wedge\Pred_2\subst{\Post_1}\implies\\
\bigvee_{a\in A}
\!\!   \left(\forall j\in J_a. \vis{\beta_{1j}}\!=\!\gamma_{ja}\land \Pred_{OT_a}
     \!\land(\Pred_{s_{11},a}\wedge\Pred_2)\subst{\Post_1\uplus\Post_{OT_a}} \right)
\end{multline*}

By replacing the inner predicate $(\Pred_{s_{11},u_a}\wedge\Pred_2)$ by the conclusion of the  statement given in case 2,  the formula becomes: 
\begin{multline*}\Pred_{s,t} \land \Pred_1)\wedge\Pred_2\subst{\Post_1}\implies\\
{\bigvee_{a\in A}}
   \!\!\Big(\forall j\in J_a. \vis{\beta_{1j}}\!\!=\!\gamma_{ja}  \land \Pred_{OT_a}
      \land \Big({\bigvee_{b\in B}}(
    \forall j\in J_{b}. \vis{\beta_{2j}}\!=\!\gamma_{jb}  \land \Pred_{OT_{b}}\!\land 
    \alpha'\!=\!\alpha_{2 b} \land\\ \Pred_{s_{21},b}\subst{\Post_2\uplus\Post_{OT_{b}}})\subst{\Post_1\uplus\Post_{OT_a}}\Big)\Big)
\end{multline*}

This can be rewritten into:
\begin{multline*}\Pred_{s,t} \land \Pred_1)\wedge\Pred_2\subst{\Post_1} \implies\\
{\bigvee_{a\in A}}\, {\bigvee_{b\in B}} 
\Big(\forall j\in J_a. \vis{\beta_{1j}}\!\!=\!\gamma_{ja} \land  \forall j \in J_{b}. \vis{\beta_{2j}}\!=\!\gamma_{jb}\subst{\Post_1\uplus\Post_{OT_a}}   \land\Pred_{OT_a} \land \\ \Pred_{OT_{b}} \subst{\Post_1\uplus\Post_{OT_a}} \land  (\alpha'\!=\!\alpha_{2 b}) \subst{\Post_1\uplus\Post_{OT_a}} \land \\ \Pred_{s_{21},b}\subst{\Post_2\shortodot\Post_1\uplus \Post_{OT_{b}}\shortodot \Post_{OT_a}}\Big)
\end{multline*}
Since $Post_1$  does not act on $\gamma_{jb}$, nor on $ \Pred_{OT_{b}}$ and $  \alpha_{2}$. As well $\Post_{OT_a}$  does not act on $\alpha'$, nor on  $\beta_{2j}$ the formula can be simplified as follows:
\begin{multline*}\Pred_{s,t} \land \Pred_1)\wedge\Pred_2\subst{\Post_1}\implies\\
\displaystyle{\bigvee_{a\in A}}\, \displaystyle{\bigvee_{b\in B}} 
\Big(\forall j\in J_a. \vis{\beta_{1j}}\!=\!\gamma_{ja} \land  \forall j \in J_{b}. \vis{\beta_{2j}}\subst{\Post_1}=\gamma_{jb} \subst{\Post_{OT_a}}  \land\Pred_{OT_a} \land \\ \Pred_{OT_{b}} \subst{\Post_{OT_a}} \land \alpha'\subst{\Post_1}=\alpha_{2 b} \subst{\Post_{OT_a}} \land\\ \Pred_{s_{21},b}\subst{\Post_2\shortodot\Post_1\uplus \Post_{OT_{b}}\shortodot \Post_{OT_a}}\Big)\end{multline*}
Finally, the conjunction with the term $\Pred_3\subst{\Post_2\shortodot\Post_1}$ of the both sides of the implication and rewriting, we get:
\begin{multline*}\Pred_{s,t} \land \Pred_1)\wedge\Pred_2\subst{\Post_1} \land \Pred_3\subst{\Post_2\shortodot\Post_1}\implies \\
\displaystyle{\bigvee_{a\in A}}\, \displaystyle{\bigvee_{b\in B}} 
\Big(\forall j\in J_a.\vis{\beta^1_j}\!=\!\gamma_{ja} \land  \forall j \in J_{b}. \vis{\beta_{2j}}\subst{\Post_1}\!=\!\gamma_{jb} \subst{\Post_{OT_a}}\land \Pred_{OT_a}\land\\\Pred_{OT_{b}}\subst{\Post_{OT_a}} \land  \alpha'\subst{\Post_1}= \alpha_{2 b} \subst{\Post_{OT_a}} \land (\Pred_{s_{21},v_b} \land  \Pred_3)\\ \subst{\Post_2\shortodot \Post_1\uplus \Post_{OT_{b}}\shortodot\Post_{OT_a}}\Big)\end{multline*} 
Again note that because of the domain  of the substitution function is independent from some predicates and expressions, we removed $\Post_1$ and  we  added the term $\Post_{OT_{b}}\shortodot\,\Post_{OT_a}$ in the substitution of the right side of the implication.\\
Finally, by replacing the predicate $(\Pred_{s_{21},b} \land  \Pred_3)$
by the conclusion of the  implication given in case 3, we get:
\begin{multline*}\Pred_{s,t} \land \underbrace{\Pred_1\land \Pred_2\subst{\Post_1} \land \Pred_3\subst{\Post_2\shortodot\Post_1}}_{\Pred_{OT}}\implies\\
 {\bigvee_{a\in A}}\, {\bigvee_{b\in B}}\, {\bigvee_{c\in C}} 
\Big(\forall\! j\in\! J_a.\vis{\beta_{1j}}\!=\!\gamma_{ja} \land  \forall j \!\in\! J_{b}.\vis{\beta_{2j}\subst{\Post_{OT_a}}}\!=\!\gamma_{jb} \subst{\Post_{OT_a}}  \land \\
 \forall j\!\in\! J_{c}. \vis{\beta_{3j} \subst{\Post_2\,\shortodot\,\Post_1}} \!=\!\gamma_{jc} \subst{\Post_{OT_{b}}\shortodot\Post_{OT_a}}  \land\\
   \underbrace{\Pred_{OT_a}\land\Pred_{OT_{b}} \subst{\Post_{OT_a}} \land 
  \Pred_{OT_{c}}\subst{\Post_{OT_{b}}\shortodot\Post_{OT_a}}}_{\Pred} \land  \underbrace{\alpha'\subst{\Post_1}}_{\alpha}=\underbrace{\alpha_{2 b} \subst{\Post_{OT_a}}}_{\alpha''} \\\land \Pred_{s_{31},c}  \subst{\underbrace{\Post_3\shortodot\Post_2\shortodot\Post_1}_{\Post_{OT}}\uplus \underbrace{\Post_{OT_{c}}\shortodot \Post_{OT_{b}}\shortodot\Post_{OT_a}}_{\Post}}\Big)
\end{multline*}
The three for all statements (on $J_a$, $J_b$ and $J_c$) can be concatenated using $\dotcup$, the list union lifted to indexed sets (if $\gamma=\gamma'$ and $\gamma''=\gamma'''$ then $\gamma\dotcup\gamma''=\gamma'\dotcup\gamma'''$).
\begin{multline*}\forall j\in J_a\uplus J_b\uplus J_c.  \vis{\beta_{1j}}\dotcup \vis{\beta_{2j}\subst{\Post_{OT_a}}}\dotcup \vis{\beta_{3j} \subst{\Post_2\shortodot\Post_1}}=\\ ~\hspace{5.5cm} \gamma_{j a} \dotcup \gamma_{jb} \subst{\Post_{OT_a}}  \dotcup \gamma_{jc} \subst{\Post_{OT_{b}}\shortodot\Post_{OT_a}}
\end{multline*}
We have  $s_{31}=s'$, so can rewrite the formula:\\
$\Pred_{s,t} \land \Pred_{OT}\implies
\displaystyle{\bigvee_{a\in A}}\, \displaystyle{\bigvee_{b\in B}}\, \displaystyle{\bigvee_{c\in C}} 
\Big(\forall j\in J.\gamma_j'=\gamma_j \land  \Pred \land \alpha=\alpha''  \land \Pred_{s',c}\subst{\Post_{OT}\uplus\Post\,}\Big)$\\

All the combinations of elements in $A$, $B$, and $C$ provide a set $X$ of weak open transitions (each combination of one transition in $A$, one in $B$, and one in $C$ provides one weak open transition in the set $X$, i.e. each $x\in X$ corresponds to a triple $(a,b,c)\in A\times B\times C$); this defines a set of weak open transitions indexed over $X$;  each such open transition leads to a $w_c$ that we call $t_x$. This re-indexing allows us to conclude:\\
$\Pred_{s,t} \land \Pred_{OT}\implies
 \displaystyle{\bigvee_{x\in X}} 
\Big(\forall j\in J.\gamma_j'=\gamma_j \land  \Pred \land \alpha=\alpha''  \land \Pred_{s',x}\subst{\Post_{OT}\uplus\Post\,}\Big)$
\qedhere

\fussy
\noindent{\bf Theorem~\ref{thm-weak-equiv}}. \emph{Weak FH-Bisimulation is an equivalence}.
 Suppose $\mathcal{R}$ 
is a weak FH-bisimulation. Then $\mathcal{R}$ is an equivalence, that is, $\mathcal{R}$ is 
reflexive, symmetric and transitive.
\sloppy

With the above lemma, we can use the same technique as for Theorem \ref{thm-equiv}  to prove that a weak FH-bisimulation is an equivalence. Indeed, we essentially use the same proof-scheme the main difference concerns  $\beta$ and  $\gamma$. Indeed, while the schema of the proof of transitivity was not directly applicable on the definition of weak bisimulation,  Lemma~\ref{lem-alternative-weak-bisim} provides a characterization of weak bisimulation similar to the definition of strong bisimulation, and thus the same proof scheme is directly applicable.

\subsection{Composition properties}\label{sec:app-composition}
This section gives decomposition/composition lemmas and their proofs, these are the equivalent of the composition lemmas for open transitions, but applied to weak open automata.
\begin{lem}[Weak open transition decomposition]\label{lem-decomposeWOT} 
	Let $\Leaves(Q)=pLTS_l^{l\in L_Q}$; suppose\footnote{Note that the hypotheses of the 
	lemma imply that $Q$ is 
	not a pLTS but a similar lemma can be proven for a pLTS $Q$}:
	\[ P[Q]_{j_0}  
		\models
		{\openrule
			{
				\gamma_j^{j\in J}, \Pred,  
				\Post}
			{\ostate{s_i^{i\in L}} \OTWeakarrow {\alpha}
				\ostate{s_i'^{\, i\in L}}}
		}
	\]
		with  $J\cap\Holes(Q)\neq\emptyset$ or $\exists i\in L_Q.\,s_i\neq s'_i$, i.e. $Q$ takes part in the reduction.  
		 Then there exist $n$, $\Pred\,'$,  
		$\Post\,'$,   and for all $p\in[0..n]$ there exist $\beta_p$, $\alpha_p$, $\Pred_p$, $\Post_p$ and a family $\gamma_{p j}^{j\in J_p}$ and for all $p\in[0..n+1]$ $s_{p i}^{\,i\in L_Q}$. s.t.:\\[-2ex]
		\begin{mathpar}
		P\models{\openrule
			{
				\gamma_j^{j\in (J_p\setminus \Holes(Q)) \cup \{j_0\}}, 
				\Pred\,',  
				\Post\,'}
			{\ostate{s_i^{i\in L\setminus L_Q}} \OTWeakarrow {\alpha}
				\ostate{s_i'^{\,i\in L\setminus L_Q}}}
		}
\text{\qquad and \qquad}\gamma_{j_0}=[\beta_0..\beta_n]
	\vspace{-2.2ex}\\\text{and for all $p\in[0..n]$~~}
		Q\models{\openrule
			{
				\gamma_{p j}^{j\in J_p}, \Pred_p,  
				\Post_p}
			{\ostate{s_{p i}^{i\in L_Q}} \OTWeakarrow {\alpha_p}
				\ostate{s_{(p+1) i}^{\,i\in L_Q}}}
		}
\\
		\text{such that~\qquad}  \bigcup_{p=0}^nJ_{p} = J\cap\Holes(Q) 
\text{, }
 \gamma_j^{j\in J\cap\Holes(Q)}= \mybigdotcup_{p=0}^n (\gamma_{p j}^{j\in J_p})\subst{\bigodot_{i=p-1}^{0}\Post_i}, \\
{~\hspace{2cm}}\Pred \iff \Pred\,'
		\land \!\!\bigwedge_{p=0}^n(\alpha_p\subst{\bigodot_{i=p-1}^{0}\Post_i} = \beta_p\land \Pred_p\subst{\bigodot_{i=p-1}^{0}\Post_i} ),
\\
{~\hspace{2cm}}\Post=\Post\,' \uplus \bigodot_{p=n}^0
		\Post_p,   \text{and\quad} \forall i \in L_Q.\, s_{(n+1) i} = s_i'\land s_{0 i} = s_i\\
		\end{mathpar} 
where for any $p$, $\Post_p$ only acts upon  variables $\vars(Q)$.
\end{lem}

\proof Suppose that we have: 
\begin{mathpar}
 P[Q]_{j_0} \models \openrule
			{
				{\gamma_j^{j\in J}}, \Pred,  
				\Post}
			{\ostate{s_l^{l\in L}} \OTWeakarrow {\alpha}
				\ostate{s_l'^{\, l\in L}}} 
\end{mathpar}
By Lemma \ref{lem-rel-OT-WOT} this implies the following: \\			
$ \forall p\! \in [0..m_1]\,		
	P[Q]_{j_0}\!\models		
\openrule
    {
       \set{\beta_{1p}},\Pred_{1p},\Post_{1p}   }
         {\ostate{s^{l\in L}_{pl}} \OTarrow {\tau} \ostate{s^{l\in L}_{(p+1)l}}}$,  $\qquad  P[Q]_{j_0}	\models \openrule
         {
           \set{\beta_2},\Pred_2,\Post_2 }
         {\ostate{t_l^{l\in L}} \OTarrow {\alpha'} \ostate{{t'}_l^{l\in L}}}
$ 
\\
and 
 $\forall p \in [0..m_2] \,\, P[Q]_{j_0}\!\models \openrule
         {
           \set{\beta_{3p}},\Pred_{3p}, \Post_{3p}   }
         {\ostate{u^{l\in L}_{pl}} \OTarrow {\tau} \ostate{u^{l\in L}_{(p+1)l}}}
$\\
where\\
{\small \begin{align*}
\forall l\in L.\,&s_l=s_{0 l} \wedge s_{(m_1+1) l}=t_l \wedge t'_l = u_{0 l} \wedge u_{(m_2+1) l}=s'_l\\
\alpha=&\alpha'{\subst{\bigodot_{j=m_1}^{0}\!\!\Post_{1j}}}\\
\gamma_j^{j\in J}=&\mybigdotcup_{i=0}^{m_1} \vis{\set{\beta_{1i}}\subst{\bigodot_{j=i-1}^{0}\!\! \Post_{1j}}}\dotcup\  \vis{\set{\beta_2}\subst{\bigodot_{j=m_1}^{0}\Post_{1j}}} \dotcup  
{\mybigdotcup_{i=0}^{m_2} \vis{\set{\beta_{3i}} \subst{\bigodot_{j=i-1}^{0}\!\Post_{3j}\shortodot\Post_2\shortodot\bigodot_{j=m_1}^{0}\!\!\Post_{1j}} }}
\\
\Pred=&\bigwedge_{i=0}^{m_1}\Pred_{1i}\subst{\bigodot_{j=i-1}^{0}\!\!\Post_{1j}}\land\Pred_2 \subst{\bigodot_{j=m_1}^{0}\!\!\Post_{1j}}\land
\bigwedge_{i=0}^{m_2}\Pred_{3i}\subst{\bigodot_{j=i-1}^{0}\!\!\Post_{3j}\shortodot\Post_2\shortodot\bigodot_{j=m_1}^{0}\!\!\Post_{1j}}\\
\Post=&\bigodot_{j=m_2}^{0}\!\!\Post_{3j}\shortodot\Post_2\shortodot\bigodot_{j=m_1}^{0}\!\!\Post_{1j}
\end{align*}}

\noindent We can apply Lemma \ref{lem-decompose} on each $\symb{OT}$:
\begin{enumerate}
\item For each open transition  $\symb{OT}_p$ in the form ($\set{\beta_{1p}}=\beta_{1pj}^{j\in J_{1p}}$):\\ 
\[P[Q]_{j_0}\!\models		
\openrule
    {
       {\beta_{1pj}^{j\in J_{1p}}},\Pred_{1p},\Post_{1p}   }
         {\ostate{s^{l\in L}_{pl}} \OTarrow {\tau} \ostate{s^{l\in L}_{(p+1)l}}}\] 
If $Q$ moves then we obtain by Lemma \ref{lem-decompose}:\\
\begin{small}$P\models{\openrule
			{
				(\beta_{1pj})^{j\in (J_{1p}\setminus \Holes(Q)) \cup \{j_0\}}, 
				{\Pred\,'_{1p}},  
				{\Post\,'_{1p}}}
			{\ostate{s_{p l}^{l\in L\setminus L_Q}} \OTarrow {\tau}
					\ostate{({s_{(p+1) l}})^{l\in L\setminus L_Q}}}
		}
	\text{~~and~~} 
		Q\models{\openrule
			{
				(\beta_{1pj})^{j\in J_{1p}\cap\Holes(Q)}, {\Pred\,''_{1p}},  
				{\Post\,''_{1p}}}
				{\ostate{s_{pl}^{l\in L_Q}} \OTarrow {\alpha_{1p}}
				\ostate{s_{(p+1)l}^{l\in L_Q}}}
				}\\
$
\end{small}
such that  \\
$\Pred_{1p} \!\!\iff\!\! {\Pred\,'_{1p}}\land {\Pred\,''_{1p}}\land\alpha_{1p}\!=\!\beta_{1pj_0}$, $\Post_{1p}\!=\!{\Post\,'_{1p}}\uplus{\Post\,''_{1p}}$  where ${{\Post\,''_{1p}}}$ is the restriction of $\Post_{1p}$ over   $\vars(Q)$.\\
Else $Q$ does not move and we have: \\
\begin{small}
$P\models{\openrule
			{
				(\beta_{1pj})^{j\in (J_{1p}\setminus \Holes(Q)) }, 
				{\Pred\,'_{1p}},  
				{\Post\,'_{1p}}}
			{\ostate{(s_{p l})^{l\in L\setminus L_Q}} \OTarrow {\tau}
				\ostate{({s}_{(p+1) l})^{l\in L\setminus L_Q}}}
		}$~~and~~ $\ostate{(s_{p})_l^{l\in L_Q}} = \ostate{({s}_{(p+1) l})^{l\in L_Q}} $
\end{small}
\item Similarly, we have similar open transitions on states $u_{pl}$ (for the final $\tau$ transitions).

\item Finally, for the open transition in the form ($\set{\beta_{2}}=\beta_{2j}^{j\in J_2}$):\\ 
\[P[Q]_{j_0}	\models \openrule
         {
           \beta_{2j}^{j\in J_2},\Pred_2,\Post_2 }
         {\ostate{t_l^{l\in L_Q}} \OTarrow {\alpha'} \ostate{(t'_l)^{l\in L_Q}}}\]
If $Q$ moves then we  obtain by Lemma \ref{lem-decompose}:\\
\begin{small}$
		P\models{\openrule
			{
				(\beta_{2j})^{j\in (J_2\setminus \Holes(Q)) \cup \{j_0\}}, 
				{\Pred\,'_2},  
				{\Post\,'_2}}
			{\ostate{t_l^{l\in L\setminus L_Q}} \OTarrow {\alpha'}
				\ostate{t_l'^{\,l\in L\setminus L_Q}}}
		}
	\,\text{~~and~~}
	Q\models{\openrule
			{
				(\beta_{2j})^{j\in J_2\cap\Holes(Q)}, {\Pred\,''_2},  
				{\Post\,''_2}}
			{\ostate{t_l^{l\in L_Q}} \OTarrow {\alpha_{20}}
				\ostate{t_l'^{\,l\in L_Q}}}
		}
$
\end{small}\\
such that  $\Pred_2 \iff {\Pred\,'_2}
		\land {\Pred\,''_2}\land \alpha_2=\beta_{2j_0}$, $\Post_2={\Post\,'_2}\uplus 
		{\Post\,''_2}$ where ${\Post\,''_2}$ is the restriction of $\Post_2$ over variables  $\vars(Q)$.

Else $Q$ does not move and we have: \\
$\begin{small} P\models{\openrule
			{
				(\beta_{2j})^{j\in (J_2\setminus \Holes(Q)) \cup \{j_0\}}, 
				{\Pred\,'_2},  
				{\Post\,'_2}}
			{\ostate{t_l^{l\in L\setminus L_Q}} \OTarrow {\alpha'}
				\ostate{(t_l)'^{\,l\in L\setminus L_Q}}}
		}
	\,\text{~~and~~} \ostate{t_l^{l\in L_Q}} = 
				\ostate{(t_l')^{\,l\in L_Q}} \end{small}$

\end{enumerate}
By using  Lemma \ref{lem-rel-OT-WOT}, and denoting $\displaystyle{J=\bigcup_{i=0}^{m_1}J_{1i}\cup J_2\cup \bigcup_{i=0}^{m_2}J_{3i}}$, we can conclude from cases (1), (2) and (3) that we have:
\begin{mathpar}
		P\models{\openrule
			{
				(\gamma_j')^{j\in (J\setminus \Holes(Q)) \cup \{j_0\}}, 
				\Pred\,',  
				\Post'}
			{\ostate{s_l^{l\in L\setminus L_Q}} \OTWeakarrow {\alpha''}
				\ostate{s_l'^{\,l\in L\setminus L_Q}}}
		}
\end{mathpar}
where
$\alpha''=\alpha'\displaystyle{\subst{\bigodot_{j=m_1}^{0}\!\!\Post\,'_{1j}}}$\\
On the other hand, we have:
$\alpha=\alpha'\displaystyle{\subst{\bigodot_{j=m_1}^{0}\!\!\Post\,'_{1j}}}\displaystyle{\subst{\bigodot_{j=m_1}^{0}\!\!\Post\,''_{1j}}}=\alpha''\displaystyle{\subst{\bigodot_{j=m_1}^{0}\!\!\Post\,''_{1j}}}$.\\
As $\displaystyle{\subst{\bigodot_{j=m_1}^{0}\!\!\Post\,''_{1j}}}$ has no effect on variables of $P$ and thus on variables of $\alpha''$, so we have $\alpha=\alpha''$.
\\
{\small \begin{align*}
\forall l\in L.\,&s_l=s_{0l} \wedge s_{(m_1+1)l}=t_l \wedge t'_l = u_{0l} \wedge u_{(m_2+1)l}=s'_l\\
{(\gamma_j')}^{j\in (J\setminus \Holes(Q)) \cup \{j_0\}}=&\displaystyle{\mybigdotcup_{i=0}^{m_1} \vis{\set{\beta_{1i}}\subst{\bigodot_{j=i-1}^{0}\!\!\Post\,'_{1j}}}\dotcup  \vis{\set{\beta_{2}}\subst{\bigodot_{j=m_1}^{0}\Post\,'_{1j}}}} \dotcup\\
& \mybigdotcup_{i=0}^{m_2}
\vis{\set{\beta_{3i}} \subst{\bigodot_{j=i-1}^{0}\!\Post\,'_{3j}\shortodot\Post\,'_2\shortodot\bigodot_{j=m_1}^{0}\!\!\Post\,'_{1j}} }
\\
\Pred\,'=&\bigwedge_{i=0}^{m_1}\Pred\,'_{1i}\subst{\bigodot_{j=i-1}^{0}\!\!\Post\,'_{1j}}\land\Pred\,'_2 \subst{\bigodot_{j=m_1}^{0}\!\!\Post\,'_{1j}}\land\\ 
&\bigwedge_{i=0}^{m_2}\Pred\,'_{3i}\subst{\bigodot_{j=i-1}^{0}\!\!\Post\,'_{3j}\shortodot\Post\,'_2\shortodot\bigodot_{j=m_1}^{0}\!\!\Post\,'_{1j}}\\
\Post'=&\bigodot_{j=m_2}^{0}\!\!\Post\,'_{3j}\shortodot\Post\,'_2\shortodot\bigodot_{j=m_1}^{0}\!\!\Post\,'_{1j}
\end{align*}}

Note that for all $j\in J\setminus \Holes(Q)$, $\gamma_j' = \gamma_j$ because for all $l$ $\Post\,'_{1l}$ coincides with $\Post_{1l}$ on the variables of $\beta_{1ij}$, and similarly for $\Post\,'_2$ and $\Post\,'_{3l}$.

We introduce the following predicate (we will need it for reasoning on the global predicate and will reason on it along the proof):\\
$\begin{array}{l}\displaystyle{
\Pred_\beta=\bigwedge_{p=0}^{m_1}(\beta_{1pj_0}=\alpha_{1p})\subst{\bigodot_{j=p-1}^{0}\!\!\Post_{1j}} \wedge (\beta_{2j_0}=\alpha_{20})\subst{\bigodot_{j=m_1}^{0}\!\!\Post_{1j}}} \wedge \\\qquad\qquad
\displaystyle{
\bigwedge_{p=0}^{m_2}(\beta_{3pj_0}=\alpha_{3p})
\subst{\bigodot_{j=p-1}^{0}\!\!\Post_{3j}\shortodot\Post_2\shortodot\bigodot_{j=m_1}^{0}\!\!\Post_{1j}}}
\end{array}$

Concerning $Q$, we reduce the sequence of OTs to a path for which it moves in all steps. In other words, if $Q$ does not move at step $q$, then we have $\ostate{s_{q_l}^{l\in L_Q}} = \ostate{{s}_{(q+1) l}^{l\in L_Q}}$ , then we skip the state $\ostate{{s}_{(q+1) l}^{l\in L_Q}}$, i.e.  we rename all the following states  $\ostate{s_{p l}^{l\in L_Q}}$ where $p\geq q+1$ into $\ostate{{s}_{(p-1) l}^{l\in L_Q}}$. Note that self-loops where $Q$ does an action but stays at the same state are not removed. We proceed in the same way  for states named $u$. To simplify the proof, we suppose that in case 3, $Q$  moves, else transition 3 of $Q$ should be skipped and the last $s_{p l}$ are equal to the first $u_{0 l}$. So we have: \\ 
\begin{small}
$\forall p\! \in\! [0..n_1]~ Q\models{\openrule
			{
				(\beta_{1pj}')^{j\in J\cap\Holes(Q)}, {{\Pred\,''_{1p}}},  
				{{\Post\,''_{1p}}}}
				{\ostate{s_{pl}^{l\in L_Q}} \OTarrow {\alpha_{1p}}
				\ostate{s_{(p+1)l}^{l\in L_Q}}}
				}
,~~$  
$ Q\models{\openrule
			{
				(\beta'_{2 j})^{j\in J\cap\Holes(Q)}, {\Pred\,''_2},  
				{\Post\,''_2}}
			{\ostate{t_l^{l\in L_Q}} \OTarrow {\alpha_{20}}
				\ostate{t_l'^{\,l\in L_Q}}}
		}
$
and \\
$\forall p \in [0..n_2]~ Q\models{\openrule
			{
				(\beta'_{3pj})^{j\in J\cap\Holes(Q)}, {{\Pred\,''_{3p}}},  
				{{\Post\,''_{3p}}}}
				{\ostate{u_{pl}^{l\in L_Q}} \OTarrow {\alpha_{3p}}
				\ostate{u_{(p+1)l}^{l\in L_Q}}}
				}
$
\end{small}
\\ 
such that $n_1 \leq m_1$ and $n_2 \leq m_2$.

By renaming all state names ($s$, $u$ and $t$) with the same state name $v$. We have: 

$\forall p \in [0..(n_1\!+\!n_2\!+\!2)]\,\, Q\models{\openrule
			{
				\beta_{pj}^{j\in J\cap\Holes(Q)}, {{\Pred\,''_p}},  
				{{\Post\,''_p}}}
				{\ostate{v_{pl}^{l\in L_Q}} \OTarrow {\alpha'_{p}}
				\ostate{v_{(p+1)l}^{l\in L_Q}}}
				}
$\\ 
In this equation, and using case 1 above for all $k\in [0\ldots n_1]$ there is a $p\in [0..m_1]$ such that $\alpha_{1p}=\alpha'_k$ (following the re-indexing done in the removal of steps where $Q$ does not move), we know that $\Pred_{1p}$ contains the predicate $( \alpha_{1p}=\beta_{1 p j_0})$. Because $\beta_{1 p j_0}$ only contains variables of $P$ and $\alpha'_k$ only variables of $Q$, we have:
\begin{equation*}
\begin{split}
(\alpha_{1p}=\beta_{1pj_0})\subst{\bigodot_{j=p-1}^{0}\!\!\Post_{1j}}&\iff
\alpha_{1p}\subst{\bigodot_{j=p-1}^{0}\!\!\Post\,''_{1j}}
= \beta_{1 p j_0}\subst{\bigodot_{j=p-1}^{0}\!\!\Post\,'_{1j}} \\
&\iff\alpha'_k\subst{\bigodot_{j=k-1}^{0}\!\!\Post\,''_{j}}= \beta_{1 p j_0}\subst{\bigodot_{j=p-1}^{0}\!\!\Post\,'_{1j}} 
\end{split}
\end{equation*}
We can obtain similar equations for $\alpha'_{n_1+1}$ related with $\beta_{2 j_0}$ and the  $\alpha'_k$ for $k\geq n_1+2$ related with $\beta_{3 p j_0}$ for some $p$. Note that the substitutions are however more complex in the other cases. 
Overall we obtain (we skip here the details about the three cases 1, 2, and 3 above that all fall into the same equation because of the re-indexing we perform): 
\begin{equation}\label{eqproofbeta}
\Pred_\beta \iff \gamma_{j_0}=
\vis{[\alpha'_{p}\subst{\bigodot_{j=p-1}^{0}\!\!\Post\,''_{j}}|p\in[0..n_1+n_2+2]]}
\end{equation}

%

Let us consider the sequence of $(n_1+n_2+3)$  actions $\alpha'_p$ some of them may be non-observable (they are $\tau$ transitions). By considering the sequence of $\tau$ and non-$\tau$ actions we split the sequence of actions into $n+1$ sub-sequences, such that each sub-sequence is a sequence of actions containing  only one observable action that will be named $\alpha_p$, and possibly many non-observable ($\tau$) ones. 

\begin{figure}
\begin{center}
\includegraphics[width=6cm]{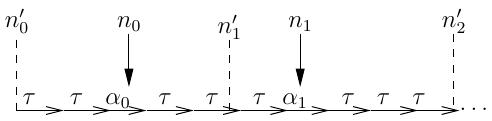}
\end{center}\caption{Composition of the subsequences}\label{figproof}
\end{figure}

 We can decompose each of the $n+1$ sub-sequences in the following way (see Figure~\ref{figproof}).
For $k\in [0..n]$ the position of the $k^{\text{th}}$ visible action is  $n_k$. For $l\in [1..n]$,  $n'_l$ is any index between $n_{l-1}$ and $n_l$, additionally $n'_0=0$ and $n'_{n+1}=n_1+n_2+3$. We obtain $n+1$ sub-sequences  made of the following OTs, for all $k\in [0..n]$ :\\
\begin{small}
$\forall p\! \in\! [n'_{k}..(n_k\!-\!1)]~ Q\models{\openrule
			{
				\beta_{pj}^{j\in J\cap\Holes(Q)}, {{\Pred\,''_{p}}},  
				{{\Post\,''_{p}}}}
				{\ostate{v_{pl}^{l\in L_Q}} \OTarrow {\tau}
				\ostate{v_{(p+1)l}^{l\in L_Q}}}
				}
,\quad
Q\models{\openrule
			{
				\beta_{n_k j}^{j\in J\cap\Holes(Q)}, {{\Pred\,''_{n_k}}},  
				{{\Post\,''_{n_k}}}}
				{\ostate{v_{n_kl}^{l\in L_Q}} \OTarrow {\alpha'_{{n_k}}}
				\ostate{v_{(n_k+1)l}^{l\in L_Q}}}
				}
$
\end{small}
and\\
\begin{small}
$\forall p \in [(n_k\!+\!1)..(n'_{k+1}\!-\!1)]\,\, Q\models{\openrule
			{
				\beta_{pj}^{j\in J\cap\Holes(Q)}, {{\Pred\,''_{p}}},  
				{{\Post\,''_{p}}}}
				{\ostate{v_{pl}^{l\in L_Q}} \OTarrow {\tau}
				\ostate{v_{(p+1)l}^{l\in L_Q}}}
				}
$
\end{small}
\\ 
Thereafter, by Lemma \ref{lem-rel-OT-WOT} we can deduce the following weak open transition:
\begin{mathpar}
Q\models{\openrule
			{
			(\gamma_{kj})^{j\in J\cap\Holes(Q)},  {{\Pred_k}},  
				{{\Post_k}}}
			{\ostate{v_{kl}^{l\in L_Q}} \OTWeakarrow {\alpha_{k}}
				\ostate{(v_{kl}')^{l\in L_Q}}}
		}
\end{mathpar}
with:
{\small
\begin{align*}
\forall l \in L_Q. &v_{kl}=v_{(n'_{k})l} \wedge v_{kl}'=v_{(n'_k)l}\\
\alpha_k=&\alpha'_{n_k}{\subst{\bigodot_{j=n_k-1}^{n'_{k}}\!\!\Post\,''_{j}}}\\
{\gamma_{kj}^{j\in J\cap\Holes(Q)} }=&{\mybigdotcup_{i=n'_{k}}^{(n_k\!-\!1)} \vis{{\beta_{ij}^{j\in J\cap\Holes(Q)}}\subst{\bigodot_{l=i-1}^{n'_{k}}\!\!\!\Post\,''_{l}}}\dotcup \vis{{\beta_{n_k j}^{j\in J\cap\Holes(Q)}}\subst{\bigodot_{l=n_k-1}^{n'_{k}}\!\!\!\Post\,''_{l}}}} \dotcup\\
&\mybigdotcup_{i=n_k+1}^{n'_{k+1}-1}
\vis{{\beta_{ij}^{j\in J\cap\Holes(Q)}} \subst{\bigodot_{l=i-1}^{n_k\!+\!1}\!\Post\,''_{l}\shortodot\Post\,''_{n_k}\shortodot\!\!\bigodot_{l=n_k-1}^{n'_{k}}\!\!\!\Post\,''_{l}} }\\
\Pred_k=&\bigwedge_{i=n'_{k}}^{n_k\!-\!1}\!\!\Pred\,''_{i}\subst{\bigodot_{j=i-1}^{n'_{k}}\!\!\Post\,''_{j}}\land\Pred\,''_{n_k} \subst{\bigodot_{j=n_k-1}^{n'_{k}}\!\!\Post\,''_{j}}\land\\ 
&\bigwedge_{i=n_k+1}^{n'_{k+1}-1}\!\!\!\Pred\,''_{i}\subst{\bigodot_{j=i-1}^{n'_{k}}\!\!\!\Post\,''_{j}\shortodot\Post\,''_{n_k}\shortodot\!\!\bigodot_{j=n_k-1}^{n'_{k}}\!\!\!\Post\,''_{j}}
\\
\Post_k= &\bigodot_{j=n'_{k+1}-1}^{n'_{k}}\!\!\Post\,''_{j}
\end{align*}
}

\noindent Note that for all $k\in[0..n-1]$, $v_{kl}'=v_{(k+1)l}$,  $v_{0l}=s_{0l}=s_l$, and $v'_{nl}=v_{(n_1+n_2+3) l}=u_{(n_2+1) l} = s'_l$.

\noindent By definition of $\Post_k$, we have $\displaystyle{\bigodot_{j=n'_{k}-1}^{0}\!\!\Post\,''_{j} = \bigodot_{j=k-1}^{0}\!\!\Post_{j}}$.
Consequently,  we have:
\[\alpha'_{n_k}{\subst{\!\bigodot_{j=n_k-1}^{0}\!\!\!\Post\,''_{j}}} =
\alpha'_{n_k}\displaystyle{\subst{\!\bigodot_{j=n_k-1}^{n'_{k}}\!\!\!\Post''_{j}\otimes\!\!\!\bigodot_{j=n'_{k}-1}^{0}\!\!\!\Post\,''_{j}}} = \alpha_k\displaystyle{\subst{\!\bigodot_{j=n'_{k}-1}^{0}\!\!\!\Post\,''_{j}}} = 
\alpha_k\displaystyle{\subst{\!\bigodot_{j=k-1}^{0}\!\!\Post_{j}}}\]
From equation~\ref{eqproofbeta}, we obtain the following equation (we recall that the actions $\alpha_k$ are the actions $\alpha'_{p}$ that are observable): 
\begin{eqnarray*}
\Pred_\beta &\iff& \gamma_{j_0}=
\mybigdotcup_{p=0}^{p=n_1+n_2+2}
\vis{\alpha'_{p}\subst{\bigodot_{j=p-1}^{0}\!\!\Post_{j}}}\\
&\iff& \gamma_{j_0}=
[\alpha_{p}\subst{\bigodot_{j=p-1}^{0}\!\!\Post_{j}}|p\in[0..n]]
\end{eqnarray*}

We need now to show that the set of WOT obtained above verifies the conditions of the lemma, i.e. it is a set of WOT of the form:
\[	Q\models{\openrule
			{
				\gamma_{p j}^{j\in J_p}, \Pred_p,  
				\Post_p}
			{\ostate{s_{p i}^{i\in L_Q}} \OTWeakarrow {\alpha_p}
				\ostate{s_{(p+1) i}^{\,i\in L_Q}}}
		}\]
with
\[  \bigcup_{p=0}^nJ_{p} = J\cap\Holes(Q) \qquad\text{trivial}\]
\[ \gamma_j^{j\in J\cap\Holes(Q)}= \mybigdotcup_{p=0}^n (\gamma_{p j}^{j\in J_p})\subst{\bigodot_{i=p-1}^{0}\Post_i)}\]
Indeed we have:
\begin{equation*}
\begin{split}
\gamma_j^{j\in J}=&\mybigdotcup_{i=0}^{m_1} \vis{\set{\beta_{1i}}\subst{\bigodot_{j=i-1}^{0}\!\! \Post_{1j}}}\dotcup  \vis{\set{\beta_2}\subst{\bigodot_{j=m_1}^{0}\Post_{1j}}} \dotcup\\& \mybigdotcup_{i=0}^{m_2} \vis{\set{\beta_{3i}} \subst{\bigodot_{j=i-1}^{0}\!\Post_{3j}\shortodot\Post_2\shortodot\bigodot_{j=m_1}^{0}\!\!\Post_{1j}} }\\
\end{split}\end{equation*}

And thus, because $\beta_{pj}^{j\in J\cap \Holes(Q)}$ are equal to the concatenation of  $(\beta'_{1pj})^{j\in J\cap \Holes(Q)}$, $(\beta'_{2j})^{j\in J\cap \Holes(Q)}$, and $(\beta_{3pj}')^{j\in J\cap \Holes(Q)}$ (re-indexed because we skipped some transitions), and additionally $(\beta_{1pj}')^{j\in J\cap \Holes(Q)}$, $(\beta_{2j}')^{j\in J\cap \Holes(Q)}$, and $(\beta_{3pj}')^{j\in J\cap \Holes(Q)}$ are identical to the hole labels $\beta_{1kj}^{j\in J\cap \Holes(Q)}$, $\beta_{2j}^{j\in J\cap \Holes(Q)}$, and $\beta_{3kj}^{j\in J\cap \Holes(Q)}$ (re-indexed) when $Q$ moves\footnote{more precisely, when $Q$ moves either $\beta_{1kj}^{j\in J\cap \Holes(Q)}$ is not empty  and thus $(\beta_{1pj}')^{j\in J\cap \Holes(Q)}=\beta_{1kj}^{j\in J\cap \Holes(Q)}$, or both are empty if the holes of $Q$ perform no action}. We can assert a similar equality on post-conditions, i.e. between $\Post\,''_p$ and $\Post\,''_{1k}$, $\Post\,''_{2}$, $\Post\,''_{3k}$ where $\Post\,''_{1p}$ is the restriction of $\Post_{1p}$ over $\vars(Q)$ (see initial decomposition, case 1, 2, and 3 above). Overall, we have
$\forall i \in L_Q.\, s_{(n+1) i} = s_i'\land s_{0 i} = s_i$ (see above):\\
{\scriptsize \begin{equation*}
\begin{split}
\gamma_j^{j\in J\cap \Holes(Q)}&=\mybigdotcup_{i=0}^{m_1} \vis{\set{(\beta_{1i}')}\subst{\bigodot_{j=i-1}^{0}\!\! \Post''_{1j}}}\dotcup  \vis{\set{\beta'_2}\subst{\bigodot_{j=m_1}^{0}\Post''_{1j}}} \dotcup \mybigdotcup_{i=0}^{m_2} \vis{\set{\beta'_{3i}} \subst{\bigodot_{j=i-1}^{0}\!\Post''_{3j}\shortodot\Post''_2\shortodot\bigodot_{j=m_1}^{0}\!\!\Post''_{1j}} }\\
&=\mybigdotcup_{k=0}^n \left( \mybigdotcup_{i=n'_k}^{n'_{k+1}-1}\vis{\beta_{ij}^{j\in J\cap \Holes(Q)}\subst{\bigodot_{j=i-1}^{n'_k}\!\!\Post''_{j}\bigodot_{j=n'_k-1}^{0}\!\!\Post''_{j}}}\right)\\
&=\mybigdotcup_{k=0}^n \left( \gamma_{kj}^{j\in J\cap \Holes(Q)}\subst{\bigodot_{j=k-1}^{0}\!\!\Post_{k}}\right)
\end{split}\end{equation*}
}
Next, we have:
\[\Pred = \displaystyle{\Pred\,'
		\land \!\!\bigwedge_{p=0}^n\Big((\alpha_p\subst{\bigodot_{i=p-1}^{0}\Post_i}) = \beta_p\land (\Pred_p\subst{\bigodot_{i=p-1}^{0}\Post_i}) \Big)}\]
Indeed we have:
\begin{scriptsize}
\begin{align*}
\Pred&\iff\bigwedge_{i=0}^{m_1}\Pred_{1i}\subst{\bigodot_{j=i-1}^{0}\!\!\Post_{1j}}\land\Pred_2 \subst{\bigodot_{j=m_1}^{0}\!\!\Post_{1j}}\land
\bigwedge_{i=0}^{m_2}\Pred_{3i}\subst{\bigodot_{j=i-1}^{0}\!\!\Post_{3j}\shortodot\Post_2\shortodot\bigodot_{j=m_1}^{0}\!\!\Post_{1j}}\\
&\iff\bigwedge_{i=0}^{m_1}\left(\Pred\,'_{1i}
		\land {\Pred\,''_{1i}}\land \alpha_{1i}=\beta_{1ij_0}\right)\subst{\bigodot_{j=i-1}^{0}\!\!\Post_{1j}} \\
&\qquad\land\left(\Pred\,'_{2}
		\land {\Pred\,''_{2}}\land \alpha_{20}=\beta_{2j_0}\right)\subst{\bigodot_{j=m_1}^{0}\!\!\Post_{1j}}\\
&\qquad\land\bigwedge_{i=0}^{m_2} \left(\Pred\,'_{3i}
		\land {\Pred\,''_{3i}}\land \alpha_{3i}=\beta_{3ij_0}\right) \subst{\bigodot_{j=i-1}^{0}\!\!\Post_{3j}\shortodot\Post_2\shortodot\bigodot_{j=m_1}^{0}\!\!\Post_{1j}}\displaybreak[0]\\
&\iff\bigwedge_{i=0}^{m_1}\left((\Pred\,'_{1i}\subst{\bigodot_{j=i-1}^{0}\!\!\Post\,'_{1j}})
		\land ({\Pred\,''_{1i}}\subst{\bigodot_{j=i-1}^{0}\!\!\Post\,''_{1j}})\land (\alpha_{1i}=\beta_{1ij_0})\subst{\bigodot_{j=i-1}^{0}\!\!\Post_{1j}}\right) \\
&\qquad\land \left(\Pred\,'_{2}\subst{\bigodot_{j=m_1}^{0}\!\!\Post_{1j}}
		\land {\Pred\,''_{2}} \subst{\bigodot_{j=m_1}^{0}\!\!\Post_{1j}} \land (\alpha_{20}=\beta_{2j_0})\subst{\bigodot_{j=m_1}^{0}\!\!\Post_{1j}} \right) \\
&\qquad\land\bigwedge_{i=0}^{m_2} \Bigg(\Pred\,'_{3i} \subst{\bigodot_{j=i-1}^{0}\!\!\Post_{3j}\shortodot\Post_2\shortodot\bigodot_{j=m_1}^{0}\!\!\Post_{1j}} 
		\land {\Pred\,''_{3i}} \subst{\bigodot_{j=i-1}^{0}\!\!\Post_{3j}\shortodot\Post_2\shortodot\bigodot_{j=m_1}^{0}\!\!\Post_{1j}}  \\
		 & \qquad \land (\alpha_{3i}=\beta_{3ij_0}) \subst{\bigodot_{j=i-1}^{0}\!\!\Post_{3j}\shortodot\Post_2\shortodot\bigodot_{j=m_1}^{0}\!\!\Post_{1j}}\Bigg)  \\
&\iff \Pred\,'\land\bigwedge_{k=0}^n\Pred_k\subst{\bigodot_{j=k-1}^{0}\!\!\Post_{j}}\land\Pred_\beta
\\&\iff \Pred\,'\land\bigwedge_{k=0}^n\Pred_k\subst{\bigodot_{j=k-1}^{0}\!\!\Post_{j}}\land
(\gamma_{j_0}=[\alpha_{i}\subst{\bigodot_{j=i-1}^{0}\!\!\Post_{j}}|i\in[0..n]])
\end{align*}
\end{scriptsize}
which is exactly what is needed with $\gamma_{j_0}=[\beta_0..\beta_n]$.

Finally we have~~
$\Post=\displaystyle{\Post\,' \uplus \bigodot_{p=n}^0
		\Post_p}$\quad
because
{\small \begin{align*}
\Post&=\bigodot_{j=m_2}^{0}\!\!\Post_{3j}\shortodot\Post_2\shortodot\bigodot_{j=m_1}^{0}\!\!\Post_{1j}\\
&=\bigodot_{j=m_2}^{0}\!\!\Post\,'_{3j}\shortodot\Post\,'_2\shortodot\bigodot_{j=m_1}^{0}\!\!\Post\,'_{1j} \uplus \bigodot_{j=m_2}^{0}\!\!\Post\,''_{3j}\shortodot\Post\,''_2\shortodot\bigodot_{j=m_1}^{0}\!\!\Post\,''_{1j}\\
&=\Post\,' \uplus \bigodot_{j=n'_{n+1}-1}^{0}\!\!\Post\,''_{j}
\end{align*}}
Which concludes because  we have $\displaystyle{\bigodot_{j=n'_{k}-1}^{0}\!\!\Post\,''_{j} = \bigodot_{j=k-1}^{0}\!\!\Post_{j}}$.
\qedhere

\begin{lem}[Weak open transition composition]\label{lem-Weakcompose1} Suppose that we have a weak open automaton such that the WOTs cannot observe silent actions (see Definition \ref{def:Non-ObsTau}).
	Suppose $j_0\in J$  and:\\[-1ex]
\begin{mathpar}
P\models{\openrule
	{
		\beta_j^{j\in J}, 
		\Pred,  
		\Post}
	{\ostate{s_i^{i\in L}} \OTarrow {\alpha}
		\ostate{(s_i')^{\, i\in L}}}
}
\quad\text{~~and~~}\quad
Q\models{\openrule
	{
		\set{\gamma},
		 \Pred_Q,  
		\Post_Q}
	{\ostate{s_{i}^{i\in L_Q}} \OTWeakarrow {\alpha_Q}
		\ostate{(s'_{i})^{\, i\in L_Q}}}
}
\end{mathpar}

Let 
\qquad
$\Pred\,'=\Pred\land (\beta_{j_0}=\alpha_Q\land \Pred_Q) \quad\text{~~and~~}\quad
\Post\,'=\Post\uplus\Post_Q
$

Then, we have:
{\small	\[ P[Q]_{j_0}  
	\models
	{\openrule
		{
			\set{\gamma}\uplus\vis{\beta_j^{j\in J\setminus \{j_0\}}}, 
			\Pred\,',  \Post\,'
			 }
		{\ostate{s_i^{i\in L\uplus L_Q}} \OTWeakarrow {\alpha}
			\ostate{(s_i')^{\, i\in L\uplus L_Q}}}
	}
	\]}
\end{lem}
\proof
By  Lemma \ref{lem-rel-OT-WOT} we can decompose the WOT of $Q$ into  a series of $k+1$ and $k'+1$ tau open transitions and an $\alpha'_Q$ open transition (observable or not depending on $\alpha_Q$):
\begin{mathpar}
\forall h\!\in\![0..k].Q\models\openrule
    {\set{\beta_{1h}},\Pred_{1h},\Post_{1h}   }
         {\ostate{{(s_{1h})}} \OTarrow {\tau} \ostate{{(s_{1(h+1)})}}}, \quad
Q\models{\openrule
	{
		\set{\beta_2},
		 \Pred_2,  
		\Post_2}
	{\ostate{s_{20}}\OTarrow {\alpha'_Q}
		\ostate{s_{21}}}
}, \\ \quad\text{~~and~~}\quad 
\forall h \in [0..k'].Q\models\openrule
         {
           \set{\beta_{3h}},\Pred_{3h}, \Post_{3h}}
         { \ostate{(s_{3h})} \OTarrow {\tau} \ostate{(s_{3(h+1)})}}
\end{mathpar}
such that\\

\begin{small}
$s_i^{i\in L_Q}=s_{10} \wedge s_{1(k+1) i}=s_{20} \wedge  s_{21}= s_{30} \wedge s_{3(k'+1) i}={s'}_i^{i\in L_Q}$
\begin{align*}
\alpha_Q=&\alpha'_Q\displaystyle{\subst{\bigodot_{j=k}^{0}\Post_{1j}}}\\
\set{\gamma}=&\mybigdotcup_{h=0}^k \vis{\set{\beta_{1h}}\subst{\!\!\bigodot_{j=h-1}^{0}\!\! \Post_{1j} } }  \dotcup  \vis{\set{\beta_2}\subst{\bigodot_{j=k}^{0}\!\!\Post_{1j}}} \dotcup
 \mybigdotcup_{h=0}^{k'} \vis{\set{\beta_{3h}} \subst{\!\!\bigodot_{j=h-1}^{0}\!\!\Post_{3j}\shortodot\Post_2\shortodot\bigodot_{j=k}^{0}\Post_{1j}} }\\
\Pred_Q=&\bigwedge_{h=0}^{k}\!\Pred_{1h}\subst{\!\!\bigodot_{j=h-1}^{0}\!\!\!\!\Post_{1 j}}\land \Pred_2\subst{\bigodot_{h=k}^{0}\Post_{1h}}\land
	\!\bigwedge_{h=0}^{k'}\!\Pred_{3h}\subst{\!\! 	\bigodot_{j=h-1}^{0}\!\!\!{\Post_{3j}}\otimes{\Post_2}  \otimes\bigodot_{h=k}^{0}\!\!\Post_{1h}}\\
\Post_Q=&\bigodot_{h=k'}^{0}\!\!\Post_{3h}\otimes\Post_2\otimes\bigodot_{j=k}^{0}\!\!\Post_{1j}
\end{align*}

\end{small}
\begin{enumerate}

\item For the first $k$ open tau transitions, by Definition \ref{def:Non-ObsTau} $P$ can necessarily make a tau open transition if the hole indexed $j_0$ makes a tau action. So by Lemma \ref{lem-compose} we obtain $k$ open transitions in the form: 
\[P[Q]_{j_0}\!\models		
\openrule
    {
       \set{\beta_{1h}},{\Pred_{1h}},{\Post_{1h}}   }
         {\ostate{s_{1h}\uplus s^{i\in L}_{i} } \OTarrow {\tau} \ostate{s_{1(h+1)}\uplus s^{i\in L}_{i}}}\]

\item For the possibly observable open transition.  By Lemma \ref{lem-compose} with the lemma hypotheses we obtain:\\ 
	\[ P[Q]_{j_0}  
	\models
	{\openrule
		{
			{\beta_{j}}^{(j\in J\setminus\{j_0\})} \uplus \set{\beta_2}, 
			\Pred\land\Pred_2\land \alpha_Q=\beta_{j_0 },  
			\Post\uplus \Post_2}
{\ostate{s^{i\in L}_{i}\uplus s_{20}} \OTarrow {\alpha} \ostate{{s'}^{i\in L}_{i}\uplus s_{21}}}
	}
	\]

%
\item We proceed in the same way as the first item for $k'$ last weak open transitions, and we obtain $k'$ open tau transitions.
\end{enumerate}

Using Lemma \ref{lem-rel-OT-WOT}, from cases $(1)$, $(2)$ and $(3)$ we get: 
\[P[Q]_{j_0}\!\models		
\openrule
    {
\set{\gamma_c}, \Pred_c,
\Post_c}
         {\ostate{s_{10}\uplus s^{i\in L}_{i}} \OTWeakarrow {\alpha'} 
	\ostate{{{s'}^{i\in L}_{i}\uplus s_{3(k'+1)}}}}\] 
where ~~
$\alpha'=\alpha\displaystyle{\subst{\bigodot_{j=k}^{0}\Post_{1j}}}$
and $\alpha=\alpha'$  because $\Post_{1j}$ acts on variables of $Q$ and $\alpha$ contains only variables of $P$.
\begin{small}
\begin{align*}
\set{\gamma_c}=&\mybigdotcup_{h=0}^k \vis{\set{\beta_{1h}}\subst{\bigodot_{i=h-1}^{0}\!\! \Post_{1i} } }  \dotcup  \vis{({\beta_{j}}^{(j\in J\setminus\{j_0\})} \uplus \set{\beta_2})\subst{\bigodot_{i=k}^{0}\Post_{1i}}} \dotcup\\
&\mybigdotcup_{h=0}^{k'} \vis{\set{\beta_{3h}} \subst{\bigodot_{i=h-1}^{0}\!\!\Post_{3i}\shortodot\Post_2\shortodot\bigodot_{i=k}^{0}\Post_{1i}} }\\
=&\set{\gamma}\uplus\vis{\beta_j^{\in J\setminus \{j_0\}}} \quad  \text{ because $\Post_{1j}$  does not act on variables of $\beta_j$.}  
\displaybreak[0]\\
\Pred_c=&
\bigwedge_{h=0}^k{\Pred_{1h}}\subst{\bigodot_{i=h-1}^{0}\Post_{1i}}\land
\left(Pred\land\Pred_2\land \alpha'_Q=\beta_{j_0}\right) \subst{\bigodot_{i=k}^{0}{\Post_{1i}}}\land\\
&\bigwedge_{h=0}^{k'}{\Pred_{3h}}\subst{\bigodot_{i=h-1}^{0}{\Post_{3i}}\shortodot({\Post\uplus \Post_2})\shortodot\bigodot_{i=k}^{0}{\Post_{1i}}}
\\
\Post_c=&(\bigodot_{i=k'}^{0}{\Post_{3i}})\shortodot({\Post\uplus \Post_2})\shortodot\bigodot_{i=k}^{0}{\Post_{1i}}
\end{align*}
\end{small}
\noindent Note that we have $s_i^{i\in L_Q}=s_{10} \wedge s_{1(k+1) i}=s_{20} \wedge  s_{21}= s_{30} \wedge s_{3(k'+1) i}={s'}_i^{i\in L_Q}$.

Note also that $\Post$ only acts on variables of $P$ while $\Post_{1i}$ only acts on  variables of $Q$. We conclude on predicate and posts as follows\footnote{$\Post_{1i}$ only has an effect on variables of $Q$ and thus does not modify $\Pred$ or $\beta_{j_0}$}:
\begin{equation*}
\begin{split}
\Pred_c&=\Pred_Q\land\Pred\subst{\bigodot_{i=k}^{0}{\Post_{1i}}}\land
(\alpha'_Q=\beta_{j_0})\subst{\bigodot_{i=k}^{0}{\Post_{1i}}}\\
&=\Pred_Q\land\Pred\land \alpha_Q=\beta_{j_0}\\[.3ex]
\Post_c&=\Post\uplus\Post_Q \qedhere
\end{split}
\end{equation*}
\qedhere
\smallskip
\begin{lem}[Weak open transition composition]\label{lem-Weakcompose} Suppose that we have a weak open automaton such that the WOTs cannot observe silent actions (see Definition \ref{def:Non-ObsTau}). Suppose $j_0\in J$ and $\gamma_{j_0}=[\beta_0..\beta_n]$ and additionally:\\[-2ex]
\begin{mathpar}
P\models{\openrule
	{
		\gamma_j^{j\in J}, 
		\Pred,  
		\Post}
	{\ostate{s_i^{i\in L}} \OTWeakarrow {\alpha}
		\ostate{s_i'^{\, i\in L}}}
}
\qquad\text{~~and for all $p\in[0..n]$~~}\quad
Q\models{\openrule
	{
		\gamma_{pj}^{j\in J_p},
		 \Pred_p,  
		\Post_p}
	{\ostate{s_{pi}^{i\in L_Q}} \OTWeakarrow {\alpha_p}
		\ostate{s_{(p+1) i}^{\, i\in L_Q}}}
}
\end{mathpar}

Let 
\begin{mathpar}
J_Q=\displaystyle{\bigcup_{p=0}^nJ_{p}}

\forall{i\in L_Q}.\,s_i=s_{0i}~~\land~~s'_i=s_{(n+1) i}

\forall j\in J_p,\,\gamma_j= \displaystyle{\mybigdotcup_{p=0}^n \gamma_{p j}\subst{\bigodot_{k=p}^{0}{\Post_k}}}

\Pred\,'=\Pred\land \bigwedge_{p=0}^{n}(\alpha_p=\beta_p\land \Pred_p)\subst{\bigodot_{i=p-1}^{0}\Post_i} 

\Post\,'=\Post\uplus\bigodot_{p=n}^{0}
		\Post_p
\end{mathpar}
Then, we have:
	\[ P[Q]_{j_0}  
	\models
	{\openrule
		{
			\gamma_j^{j\in (J\setminus\{j_0\}) \uplus J_Q}, 
			\Pred\,',  \Post\,'
			 }
		{\ostate{s_i^{i\in L\uplus L_Q}} \OTWeakarrow {\alpha}
			\ostate{s_i'^{\, i\in L\uplus L_Q}}}
	}
	\]
\end{lem}

\proof Suppose we have:
\begin{mathpar}
P\models{\openrule
	{
		\gamma_j^{j\in J}, 
		\Pred,  
		\Post}
	{\ostate{s_i^{i\in L}} \OTWeakarrow {\alpha}
		\ostate{s_i'^{\, i\in L}}}
}
\end{mathpar}
By Lemma \ref{lem-rel-OT-WOT} this implies the following: \\			
\begin{small}
$\forall p\!\in\![0\,..\,m_1].P\models\openrule
    {{\beta_{1pj}^{j\in J_{1p}}},\Pred_{1p},\Post_{1p}   }
         {\ostate{{(s_{1pi})}^{i\in L}} \OTarrow {\tau} \ostate{{(s_{1(p+1)i})}^{i\in L}}},  \quad
P\models\openrule
         {\beta_{2j}^{j\in J_2},\Pred_2,\Post_2 }
         {\ostate{({s_{20i}})^{i\in L}} \OTarrow {\alpha'} \ostate{({s_{21i}})^{i\in L}}} $ 
\\
and  $\forall p \in [0\,..\,m_2].P\models\openrule
         {
           \beta_{3pj}^{j\in J_{3p}},\Pred_{3p}, \Post_{3p}}
         { \ostate{(s_{3pi})^{i\in L}} \OTarrow {\tau} \ostate{(s_{3(p+1)i})^{i\in L}}}$
\end{small}

 where:
{\small
\begin{align*}
\forall i\in L.&~~s_i=s_{10 i} \wedge s_{1(m_1+1) i}=s_{20i} \wedge  s_{21i}= s_{30 i} \wedge s_{3(m_2+1) i}=s'_i\\
\alpha=&\alpha'{\subst{\bigodot_{j=m_1}^{0}\Post_{1j}}} \\
\gamma_j^{j\in J}=&{\mybigdotcup_{i=0}^{m_1} \vis{{\beta_{1ij}^{j\in J_{1p}}}\subst{\bigodot_{k=i-1}^{0}\!\! \Post_{1k} } }  \dotcup \, \vis{{\beta_{2j}^{j\in J_{2}}}\subst{\bigodot_{k=m_1}^{0}\Post_{1k}}}}\, \dotcup\\
&\mybigdotcup_{i=0}^{m_2} \vis{\beta_{3ij}^{j\in J_{3p}} \subst{\bigodot_{k=i-1}^{0}\Post_{3k}\shortodot\Post_2\shortodot\bigodot_{k=m_1}^{0}\Post_{1k}}}\\
\Pred=&{\bigwedge_{p=0}^{m_1}\!\big(\Pred_{1p}\subst{\!\bigodot_{j=p-1}^{0}\!\!\Post_{1j}}\big)\land\Pred_2\subst{\!\bigodot_{p=m_1}^{0}\!\!\Post_{1p}}\land}
{\bigwedge_{p=0}^{m_2}\!\Pred_{3p}\subst{\!\bigodot_{j=p-1}^{0}\!\!\Post_{3j}\otimes\Post_2 \otimes\!\bigodot_{p=m_1}^{0}\!\!\Post_{1p}}}\\
\Post=&{\bigodot_{p=m_2}^{0}\Post_{3p}\otimes\Post_2\otimes\bigodot_{j=m_1}^{0}\Post_{1j}}
\end{align*}
}

Note that, for $l\in\{1,3\}$ if $\beta_{l p j_0}=\tau$, then, because of Definition \ref{def:Non-ObsTau}, $P$ necessarily makes a $\tau$ open transition and remains in the same state, e.g. $s_{1pi} = s_{1(p+1)i}$. Thus without loss of generality, we can bypass such an open transition and obtain another decomposition of the WOT without the open transition that requires ${\beta_{lp{j_0}}}=\tau$. We can thus  suppose that for all $p$ and $l$ we have ${\beta_{lp{j_0}}}\neq\tau$ or $j_0\not\in J_{1 p}$. To avoid a special case, we suppose that the hole $j_0$ moves during the OT $\alpha'$, i.e. $\beta_{2 j_0}=\beta_{m}$ for some $m$. Additionally, $\beta_{m}\neq \tau$, else we would have $\alpha=\alpha'=\tau$ and the $\alpha'$ OT could be also removed from the reduction, leading to a particular and simpler case.

We introduce $n_i^{i\in[0..m-1]}$, and $(n'_i)^{i\in[m+1..n]}$ the indices of the steps in 
which the hole $j_0$ moves in the 3 sets of OTs above ($\beta_m$ is the action that matches the hole $j_0$ in the OT $\alpha'$), in other words, we have for all $i$, $\beta_{1 n_i j_0}$ a visible action, as additionally:\\
{\small
\begin{align*}
\gamma_{j_0}\!=&~[\beta_0..\beta_n]\\
 =&\!\!
{\mybigdotcup_{\substack{i=0\\j_0\in J_{1i}}}^{m_1}\!\!\! \vis{{\beta_{1ij_0}}\subst{\!\bigodot_{k=i-1}^{0} \!\!\!\!\Post_{1k} } }  \,\dotcup\, \vis{{\beta_{2j_0}}\subst{\!\bigodot_{k=m_1}^{0}\!\!\!\Post_{1k}}}}\dotcup\! 
{ \!\mybigdotcup_{\substack{i=0\\j_0\in J_{3i}}}^{m_2}\!\! \vis{{\beta_{3ij_0}} \subst{\!\bigodot_{k=i-1}^{0}\!\!\!\Post_{3k}\shortodot\Post_2\shortodot
\!\!\bigodot_{k=m_1}^{0}\!\!\!\Post_{1k}}}}
\end{align*}}

\noindent We have, by definition of $n_i$ and $n'_i$:\\
$\displaystyle{\forall{i\in[0\,..\,m\!-\!1]}, \beta_{1 n_i j_0}\subst{\bigodot_{k=n_i-1}^{0}\Post_{1k}}=\beta_i}$, \qquad$\displaystyle{\beta_{2 j_0}\subst{\bigodot_{k=m_1}^{0} \Post_{1k} }=\beta_m}$, and \\
$\displaystyle{\forall{i\in[m\!+\!1\,..\,n]}, \beta_{3 n_i' j_0}\subst{\bigodot_{k=n_i'-1}^{0}\Post_{3k}\shortodot\Post_2\shortodot\bigodot_{k=m_1}^{0}\Post_{1k}}=\beta_i}$\\

\noindent Now, we compose OTs for each of the case above (depending on the OT of $P$):
\begin{enumerate}
\item For the first $\tau$ OTs, i.e. $p\in [0\,..\,m_1]$. We have: \\
Either there is $i$ such that $p=n_i$, and thus $\beta_i$ and $\beta_{1 p j_0}$ are defined. In this case by Lemma~\ref{lem-Weakcompose1},  we have: 
\[ P[Q]_{j_0}  
	\models
	{\openrule
		{
			\set{\gamma'_{1p}}, 
			\Pred\,'_{1p},  \Post\,'_{1p}
			 }
		{\ostate{s_{1pj}^{j\in L}\uplus s_{ij}^{j\in L_Q}} \OTWeakarrow {\tau}
			\ostate{s_{1(p+1)j}^{\, j\in L}\uplus s_{(i+1)j}^{j\in L_Q} } }
	}
	\]
with 
\begin{mathpar}
 \set{\gamma'_{1p}}=\gamma_{i j}^{j\in J_i}\uplus\vis{\beta_{1pj}^{j\in J_{1p}\setminus \{j_0\}}}

\Pred\,'_{1p}=\Pred_{1p}\land (\beta_{1pj_0}=\alpha_i\land \Pred_i)

 \Post\,'_{1p}=\Post_{1p}\uplus\Post_i
\end{mathpar}

Or $j_0\not \in \dom(\beta_{1 p})$ and $Q$ does not move in the composed reduction. In this case there is no $i$ such that $p=n_i$, but there is $i$ such that $p\in]n_i .. n_{i+1}[$, and
\[P[Q]_{j_0} \models\openrule
    {\set{\beta_{1p}},\Pred_{1p},\Post_{1p}   }
         {\ostate{s_{1pj}^{j\in L}\uplus s_{ij}^{j\in L_Q}} \OTarrow {\tau} \ostate{s_{1(p+1)j}^{j\in L}\uplus s_{ij}^{j\in L_Q}}}
\]
and thus we also have a weak OT by Definition \ref{def:buildweakOT} (rule (\WTDeux)):
\[P[Q]_{j_0} \models\openrule
    {\set{\gamma'_{1p}},\Pred\,'_{1p},\Post\,'_{1p}   }
         {\ostate{s_{1pj}^{j\in L}\uplus s_{ij}^{j\in L_Q}} \OTWeakarrow {\tau} \ostate{s_{1(p+1)j}^{j\in L}\uplus s_{ij}^{j\in L_Q}}}
\]
with 
$\set{\gamma'_{1p}}=\set{\vis{\beta_{1p}}}, \Pred\,'_{1p}=\Pred_{1p}, \Post\,'_{1p}=\Post_{1p}$\\

\item Similarly, for the middle OT with label $\alpha$:
\[P[Q]_{j_0}\models\openrule
         {	\set{\gamma'_2},\Pred\,'_2,\Post\,'_2 }
         {\ostate{{(s_{20j})}^{j\in L}\uplus (s_{mj})^{j\in L_Q}}\OTWeakarrow {\alpha'} \ostate{{(s_{21j})}^{j\in L}\uplus (s_{(m+1)j})^{j\in L_Q}}}\]
with
\begin{mathpar}
\set{\gamma'_2}=\gamma_{m j}^{j\in J_m}\uplus\vis{\beta_{2j}^{j\in J_2\setminus \{j_0\}}}

\Pred\,'_{2}=\Pred_{2}\land (\beta_{2j_0}=\alpha_m\land \Pred_m)

\Post\,'_{2}=\Post_{2}\uplus\Post_m
\end{mathpar}

\item For the last $\tau$ OTs, i.e. $p\in [0\,..\,m_2]$. We have similarly to the first case:\\
 Either there is $i$ such that $p=n'_i$, and thus $\beta_i$ and $\beta_{1 p j_0}$ are defined. In this case by Lemma~\ref{lem-Weakcompose1}, we have: 
	\[ P[Q]_{j_0}  
	\models
	{\openrule
		{
			\set{\gamma'_{3p}}, 
			\Pred\,'_{3p} ,  \Post\,'_{3p} 
			 }
		{\ostate{s_{3pj}^{j\in L}\uplus s_{ij}^{j\in L_Q}} \OTWeakarrow {\tau}
			\ostate{s_{3(p+1)j}^{\, j\in L}\uplus s_{(i+1)j}^{j\in L_Q} } }
	}
	\]
with
\begin{mathpar}
\set{\gamma'_{3p}}=\gamma_{i j}^{j\in J_i}\uplus\vis{\beta_{3pj}^{j\in J_{3p}\setminus \{j_0\}}}

\Pred\,'_{3p}=\Pred_{3p}\land (\beta_{3pj_0}=\alpha_i\land \Pred_i)

\Post\,'_{3p} =\Post_{3p}\uplus\Post_i
\end{mathpar}

Or $j_0\not \in \dom(\beta_{3 p})$ and $Q$ does not move in the composed reduction. In this case there is no $i$ such that $p=n'_i$, but there is $i$ such that $p\in]n'_i .. n'_{i+1}[$, and
\[P[Q]_{j_0} \models\openrule
    {\set{\beta_{3p}},\Pred_{3p},\Post_{3p}   }
         {\ostate{s_{3pj}^{j\in L}\uplus s_{ij}^{j\in L_Q}} \OTarrow {\tau} \ostate{s_{3(p+1)j}^{j\in L}\uplus s_{ij}^{j\in L_Q}}}
\]
and thus we also have a weak OT by definition \ref{def:buildweakOT} (rule \WTDeux):
\[P[Q]_{j_0} \models\openrule
    {\set{\gamma'_{3p}},\Pred\,'_{3p},\Post\,'_{3p}   }
         {\ostate{s_{3pj}^{j\in L}\uplus s_{ij}^{j\in L_Q}} \OTWeakarrow {\tau} \ostate{s_{3(p+1)j}^{j\in L}\uplus s_{ij}^{j\in L_Q}}}
\]
with 
$\set{\gamma'_{3p}}=\set{\beta_{3p}}, \Pred\,'_{3p} =\Pred_{3p}, \Post\,'_{3p} =\Post_{3p}$
\end{enumerate}

\noindent By definition of weak open transition (Definition~\ref{def:buildweakOT}, rule \WTTrois),
 we obtain:

	\[ P[Q]_{j_0}  
	\models
	{\openrule
		{
			\set{\gamma'}, 
			\Pred\,'',  \Post\,''
			 }
		{\ostate{s_{10j}^{j\in L}\uplus s_{0j}^{j\in L_Q}} \OTWeakarrow {\alpha''}
			\ostate{s_{3(m_2+1)j}^{\, j\in L}\uplus s_{(n+1)j}^{j\in L_Q} } }
	}
	\]
where
{\small 
\begin{align*}
\alpha''=&\alpha'{\subst{\bigodot_{j=m_1}^{0}\Post'_{1j}}}\\
\set{\gamma'}=&
 \mybigdotcup_{i=0}^{m_1}\set{\gamma'_{1i}}\subst{\bigodot_{k=i-1}^{0}\Post\,'_{1k}}\,\dotcup\,
\set{\gamma'_{2}}\subst{\bigodot_{k=m_1}^{0}\Post\,'_{1k}}\,\dotcup
 \mybigdotcup_{i=0}^{m_2}\set{\gamma'_{3i}}\subst{\bigodot_{k=i-1}^{0}\Post\,'_{3k}\shortodot\Post\,'_2\shortodot\bigodot_{k=n}^{0}\Post\,'_{1k}}
\displaybreak[0]\\
\Pred\,''=&\bigwedge_{i=0}^{m_1}\Pred\,'_{1i}\subst{\bigodot_{j=i-1}^{0}\Post\,'_{1j}}\land\Pred\,'_2 \subst{\bigodot_{j=m_1}^{0}\Post\,'_{1j}}\land\\ 
&\bigwedge_{i=0}^{m_2}\Pred\,'_{3i}\subst{\bigodot_{j=i-1}^{0}\Post\,'_{3j}\shortodot\Post\,'_2\shortodot\bigodot_{j=m_2}^{0}\Post\,'_{1j}}\displaybreak[0]\\
\Post\,''=&\bigodot_{j=m_2}^{0}\Post\,'_{3j}\shortodot\Post\,'_2\shortodot\bigodot_{j=m_1}^{0}\Post\,'_{1j}
\end{align*}
}

However it must be noticed that in steps 1 and 3, we have two kinds of WOTs with different signatures (depending on whether $Q$ moves or not). It is still possible to glue them together in a global rule with two more terms for $\Pred$ and $\Post$ terms. This global merge is possible because the post-conditions of $P$ only act on variables of $P$ and those of $Q$ on variables of $Q$ (for example $\Post_i$ has no effect on $\Pred_{1p}$ and thus does not need to be taken into account when dealing with WOTs where $Q$ does not move).

We now  compare each element of the obtained WOT with the conclusion of the lemma:
{\small \begin{align*}
\alpha''&=\alpha'\displaystyle{\subst{\bigodot_{j=m_1}^{0}\Post\,'_{1j}}}\\
& = \alpha'\displaystyle{\subst{\bigodot_{j=m_1}^{0}\Post_{1j}}}&\text{$\alpha'$ only contains variables of $P$ untouched by $\Post_i$}\\
& = \alpha
\end{align*}}

For $\set{\gamma'}$ we distinguish elements in the holes of $P$ and of $Q$.

 First suppose $j\in J\setminus\{j_0\}$ we have $\gamma'_j=\gamma_j$ because $\Post^{\,\prime}_{ij}$ has no effect on variables of $P$ and on $\beta_{1pj}$, consequently we have:\\
{\small \begin{equation*}
\begin{split}\gamma'_j=&\mybigdotcup_{i=0}^{m_1} \vis{{\beta_{1ij}}\subst{\bigodot_{k=i-1}^{0}\! \Post_{1k} } }  \dotcup  \vis{{\beta_{2j}}\subst{\bigodot_{k=m_1}^{0}\!\Post_{1k}}}\, \dotcup
\mybigdotcup_{i=0}^{m_2} \vis{\beta_{3ij} \subst{\!\bigodot_{k=i-1}^{0}\!\!\!\Post_{3k}\shortodot\Post_2\shortodot\bigodot_{k=m_1}^{0}\!\!\Post_{1k}}}
\end{split}
\end{equation*}
}

Second, when $j\in J_t$ for some $t$,  $\gamma'_j$ is the concatenation of elements of $\gamma'_{1ij}$, $\gamma'_{2 j}$, $\gamma'_{3ij}$  that are not empty. By construction the concatenation of these elements is $\gamma_{tj}$, for $t\in[0..n]$. $\Post_{ik}$ has no effect on $\gamma_{tj}$ but  $\Post_{k}$ has.
We obtain:
{\small\begin{equation*}
\begin{split}
\gamma'_j =&
\mybigdotcup_{i=0}^{m_1}{\gamma'_{1ij}}\subst{\bigodot_{k=i-1}^{0}\Post\,'_{1k}}\dotcup
 {\gamma'_{2 j}}\subst{\bigodot_{k=m_1}^{0}\Post\,'_{1k}}\,\dotcup\, \mybigdotcup_{i=0}^{m_2}{\gamma'_{3ij}}\subst{\bigodot_{k=i-1}^{0}\Post\,'_{3k}\shortodot\Post\,'_2\shortodot\bigodot_{k=n}^{0}\Post\,'_{1k}}\\
=&
{\mybigdotcup_{t=0}^{n} {\gamma_{tj}}\subst{\bigodot_{k=t-1}^{0} \Post_{k}}}   
\end{split}
\end{equation*}
}

Concerning predicates, we also separate predicates on $P$ from predicates on $Q$, and from the equality on the action filling the hole:
\begin{footnotesize}
\begin{align*}
\Pred\,''&=\Big(\bigwedge_{i=0}^{m_1}\Pred\,'_{1i}\subst{\bigodot_{j=i-1}^{0}\!\!\Post\,'_{1j}}\land\Pred\,'_2 \subst{\bigodot_{j=m_1}^{0}\!\!\Post\,'_{1j}}\land
\bigwedge_{i=0}^{m_3}\Pred\,'_{3i}\subst{\bigodot_{j=i-1}^{0}\!\!\Post\,'_{3j}\shortodot\Post\,'_2\shortodot\bigodot_{j=m_1}^{0}\!\Post\,'_{1j}}\Big)\\
&=\Big(\bigwedge_{p=0}^{m_1}\!\Pred_{1p}\subst{\!\!\bigodot_{j=p-1}^{0}\!\!\Post_{1j}}\land\Pred_2\subst{\!\!\bigodot_{p=m_1}^{0}\!\!\Post_{1p}}\land \bigwedge_{p=0}^{m_2}\!\Pred_{3p}\subst{\!\!\bigodot_{j=p-1}^{0}\!\!\Post_{3j}\otimes\Post_2 \otimes\!\!\bigodot_{p=m_1}^{0}\!\!\Post_{1p}}\Big)\\&\qquad
\land \bigwedge_{t=0}^{n} \Pred_t\bigodot_{i=t-1}^{0}\!\!\Post_i
\land \Big( \bigwedge_{i=0}^{m-1}(\beta_{1{n_i}{j_0}}=\alpha_i)\subst{\bigodot_{j={n_i}-1}^{0}\Post\,'_{1j}}\land(\beta_{2{j_0}}=\alpha_m) \subst{\bigodot_{j=m_1}^{0}\!\!\Post\,'_{1j}}\land\\ &\qquad
~\quad\bigwedge_{i=m+1}^{n}(\beta_{3{n'_i}{j_0}}=\alpha_i)\subst{\bigodot_{j=n'_i-1}^{0}\Post\,'_{3j}\shortodot\Post\,'_2\shortodot\bigodot_{j=m_1}^{0}\!\!\Post\,'_{1j}}\Big)\displaybreak[0]\\
&=\Big(\bigwedge_{p=0}^{m_1}\!\Pred_{1p}\subst{\!\!\bigodot_{j=p-1}^{0}\!\!\Post_{1j}}\big)\land\Pred_2\subst{\!\!\bigodot_{p=m_1}^{0}\!\!\Post_{1p}}\land \bigwedge_{p=0}^{m_2}\!\Pred_{3p}\subst{\!\!\bigodot_{j=p-1}^{0}\!\!\Post_{3j}\otimes\Post_2 \otimes\!\!\bigodot_{p=m_1}^{0}\!\!\Post_{1p}}\Big)\\&\qquad 
\land \bigwedge_{t=0}^{n} \Pred_t\bigodot_{i=t-1}^{0}\!\!\Post_i
\land \Big(\bigwedge_{i=0}^{m-1}(\beta_{i}=\alpha_i)\subst{\bigodot_{j={i-1}}^{0}\!\!\Post_{j}}\land(\beta_m=\alpha_m )\subst{\bigodot_{j=m}^{0}\Post_{j}}\land\\ &\qquad
~\quad\bigwedge_{i=m+1}^{n}(\beta_i=\alpha_i)\subst{\bigodot_{j=i-1}^{m}\!\!\Post_j\shortodot\Post_m\shortodot\bigodot_{j=m-1}^{0}\!\!\Post_{j}}\Big)\\
&=\Pred
\end{align*}
\end{footnotesize}

Finally, concerning post-conditions:
{\footnotesize \begin{equation*}
\begin{split}
\Post\,''&=\bigodot_{j=m_2}^{0}\Post\,'_{3j}\shortodot\Post\,'_2\shortodot\bigodot_{j=m_1}^{0}\Post\,'_{1j}\\
&=\left(\bigodot_{j=m_2}^{0}\Post\,'_{3j}\shortodot\Post\,'_2\shortodot\bigodot_{j=m_1}^{0}\Post\,'_{1j}\right)\uplus \bigodot_{j=n}^{0}\Post_j\\
&=\Post \uplus \bigodot_{j=n}^{0}\Post_j\
\end{split}
\end{equation*}
}
This allows us to conclude concerning the lemma. 
\qed

\noindent
{\bf Theorem~\ref{weak-thm-congr-eq}}. \emph{Congruence}.
	Consider an open pNet:
	$\pNet = \mylangle \pNet_i^{i\in I}, \Sort_j^{j\in J}, 
	\set{\symb{SV}}\myrangle$.
	Let $j_0\in J$ be a hole. Let $\pNetQ$ and $\pNetQ'$ be two weak FH-bisimilar pNets such that 
	$\Sortop(\pNetQ)=\Sortop(\pNetQ')=\Sort_{j_0}$. Then 
	$\pNet[\pNetQ]_{j_0}$ and 
	$\pNet[\pNetQ']_{j_0}$ are weak FH-bisimilar.
\proof  Consider $Q$ weak FH-bisimilar to $Q'$.  It means that there exists an FH-bisimulation $\mathcal{R}_{Q,Q'}$ relating the two pNets $Q$ and $Q'$. We define a relation $\mathcal{R}$ relating states of $P[Q]_{j_0}$ with states of $P[Q']_{j_0}$: 
\[\mathcal{R} = \{(\ostate{S_P\uplus S_Q},\ostate{S_P\uplus S_{Q'}}, \Pred_{Q,Q'})|\,(S_Q,S_{Q'}, \Pred_{Q,Q'})\in\mathcal{R}_{Q,Q'}\}\]

To prove weak FH-bisimulation of $\pNet[\pNetQ]_{j_0}$ and 
	$\pNet[\pNetQ']_{j_0}$, we consider  an open transition $OT$ of $\pNet[\pNetQ]_{j_0}$, and an equivalent state of $\pNet[\pNetQ']_{j_0}$, and we try to find a family of WOT of 	$\pNet[\pNetQ']_{j_0}$ that simulates $OT$.
Consider an OT of  $\pNet[\pNetQ]_{j_0}$ it is of the form (notations introduced to prepare the decomposition):
\[
\pNet[\pNetQ]_{j_0}\models\openrule
	{
		\beta_j^{j\in( J_P\uplus J_Q)}, 
		\Pred_P\land \Pred_Q,  
		\Post_P\uplus \Post_Q}
	{\ostate{S_P\uplus S_Q} \OTarrow {\alpha}
		\ostate{S'_P\uplus S'_Q}
}
\]

By the decomposition lemma for OTs (Lemma~\ref{lem-decompose}), we obtain the 2 following OTs (equality side-conditions have been unlined for clarity):
\begin{mathpar}
\pNet\models{\openrule
	{
		\beta_j^{j\in J_P}\uplus(j_0\mapsto \alpha_Q), 
		\Pred_P,  
		\Post_Q}
	{\ostate{S_P} \OTarrow {\alpha}
		\ostate{S'_P}}
}
\quad\text{~~and~~}\quad
Q\models{\openrule
	{
		\beta_j^{j\in J_Q},
		 \Pred_Q,  
		\Post_Q}
	{\ostate{S_Q} \OTarrow {\alpha_Q}
		\ostate{S'_Q}}
}

\end{mathpar}

By definition of $\mathcal{R}$ we have 
$(S_Q,S_{Q'}| \Pred_{Q,Q'})\in\mathcal{R}_{Q,Q'}$. And thus, by definition of weak FH-bisimulation, there exists a family of weak open transitions $WOT_{x}$:
 \begin{mathpar}
    \openrule
         {
           \gamma_{j x}^{j\in J_Q}, \Pred_{Q' x},\Post_{Q' x}}
         {\ostate{S_{Q'}}\OTWeakarrow {\alpha_{x}} \ostate{S'_{Q' x}}}
\end{mathpar}

where  
\[\forall x.\, (S'_Q,S'_{Q' x}| \Pred_{Q,Q' x})\in\mathcal{R}_{Q,Q'}\]

and
{\small
\begin{multline*}
\Pred_{Q,Q'}\land \Pred_Q \implies \\
\Big(\bigvee_{x\in X}
(\forall j\in J_Q.\, \vis{\beta_j}=\gamma_{j x})\Rightarrow (\Pred_{Q_x}\land
\alpha_Q=\alpha_x \land \Pred_{Q,Q' x}\subst{\Post_{Q' x}\uplus \Post_{Q}}) \Big)
\end{multline*}
} 

Composing the OT of $P$ with the WOTs of $Q'$ by Lemma~\ref{lem-Weakcompose1} we obtain:
\[
\pNet[\pNetQ']_{j_0}\models\openrule
	{
		\vis{\beta_j^{j\in J_P}}\uplus\gamma_{j x}^{j\in J_Q}, 
		\Pred_P\land \Pred_{Q' x},  
		\Post_P\uplus \Post_{Q' x}}
	{\ostate{S_P\uplus S_{Q'}} \OTWeakarrow {\alpha}
		\ostate{S'_P\uplus S'_{Q' x}}
}
\]
with $\displaystyle{\bigvee_{x\in X}
\big(\forall j\in J_Q.\, \vis{\beta_j}=\gamma_{j x} \implies \alpha_Q=\alpha_x\big)}$ that ensures that the open transitions can be recomposed when the OT fires.

Side conditions necessary to prove weak-FH bisimulations are:
\[\forall x.\, (S'_P\uplus S'_Q,S'_P\uplus S'_{Q' x}| \Pred_{Q,Q' x})\in\mathcal{R}\]
which is true, and
\begin{multline*}
\Pred_{Q,Q'}\land \Pred_P\land \Pred_Q \implies \\
\Big(\bigvee_{x\in X}
(\forall j\in J_Q.\, \vis{\beta_j}=\gamma_{j x}\land \forall j\in J_P.\, \vis{\beta_j}=\vis{\beta_j}))\Rightarrow \\(\Pred_P\land \Pred_{Q' x}\land \alpha=\alpha \land \Pred_{Q,Q' x}\subst{\Post_P\uplus\Post_{Q' x}\uplus \Post_{Q}}) \Big)
\end{multline*}
We conclude by observing that $\Post_P$ has no effect on variables of $Q$ and $Q'$, and thus on $\Pred_{Q,Q' x}$.
\qed

\noindent {\bf Theorem~\ref{weak-thm-ctxt-eq}}. \emph{Context equivalence}.
	Consider two FH-bisimilar open pNets:\\
	$\pNet = \mylangle \pNet_i^{i\in I}, \Sort_j^{j\in J}, 
	\set{\symb{SV}}\myrangle$ and 	$\pNet' = \mylangle {\pNet'}_i^{i\in I}, 
	\Sort_j^{j\in 
	J}, 	\set{\symb{SV'}}\myrangle$ 
	(recall they must have the same holes to be bisimilar).
	Let $j_0\in J$ be a hole, and $Q$ be a pNet such that $\Sortop(Q)=\Sort_{j_0}$. Then 
	$\pNet[Q]_{j_0}$ and 
	$\pNet'[Q]_{j_0}$ are FH-bisimilar.

\proof  Consider $P$ weak FH-bisimilar to $P'$.  There exists an FH-bisimulation $\mathcal{R}_{P,P'}$ relating $P$ and $P'$. We define a relation $\mathcal{R}$ relating states of $P[Q]_{j_0}$ with states of $P'[Q]_{j_0}$: 
\[\mathcal{R} = \{(\ostate{S_P\uplus S_Q},\ostate{S_{P'}\uplus S_Q}, \Pred_{P,P'})|\,(S_P,S_{P'}, \Pred_{P,P'})\in\mathcal{R}_{P,P'}\}\]

To prove weak FH-bisimulation of $\pNet[\pNetQ]_{j_0}$ and 
	$\pNet'[\pNetQ]_{j_0}$, we consider  an open transition $OT$ of $\pNet[\pNetQ]_{j_0}$, and an equivalent state of $\pNet'[\pNetQ]_{j_0}$, and we try to find a family of WOT of 	$\pNet'[\pNetQ]_{j_0}$ that simulates $OT$.
Consider an OT of  $\pNet[\pNetQ]_{j_0}$ it is of the form (notations introduced to prepare the decomposition):
\[
\pNet[\pNetQ]_{j_0}\models\openrule
	{
		\beta_j^{j\in( J_P\uplus J_Q)}, 
		\Pred_P\land \Pred_Q \land \Pred,  
		\Post_P\uplus \Post_Q}
	{\ostate{S_P\uplus S_Q} \OTarrow {\alpha}
		\ostate{S'_P\uplus S'_Q}
}
\]

By the decomposition lemma for OTs (Lemma~\ref{lem-decompose}), we obtain the 2 following OTs (equality side-conditions have been unlined for clarity):
\begin{mathpar}
\pNet\models{\openrule
	{
		\beta_j^{j\in J_P}\uplus(j_0\mapsto \alpha_Q), 
		\Pred_P,  
		\Post_Q}
	{\ostate{S_P} \OTarrow {\alpha}
		\ostate{S'_P}}
}
\quad\text{~~and~~}\quad
Q\models{\openrule
	{
		\beta_j^{j\in J_Q},
		 \Pred_Q,  
		\Post_Q}
	{\ostate{S_Q} \OTarrow {\alpha_Q}
		\ostate{S'_Q}}
}

\end{mathpar}

With $\Pred \iff \alpha_Q=\beta_{j_0}$

By definition of $\mathcal{R}$ we have 
$(S_P,S_{P'}, \Pred_{P,P'})\in\mathcal{R}_{P,P'}$. And thus, by definition of weak FH-bisimulation, there exists a family of weak open transitions $WOT_{x}$:
 \begin{mathpar}
    \openrule
         {
           \gamma_{j x}^{j\in J_P\uplus\{j_0\}}, \Pred_{P' x},\Post_{P' x}}
         {\ostate{S_{P'}}\OTWeakarrow {\alpha_{x}} \ostate{S'_{P' x}}}
\end{mathpar}

where  
\[\forall x.\, (S'_P,S'_{P' x}, \Pred_{P,P' x})\in\mathcal{R}_{P,P'}\]
and
\begin{equation*}
\begin{split}
\Pred_{P,P'}\land \Pred_P \implies &
\Big(\bigvee_{x\in X}
(\forall j\in J_P.\, \vis{\beta_j}=\gamma_{j x} \land \vis{\alpha_Q}=\gamma_{j_0})\Rightarrow\\ & \quad (\Pred_{P'_x}\land \alpha=\alpha_x \land \Pred_{P,P' x}\subst{\Post_{P' x}\uplus \Post_{P}}) \Big)
\end{split}
\end{equation*}

We here need a special case of Lemma~\ref{lem-Weakcompose} where the inner pNet $Q$ does a simple OT. This is just a particular case of the theorem but where notations get simplified because the inner pNet does a single transition.
 This way we can compose the WOTs of $P'$ with the OT of $Q$ and obtain, with $\gamma_{j_0}=[\beta] $:
\[
\pNet'[\pNetQ]_{j_0}\models\openrule
	{
		\vis{\beta_j^{j\in J_Q}}\uplus\gamma_{j x}^{j\in J_P}, 
		\Pred_{P' x}\land \Pred_Q \land \alpha_Q=\beta,  
		\Post_{P' x}\uplus \Post_Q}
	{\ostate{S_{P'}\uplus S_{Q}} \OTWeakarrow {\alpha_x}
		\ostate{{S'_{P' x}}\uplus S'_Q}
}
\]
Side conditions necessary to prove weak-FH bisimulations are:
\[\forall x.\, (S'_P\uplus S'_Q,S'_{P' x}\uplus S'_Q, \Pred_{P,P' x})\in\mathcal{R}\]
which is true, and
\begin{multline*}
\Pred_{P,P'}\land \Pred_P\land \Pred_Q \Pred \implies \\
\Big(\bigvee_{x\in X}
(\forall j\in J_P.\, \vis{\beta_j}=\gamma_{j x} \land \forall j\in J_Q.\, \vis{\beta_j}=\vis{\beta_j} )\Rightarrow\\
 (\Pred_{P' x}\land \Pred_Q\land \alpha_Q=\beta\land  \alpha_x=\alpha \land \Pred_{P,P' x}\subst{\Post_{P' x}\uplus\Post_P\uplus \Post_{Q}}) \Big)
\end{multline*}
We conclude by observing that $\Post_Q$ has no effect on variables of $P$ and $P'$, and thus on $\Pred_{P,P' x}$ and $\Pred$ leading to the conclusion about $\alpha_Q=\beta$.
\qed 

\newpage

\section{Full details of the Simple Protocol Example}

\label{Appendix:FullExample}




The first piece of code is the textual definition of the $\symb{SimpleProtocolSpec}$
pNet, that was drawn in Figure \ref{SimpleProt:Impl}, page \pageref{SimpleProt:Impl}. This code should be intuitive enough to read, with the following
language conventions, that brings some user-friendly features, mapped by the editor into pure pNet constructs.

\begin{itemize}
  \item Constants of any type (including Action) must be declared as
    ``const''. They are used either as functions with argument, as
    typically \texttt{in(msg)}, or constants without argument, typically as "tau()".
  \item Variables can be declared as global variables of a pLTS (e.g. \texttt{m\_msg} in \texttt{PerfectBuffer}), or a pNet Node in the case of synchronisation vector variables (e.g. \texttt{p\_a}),
    or as input variables in a pLTS, as \texttt{?msg} in \texttt{PerfectBuffer}.
  \item The variables in the guards of synchronisation vectors (e.g. in SV1) do not need to be explicitly quantified: by convention, all variables in a guard that do not appear inside the vector actions will be recognised as bound by a \emph{forall} quantifier inside the guard.
    \item The tools will check that everything is correctly declared, that variables are used properly and do not conflict between different objects, that vectors have coherent length, etc.
\end{itemize}
\bigskip

\begin{lstlisting}[basicstyle=\scriptsize\ttfamily, language=java, frame=single]
SimpleProtocolSpec:
import "Data_Alg.algp"
root SimpleProtocolSpec
const in, out:Action
const p_send, q_recv: Action
const tau:Action

pLTS PerfectBuffer
initial b0 
vars ?m:Data
vars b_msg:Data b_ec:Nat

state b0
transition in(m) -> b1 {b_msg:=m, b_ec:=0}

state b1
transition out(b_msg, b_ec) -> a0
transition synchro(tau())  -> b1 {b_ec:=b_ec+1}

pNet SimpleProtocolSpec
holes P,Q
subnets P,PerfectBuffer,Q
vars p_a,q_b:Action m:Data ec:Nat

vector SV0 <p_send(m),in(m),_>->synchro(in(m))
vector SV1 <p_a,_,_>->p_a [p_a != p_send(x)]
vector SV2 <_,out(m,ec),q_recv(m,ec)>->synchro(out(m,ec))
vector SV3 <_,_,q_b>->q_a [q_b != q_recv(x,y)]

  \end{lstlisting}

\bigskip
The corresponding generated Open Automaton was given in Figure \ref{SimpleProtCounter:SpecOA},
page \pageref{SimpleProtCounter:SpecOA}.\\

Next is the code for the \symb{SimpleProtocolImpl} pNet:
\medskip
\begin{lstlisting}[basicstyle=\scriptsize\ttfamily, language=java, frame=single]
SimpleProtocolImpl:
  import "Data_Alg.algp"
  root SimpleProtocolImpl
const in,out:Action
const tau,p_send,q_recv,m_recv,m_send,m_error: Action
const s_recv,s_send,s_ack,s_error,r_recv,r_ack,r_send: Action

pLTS Sender
  initial s0
  vars ?m:Data
  vars  s_msg:Data s_ec:Nat
state s0
  transition s_recv(m) -> s1 {s_msg:=m, s_ec:=0}
state s1
  transition s_send(s_msg, s_ec) -> s2 
state s2
  transition s_ack() -> s0
  transition s_error() -> s1 {s_ec:=s_ec+1}

pLTS Medium
  initial m0
  vars ?m:Data ?ec:Nat
  vars m_msg:Data m_ec:Nat
state m0
  transition m_recv(m,ec) -> m1 {m_msg:=m, m_ec:=ec}
state m1
  transition m_send(m_msg, m_ec) -> m0 
  transition synchro(tau()) -> m2
state m2
  transition m_error() -> m0

pLTS Receiver
  initial r0
  vars ?m:Data ?ec:Nat
  vars r_msg:Data r_ec:Nat
state r0
  transition r_recv(m,ec) -> r1 {r_msg:=msg, r_ec:=ec}
state r1
  transition r_send(r_msg, r_ec) -> r2
state r2
  transition r_ack() -> r0

pNet SimpleProtocol
  subnets Sender,Medium,Receiver
  vars m:Data c:Nat
vector SV0 <s_recv(m),_, _>->in(m)
vector SV1 <s_send(m,ec),m_recv(m,ec),_>->synchro(tau())
vector SV2 <_,m_send(m,ec),r_recv(m,ec)>->synchro(tau())
vector SV3 <s_ack(),_,r_ack()>->synchro(tau())
vector SV4 <s_error(),m_error(),_>->synchro(tau())
vector SV5 <_,_,r_send(m,ec)>->out(m,ec)

pNet SimpleProtocolImpl
  holes P,Q
  subnets P,SimpleProtocol,Q
  vars p_a,q_a:Action  m:Data c:Nat
vector SV0 <p_send(m),in(m),_>->synchro(in(m))
vector SV1 <p_a,_,_>->p_a [p_a != p_send(x)]
vector SV2 <_,out(m,ec),q_recv(m,ec)>->synchro(out(m,ec))
vector SV3 <_,_,q_b>->q_b [q_b != q_recv(x,y)]
\end{lstlisting}

\begin{figure}[h]
   \centerline{\includegraphics[width=10cm]{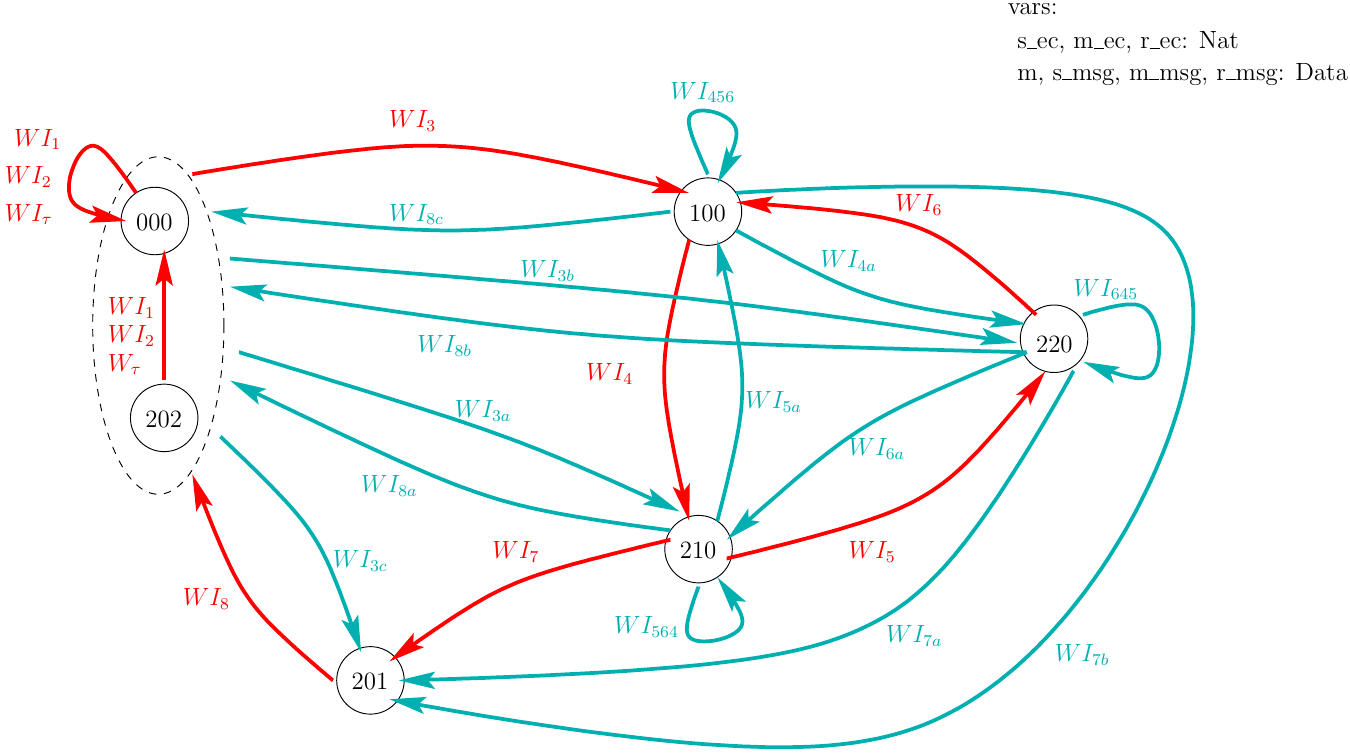}}
   \caption{Weak Open Automaton for \symb{SimpleProtocolImpl}}
   \label{Appendix:ImplOA2}
 \end{figure}

In Figure \ref{Appendix:ImplOA2} we recall the weak open automaton of \symb{SimpleProtocolImpl}. 
This drawing is based on the observation that states 202 and 000 are only linked by a "pure $\tau$" transition, and have exactly the same possible behaviours.
In this configuration we can guarantee that they are weak bisimilar, and we have merged their (incoming and outgoing) transitions in the figure. We denote this 
equivalence class of states as $\{000,202\}$.


Full details of the weak transitions is listed here:

In the first 3 weak transitions, $S$ denotes the set of all global states.

$ W_\tau = \openrule
{\{\}, True, ()}
{S \OTWeakarrow {\tau} S}$

$ WI_1 = \openrule
{\{\texttt{P}\mapsto \texttt{p-a}\}, [\forall \texttt{x}. \texttt{p-a} \neq \texttt{p-send(x)} ], ()}
{S \OTWeakarrow {\texttt{p-a}} S}$

$ WI_2 = \openrule
{\{\texttt{Q}\mapsto \texttt{q-b}\}, [\forall \texttt{x,y}. \texttt{q-b} \neq \texttt{q-recv(x,y)} ], ()}
{S \OTWeakarrow {\texttt{q-b}} S}$

All the following transitions are parameterised by an integer $n\in \texttt{Nat}$, meaning they stand for the corresponding (infinite) set of weak OTs. 
In some cases, this set is further restricted (see e.g. $WI_{7b}(n)$), in which cases we have added an explicit quantifier.
\smallskip

$ WI_3(n) = \openrule
  {\{\texttt{P}\mapsto \texttt{p-send(m)}\}, True,
    (\texttt{s\_msg}\gets \texttt{m}, \texttt{s\_ec}\gets n)}
  { \{000,202\} \OTWeakarrow {\nounderline{\texttt{in(m)}}} 100}
$

$ WI_{3a}(n) = \openrule
  {\{\texttt{P}\mapsto \texttt{p-send(m)}\}, True,
    (\texttt{m\_msg}\gets \texttt{m}, \texttt{m\_ec}\gets n, \texttt{s\_ec}\gets n)}
  { \{000,202\} \OTWeakarrow {\nounderline{\texttt{in(m)}}} 210}
$

$ WI_{3b}(n) = \openrule
  {\{\texttt{P}\mapsto \texttt{p-send(m)}\}, True,
    (\texttt{s\_ec}\gets n)}
  { \{000,202\} \OTWeakarrow {\nounderline{\texttt{in(m)}}} 220}
$

$ WI_{3c}(n) = \openrule
  {\{\texttt{P}\mapsto \texttt{p-send(m)}\}, True,
    (\texttt{r\_msg}\gets \texttt{m}, \texttt{r\_ec}\gets n)}
  { \{000,202\} \OTWeakarrow {\nounderline{\texttt{in(m)}}} 201}
$

$ WI_4(n) = \openrule
         {\{\}, True, 
   (\texttt{m\_msg}\gets \texttt{s\_msg}, \texttt{m\_ec}\gets \texttt{s\_ec}+n, \texttt{s\_ec}\gets \texttt{s\_ec}+n)}
         {100 \OTWeakarrow {\tau} 210}
$

$ WI_{4a}(n) = \openrule
         {\{\}, True, 
    (\texttt{s\_ec}\gets \texttt{s\_ec}+n)}
         {100 \OTWeakarrow {\tau} 220}
$

$ WI_5(n) = \openrule
         {\{\}, True, (\texttt{s\_ec}\gets \texttt{s\_ec}+n)}
         {210 \OTWeakarrow {\tau} 220}
         $

$ WI_{5a}(n) = \openrule
         {\{\}, True, (\texttt{s\_ec}\gets \texttt{s\_ec}+1+n)}
         {210 \OTWeakarrow {\tau} 100}
         $

$ WI_6(n) = \openrule
         {\{\}, True, (\texttt{s\_ec}\gets \texttt{s\_ec}+1+n)}
         {220 \OTWeakarrow {\tau} 100}
         $

$ WI_{6a}(n) = \openrule
         {\{\}, True, 
         (\texttt{m\_msg}\gets \texttt{s\_msg}, \texttt{m\_ec}\gets \texttt{s\_ec}+1+n, \texttt{s\_ec}\gets \texttt{s\_ec}+1+n)}
         {220 \OTWeakarrow {\tau} 210}
         $

Because 
\begin{align*}
Post_{6a}&=\ post_{4}\shortodot\ post_{456}^*\shortodot\ post_{6}\\
&= \left( (\texttt{m\_msg}\gets \texttt{s\_msg}, \texttt{m\_ec}\gets \texttt{s\_ec})
\shortodot (\texttt{s\_ec}\gets \texttt{s\_ec}+n) \right)
\shortodot (\texttt{s\_ec}\gets \texttt{s\_ec}+1) \\
&= (\texttt{m\_msg}\gets \texttt{s\_msg}, \texttt{m\_ec}\gets (\texttt{s\_ec}+1)+n, \texttt{s\_ec}\gets (\texttt{s\_ec}+1)+n)
\end{align*}

$ WI_{456*}(n) = \openrule
         {\{\}, True, 
    (\texttt{s\_ec}\gets \texttt{s\_ec}+n)}
  {100 \OTWeakarrow {\tau} 100}
        $

$ WI_{564*}(n) = \openrule
         {\{\}, True, 
    (\texttt{m\_msg}\gets \texttt{s\_msg}, \texttt{s\_ec}\gets \texttt{s\_ec}+1+n, \texttt{m\_ec}\gets \texttt{s\_ec}+1+n)}
  {210 \OTWeakarrow {\tau} 210}
        $

$ WI_{645*}(n) = \openrule
         {\{\}, True, 
    (\texttt{s\_ec}\gets \texttt{s\_ec}+1+n)}
  {220 \OTWeakarrow {\tau} 220}
        $

\medskip
$ WI_7(n) = \openrule
         {\{\}, True, (\texttt{r\_msg}\gets \texttt{s\_msg}, \texttt{r\_ec}\gets \texttt{s\_ec}+n)}
         {210 \OTWeakarrow {\tau} 201}
         $
         
$ WI_{7a}(n) = \openrule
         {\{\}, True, (\texttt{r\_msg}\gets \texttt{s\_msg}, \texttt{r\_ec}\gets \texttt{m\_ec}+n)}
         {220 \OTWeakarrow {\tau} 201}
         $
         
$ \forall n\ge 1. WI_{7b}(n) = \openrule
         {\{\}, True, (\texttt{r\_msg}\gets \texttt{m\_msg}, \texttt{r\_ec}\gets \texttt{s\_ec}+n)}
         {100 \OTWeakarrow {\tau} 201}
         $
                 
$ WI_8 = \openrule
         {\{\texttt{Q}\mapsto \texttt{q-recv(r1-msg,r1-ec)}\}, True, ()}
         {201 \OTWeakarrow {\nounderline{\texttt{out(r1-msg,r1-ec)}}} \{202, 000\}}
         $

$ \forall n\ge 1. WI_{8a}(n) = \openrule
         {\{\texttt{Q}\mapsto \texttt{q-recv(m\_msg,m\_ec}+n)\}, True, ()}
         {210 \OTWeakarrow {\nounderline{\texttt{out(m\_msg,m\_ec}+n)}} \{202, 000\}}
         $

 $ \forall n\ge 1.  WI_{8b}(n) = \openrule
         {\{\texttt{Q}\mapsto \texttt{q-recv(??\_msg,s\_ec}+n)\}, True, ()}
         {220 \OTWeakarrow {\nounderline{\texttt{out(??\_msg,m\_ec}+n)}} \{202, 000\}}
         $

 $ \forall n\ge 1.  WI_{8c}(n) = \openrule
         {\{\texttt{Q}\mapsto \texttt{q-recv(s\_msg,s\_ec}+n)\}, True, ()}
         {100 \OTWeakarrow {\nounderline{\texttt{out(s\_msg,s\_ec}+n)}} \{202, 000\}}
         $

\medskip
Then for all $\tau$ transitions above we have a similar WOT that include a non-$\tau$ move from an external action of $P$ or $Q$, like for example:

$ WI_4P(n) = \openrule
         {\{\texttt{P}\mapsto \texttt{p-a}\}, [\forall \texttt{x}. \texttt{p-a} \neq \texttt{p-send(x)} ], \\
   (\texttt{m\_msg}\gets \texttt{s\_msg}, \texttt{m\_ec}\gets \texttt{s\_ec}+n, \texttt{s\_ec}\gets \texttt{s\_ec}+n)}
         {100 \OTWeakarrow {\texttt{p-a}} 210}
$

and
$ WI_4Q(n) = \openrule
         {\{\texttt{Q}\mapsto \texttt{q-b}\}, [\forall \texttt{x,y}. \texttt{q-b} \neq \texttt{q-recv(x,y)} ],\\
   (\texttt{m\_msg}\gets \texttt{s\_msg}, \texttt{m\_ec}\gets \texttt{s\_ec}+n, \texttt{s\_ec}\gets \texttt{s\_ec}+n)}
         {100 \OTWeakarrow {\texttt{q-b}} 210}
$

but also e.g.:

$ WI_{456*}P(n) = \openrule
        {\{\texttt{P}\mapsto \texttt{p-a}\}, [\forall \texttt{x}. \texttt{p-a} \neq \texttt{p-send(x)} ], 
    (\texttt{s\_msg}\gets \texttt{s\_msg}, \texttt{s\_ec}\gets \texttt{s\_ec}+n)}
  {100 \OTWeakarrow {\texttt{p-a}} 100}
        $

\bigskip
The following table give a summary of  WOTs, when sharing their names as much as possible.

\noindent\begin{tabular}{|@{\,}l@{\,}|@{\,}l@{\,}|@{\,}c@{\,}|}
\hline
    WOT name & Pairs of source states and target states & \# WOTs \\
    \hline
    $WI_1$ $WI_2$ $WI_\tau$ & $\{(s,s)| s \in \texttt{States of WOA}\} \cup \{(202,000)\} $ & 21 \\
    $WI_3(n)$ & \{(202,100),(000,100)\} & 2 \\
    $WI_{3a}(n)$ & \{(202,210),(000,210)\} & 2 \\
    $WI_{3b}(n)$ & \{(202,220),(000,220)\} & 2 \\
    $WI_{3c}(n)$ & \{(202,201),(000,201)\} & 2 \\
    $WI_4(n)$ $WI_4P(n)$ $WI_4Q(n)$ & \{(100,210)\} & 3 \\
    $WI_{4a}(n)$ $WI_4aP(n)$ $WI_4aQ(n)$ & \{(100,220)\} & 3 \\
    $WI_{456*}(n) $ $WI_{456*}P(n) $ $WI_{456*}Q(n) $ & \{(100,100)\} & 3 \\
    $WI_5(n)$ $WI_5P(n)$ $WI_5Q(n)$ & \{(210,220)\} & 3 \\
    $WI_{5a}(n)$ $WI_{5a}P(n)$ $WI_{5a}Q(n)$ & \{(210,100)\} & 3 \\
    $WI_{564*}(n) $ $WI_{564*}P(n) $ $WI_{564*}Q(n) $ & \{(210,210)\} & 3 \\
    $WI_6(n)$ $WI_6P(n)$ $WI_6Q(n)$ & \{(220,100)\} & 3 \\
    $WI_{6a}(n)$ $WI_{6a}P(n)$ $WI_{6a}Q(n)$ & \{(220,210)\} & 3 \\
    $WI_{645*}(n) $ $WI_{645*}P(n) $ $WI_{645*}Q(n) $ & \{(220,220)\} & 3 \\
    $WI_7(n)$ $WI_7P(n)$ $WI_7Q(n)$ & \{(210,201)\} & 3 \\
    $WI_{7a}(n)$ $WI_{7a}P(n)$ $WI_{7a}Q(n)$ & \{(220,201)\} & 3 \\
    $WI_{7b}(n)$ $WI_{7b}P(n)P$ $WI_{7b}Q(n)$ & \{(100,201)\} & 3 \\
    $WI_8(n)$ & \{(201,202),(201,000)\} & 2\\ 
    $WI_{8a}(n)$ & \{(210,202),(210,000)\} & 2 \\
    $WI_{8b}(n)$ & \{(220,202),(220,000)\} & 2 \\
    $WI_{8c}(n)$ & \{(100,202),(100,000)\} & 2 \\
   
    \hline
    \end{tabular}

    \bigskip
    That makes a total of 73 WOTs  in the open automaton for \symb{SimpleProtocolImpl}.

        \subsection{Details of the Bisimulation Checking}
         
We recall here the relation $\mathcal{R}$ that is the candidate for our weak bisimulation relation:

\bigskip
\noindent  \begin{tabular}{|c|c|l|}
\hline
    \symb{SimpleProtocolSpec} states & \symb{SimpleProtocolImpl} states & Predicate\\
    \hline
    b0 & $000$ & True\\
    b0 & $202$ & True\\
    b1 & $100$ & $\texttt{b\_msg = s\_msg} \land \texttt{b\_ec = s\_ec}$\\
    b1 & $210$ & $\texttt{b\_msg = m\_msg} \land \texttt{b\_ec = m\_ec}$\\
    b1 & $220$ & $\texttt{b\_msg = s\_msg} \land \texttt{b\_ec = s\_ec}$\\
    b1 & $201$ & $\texttt{b\_msg = r\_msg} \land \texttt{b\_ec = r\_ec}$\\
    \hline
    \end{tabular}

\bigskip
Consider the first triple <b0, 000, True>, we have to prove the following 6 properties, in which $OT<<WOT$ means that the (strong) open transition $OT$ is covered, in the sense
of definition \ref{def-Weak-bisim}, by the weak transition $WOT$ (it could be a set, but this will not be used here):

\bigskip
\begin{minipage}{0.2\linewidth} 	 
$SS_1 << WI_1$

$SS_2 << WI_2$

$SS_3 << WI_3$
\end{minipage}
\hspace{1cm}
\begin{minipage}{0.4\linewidth}
$SI_1 << WS_1$

$SI_2 << WS_2$

$SI_3 << WS_3$
\end{minipage}

\bigskip

         Note that if we were using the alternative weak bisimulation relation from Appendix \ref{app-WFH-equiv}, Lemma \ref{lem-rel-OT-WOT}, that is checking strong bisimulation between the corresponding 
weak automaton, we would have a more transitions coverage to examine, as we have 4 weak transitions for $b0$ in the \symb{SimpleProtocolSpec} weak automaton, and 7 WOTs (including 4 
parameterised WOTs) from 000 in the \symb{SimpleProtocolImpl} automaton.

Preliminary remarks:
  \begin{itemize}
    \item Both pNets trivially verify the ``non-observability''
      condition: the only vectors having $\tau$ as an action of a
      sub-net are of the form ``$< -, \tau, -> -> \tau$''.
    \item We must take care of variable name conflicts: in our example, the variables of the 2 systems already have different names, but the action parameters occurring in the 
transitions (m, msg, ec) are the same, that is not correct. Recall that we disambiguate the reference to the variable $m$ into $m1$ for \symb{SimpleProtocolSpec} and $m2$ for \symb{SimpleProtocolImpl}.
    \end{itemize}

  In our running example in page \pageref{subsubsection:runnig example}, we have shown the proof for one of the
transitions of ($\text{b0},  202,  True$), namely that $SS_3$ is covered by $WI_3(0)$.
We give here another example with $SS_1 << WI_1$, from the first triple ($\text{b0},  000,  True$). It includes less trivial predicates in the OTs:

  $ SS_1 = \openrule
  {\{\texttt{P}\mapsto \texttt{p-a1}\}, 
 [\forall \texttt{m1}. \texttt{p-a1} \neq \texttt{p-send(m1)} ], ()}
  {\text{b0} \OTarrow {\nounderline{\texttt{p-a1}}} \text{b0}}
  $
  
$ WI_1 = \openrule
  {\{\texttt{P}\mapsto \texttt{p-a2}\}, 
 [\forall \texttt{m2}. \texttt{p-a2} \neq \texttt{p-send(m2)} ], ()}
{000 \OTWeakarrow {\texttt{p-a2}} 000}$

  Let us check formally the conditions:
  
\begin{itemize}
  \item Their sets of active (non-silent) holes is the same: $J' = J_x = \{\texttt{P}\}$.
  \item Triple ($\text{b0},  000,  True$) is in $\mathcal{R}$.
  \item The verification condition \\
    $\forall fv_{OT}. \{ \Pred \land \Pred_{OT}\\
\hspace{1cm} \implies\!\!\! \displaystyle{\bigvee_{x\in X}\!\!
   \left[\exists fv_{OT_x}.
  \left( \forall j\in J_x. \vis{\beta_j}\!=\!\gamma_{jx}\! \land\! \Pred_{OT_x}
     \!\land\! \alpha\!=\!\alpha_x\! \land\!  
     \Pred_{s',t_x}\subst{\Post_{OT}\!\uplus\!\Post_{OT_x}}\right)\right]\}}$

\medskip Gives us:

$\forall \texttt{p-a1}. \{True \land \forall \texttt{m1}. \texttt{p-a1} \neq \texttt{p-send(m1)}\\
 \hspace{3mm} \implies \exists \texttt{p-a2}. 
(\texttt{p-a1} = \texttt{p-a2}
\land \forall \texttt{m2}. \texttt{p-a2} \neq \texttt{p-send(m2)}
\land \nounderline{\texttt{p-a1}} = \nounderline{\texttt{p-a2}} 
\land True\}$

\medskip That is trivially true, choosing \texttt{p-a2=p-a1} for each given \texttt{p-a1}.

\end{itemize}

  \bigskip
  All others pairs from this set are just as easily proven true.

\end{document}